\documentclass[journal=mamobx,manuscript=article]{achemso}

\usepackage{chemformula} 
\usepackage[T1]{fontenc} 
\usepackage{xr} 
\usepackage{makecell} 
\usepackage{titlesec}
\usepackage{setspace} 

\SectionNumbersOn

\makeatletter
\newcommand*{\addFileDependency}[1]{
  \typeout{(#1)}
  \@addtofilelist{#1}
  \IfFileExists{#1}{}{\typeout{No file #1.}}
}
\makeatother

\newcommand*{\myexternaldocument}[1]{%
    \externaldocument{#1}%
    \addFileDependency{#1.tex}%
    \addFileDependency{#1.aux}%
}

\myexternaldocument{supporting_info}

\titlespacing\section{0pt}{0pt plus 0pt minus 0pt}{0pt plus 0pt minus 0pt}

\author{Christopher C. Walker}
\affiliation{Department of Chemical and Biological Engineering\\
University of Colorado Boulder, Boulder, CO}
\author{Theodore L. Fobe}
\affiliation{Department of Chemical and Biological Engineering\\
University of Colorado Boulder, Boulder, CO}
\author{Michael R. Shirts}
\affiliation{Department of Chemical and Biological Engineering\\
University of Colorado Boulder, Boulder, CO}
\email{michael.shirts@colorado.edu}

\title{How cooperatively folding are homopolymer molecular knots?}

\abbreviations{}
\keywords{}

\begin{document}
\begin{singlespace}
\begin{tocentry}




\centering
\includegraphics[width=3.25in]{"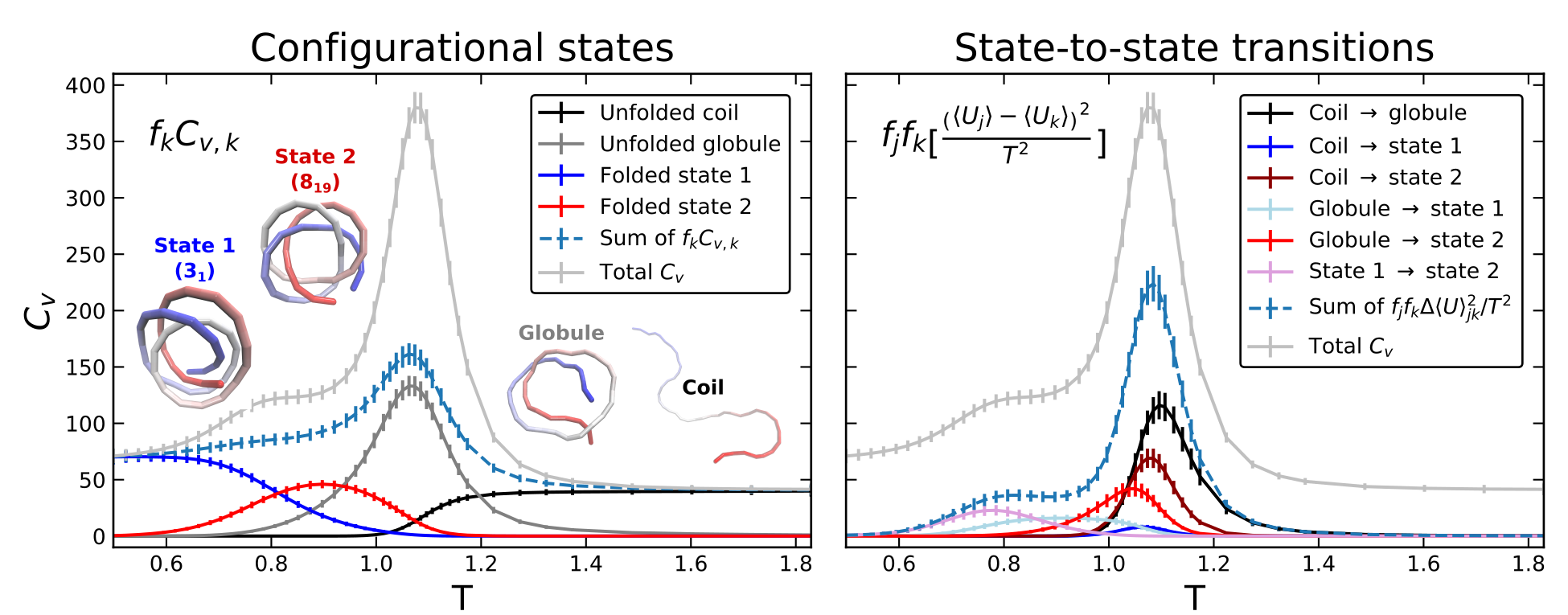"}

\end{tocentry}

\vspace{-24pt}
\begin{abstract}
Detailed thermodynamic analysis of complex systems with multiple stable configurational states allows for insight into the cooperativity of each individual transition. In this work we derive a heat capacity decomposition comprising contributions from each individual configurational state, which together sum to a baseline heat capacity, and contributions from each state-to-state transition. We apply this analysis framework to a series of replica exchange molecular dynamics simulations of linear and 1-1 coarse-grained homo-oligomer models which fold into stable, configurationally well-defined molecular knots, in order to better understand the parameters leading to stable and cooperative folding of these knots. We find that a stiff harmonic backbone bending angle potential is key to achieving knots with specific 3D structures. Tuning the backbone equilibrium angle in small increments yields a variety of knot topologies, including $3_1$, $5_1$, $7_1$, and $8_{19}$ types. Populations of different knotted states as functions of temperature can also be manipulated by tuning backbone torsion stiffness or by adding side chain beads. We find that sharp total heat capacity peaks for the homo-oligomer knots are largely due to a coil-to-globule transition, rather than a cooperative knotting step. However, in some cases the cooperativity of globule-to-knot and coil-to-globule transitions are comparable, suggesting that highly cooperative folding to knotted structures can be achieved by refining the model parameters or adding sequence specificity.
\end{abstract}


\section{Introduction}~\label{1}

Macro-scale knots have long held great practical and cultural significance.~\cite{Turner1996} In recent years, a renewed interest in molecular-scale knots has been driven by progress in synthesis approaches for increasingly complex artificial knots,~\cite{Fielden2017} and a quest to unravel the intriguing mystery of why knots exist in a now significant subset of proteins.~\cite{Dabrowski-Tumanski2019,Perego2019} Knotted molecules can display unusual chemical and mechanical properties.~\cite{Saitta1999,Arai1999} For example, knotted proteins' high resistance to enzymatic degradation likely contributed to their preservation throughout the course of evolution.\cite{Lim2015} Artificial molecular knots similarly can be tailored to retain their native structure under conditions far from native ones, either by kinetic trapping or chemically locking chain ends together by covalent bonds to form true closed knots. Potential applications for molecular knots include catalysis~\cite{Fielden2017}, ultra-stable vesicles for drug delivery~\cite{Newland2012,Coluzza2013}, and next-generation advanced materials such as molecular woven materials.~\cite{Liu2016,August2020,Leigh2021,Zhang2022}

Generally the term `molecular knot' is applied to both open and closed chains, though by definition a mathematically pure knot must be a closed loop in 3D space. The open chain homo-oligomer knots in this work are thus better classified as `physical knots'---however, we will use the term `knot' throughout for conciseness. One criteria for being a physical knot is if the chain is pulled at both ends, the knot will not become undone.~\cite{Virnau2010} Slipknots, in contrast, do come undone when pulled at both ends---for example, as in tied shoelaces. To classify open molecular knots, such as by their Alexander-Briggs type, an appropriate closure scheme must first be applied, in which the chain ends are joined together by some algorithm to form a closed loop. Nomenclature and classification of molecular knots are discussed in much greater detail in references \citenum{Virnau2010,Lim2015,horner2016,Fielden2017,Perego2019}.

Knots are commonplace in semi-flexible homopolymers---their existence has long been observed in simulations~\cite{Vologodskii1974}---but stable, well-defined, cooperatively folded structures are generally not achieved. Chain bending stiffness, or equivalently the persistence length, is known to play a key role in the propensity for knotting. In lattice simulations of self-avoiding homopolymers with 150 monomers,~\cite{Virnau2013} knotting probability was found to exhibit a maximum at intermediate bending stiffness. For weak bending stiffness, the occurrence of semi-flexible loops, which are then threaded to form knots, is inhibited by excluded volume interactions. At bending stiffness above the optimal range for knotting, chains are too extended to form loops with the necessary curvature. Knotting probability in polymers also increases with chain length, as demonstrated in early lattice Monte Carlo simulations~\cite{Vologodskii1974}, and in line with the well-known concept that chain entanglements increase with the degree of polymerization. More recently, Marenz~\cite{Marenz2016} and Majumder~\cite{Majumder2021} constructed phase diagrams in temperature and bending stiffness for homo-oligomers of 14 and 28 monomers using 2D replica exchange Monte Carlo simulations. Several knotted phases were identified, and it was found that the ratio of the bond length to equilibrium nonbonded distance is an important factor in determining whether the knotted phases occur. Wu et. al. computed a similar phase diagram in bending stiffness and temperature for longer 80 monomer homo-oligomers, which included toroidal phases.~\cite{Wu2018} In addition, Wu and coworkers performed a kinetic analysis, finding that the homo-oligomer knotting can proceed through several different pathways depending on both temperature and bending stiffness. Bending stiffness implemented through bead overlap, rather than by explicit angle potentials, was explored by Werlich et. al.~\cite{Werlich2017}. This stiffness induced by entropic effects, rather than energetic effects, also yielded rich phase behavior in temperature and degree of bead overlap (stiffness) for 20 and 40 monomer homo-oligomers, including several knotted phases.

In contrast to most homopolymer knots, knots occurring in proteins and DNA do have precise three-dimensional structures. Whereas the focus of homopolymer simulation studies has largely been identifying \textit{types} of knots, in the realm of biological polymers, the focus is instead on uncovering the folding pathway by which precise knotted motifs form. More and more knots and slipknots are being discovered in proteins by combing through the PDB database with knot-detecting software, with the KnotProt 2.0 database now approaching 2000 knot or slipknot entries.~\cite{Dabrowski-Tumanski2019} The most complex protein knot discovered thus far is a $6_1$ or `Stevedore' knot.~\cite{Bolinger2010} The existence of topologically complex knots in proteins poses a fundamental question about the evolution of proteins: why do the knots exist, when evolution might have disfavored knots due to higher chance of misfolding and much slower kinetics. For further reading on biological knots, a review by Lim and Jackson details the current understanding of the structure and function of knots in proteins, single and double-stranded DNA, as well as RNA pseudoknots.~\cite{Lim2015} A recent review by Perego and Potestio~\cite{Perego2019} provides a comprehensive summary of computational approaches used to study protein knots and other entanglements, and Fa{\'{i}}sca~\cite{Faisca2015} details key insights into protein knotting mechanisms gained through coarse-grained simulations.

Molecular modeling has already and will likely continue to play a key role in guiding the discovery of new molecular knot topologies and compatible chemistries. In simulations, the precise topology can be observed and characterized at every point in time, a significant challenge in experiments. A reduced design space in coarse-grained models compared to the all-atom level also allows for more efficiently scanning large parameter sets and uncovering fundamental principles. In one recent example, eight-crossing $8_{19}$ knots discovered from Monte Carlo simulations~\cite{Marenda2018} of the assembly of helical fragments with sticky ends were subsequently synthesized by an analogous circular triple-helicate scheme.~\cite{Danon2017} Recent achievements in the synthesis of non-natural molecular knots has allowed for increasingly intricate molecular topologies to be achieved. Most synthesis routes for molecular knots employ metal-template schemes, which are reviewed in Fielden et al.~\cite{Fielden2017} By using chiral building blocks, control over the knot handedness can be achieved,~\cite{Zhong2019} a key attribute for potential applications such as stereoselective catalysis. Molecular knots have been obtained without the use of stabilizing metal ions in only a few cases to date, using instead $\pi-\pi$ stacking or hydrogen bonding as stabilizing interactions, and these were limited to $3_1$ knots and in relatively low yields.~\cite{Leigh2021} Another successful strategy used a dynamic combinatorial library (DCL) approach based on naphthalenediimide monomer units a to achieve trefoil~\cite{Ponnuswamy2012} ($3_1$) and figure-of-eight~\cite{Ponnuswamy2014} ($4_1$) knots, relying on the hydrophobic effect to stabilize the knots. In a significant step towards achieving molecular knots through folding mechanisms alone, rather than via a series of reactions, Song et al.~designed pseudopeptides which fold into $3_1$ knots in acetonitrile in the presence of lanthanide ions.~\cite{Song2021} Moreover, Song and coworkers demonstrated that tuning linker chemistry and linker length between stretches of amino acid residues allows for control over knot tightness. Further understanding of folding principles of knots, though both simulations and experiments, will enable access to more topologies beyond the common toroidal knots such as $3_1$---for example, asymmetric knots.

We explore the use of bent angles in homo-oligomer coarse-grained models to create more stable and configurationally well-defined knotted structures than have generally been achieved previously. Few homo-oligomer or homopolymer simulation studies have considered knotting in systems with equilibrium backbone angles other than 180 degrees (fully extended), instead employing generic bending potentials with a single parameter, the stiffness.~\cite{Virnau2013,Wu2018,Marenz2016,Majumder2021} Specifically, we use stiff harmonic angle potentials in microsecond temperature replica exchange molecular dynamics (REMD) simulations, which allow for sufficient sampling of coarse-grained knotting. We study the effects of the equilibrium backbone bond-bending angle (Section~\ref{results_vary_theta}), backbone torsion stiffness (Section~\ref{results_vary_kt}), and the addition of side chain beads (Section~\ref{results_side chains}) on knot topology, stability and cooperativity through a series of simulations conducted and analyzed with the \texttt{cg\_openmm}~\cite{Walker2021} / \texttt{analyze\_foldamers}~\cite{analyzefoldamers2022} Python framework. We also herein develop clustering and state classification protocol for identifying the distinct configurational states that exist in these complex systems, which include coil, globule, and multiple knotted states, and extend the \texttt{cg\_openmm} code base to facilitate the analysis of such systems.

To assess the folding cooperativity of coarse-grained homo-oligomer systems containing multiple stable, configurationally well-defined knotted states, a more detailed thermodynamic analysis is required than for simple 2-state systems. Typically, in both experiments and simulations, cooperativity is evaluated by the characteristic width of the total heat capacity versus temperature curves,~\cite{Klimov1998} and the principal quantity of interest is the van't Hoff ratio, or `two-stateness` of the transition.~\cite{Zhou1999,Kaya2000} A broader peak signifies weakly cooperative folding, while a sharper peak signifies strongly cooperative folding. However, it has been shown that three-state, or even N-state folders can theoretically have a van't Hoff ratio of 1, the value for ideal 2-state folding.\cite{Shirley1995,Zhou1999,Bakk2004,Bakk2004a} To analyze the thermodynamic behavior of complex knotting systems with 3 or more configurational states, we deconstruct the total heat capacity curve into contributions from each individual configurational state, and contributions from each state-to-state transition. The sum of the individual state terms weighted by their population fraction represent a `baseline' heat capacity, while each transition term can be viewed as a cooperativity metric of that transition. 

For many of the knot systems studied, sharp peaks in the heat capacity versus temperature curve appear to indicate highly cooperative folding. We explore quantitatively how much of these heat capacity peaks are due to knotting steps, and how much arises from other transitions such as the coil-to-globule collapse. Through this process we uncover clues as to how to increase the cooperativity of globule-to-knot transitions.

\vspace{12pt}
\section{Methods}~\label{methods}

In Section~\ref{methods_cgmodel}, we describe the coarse-grained models and potential functions used. We briefly reiterate the temperature replica exchange protocol in Section~\ref{methods_replica_exchange}. In Section~\ref{methods_heat_capacity}, we describe a decomposition of the heat capacity versus temperature curve into contributions from each configurational state and state-to-state transition term. In Section~\ref{methods_clustering_classification}, we detail a RMSD-based clustering and classification scheme used to identify and sort structures into stable folded states and unfolded coil and globule states. Knot closure and classification is described in Section~\ref{methods_knot_classification}.

\subsection{Coarse-grained models}~\label{methods_cgmodel}
\begin{figure}[H]
\includegraphics[width=2.50in]{"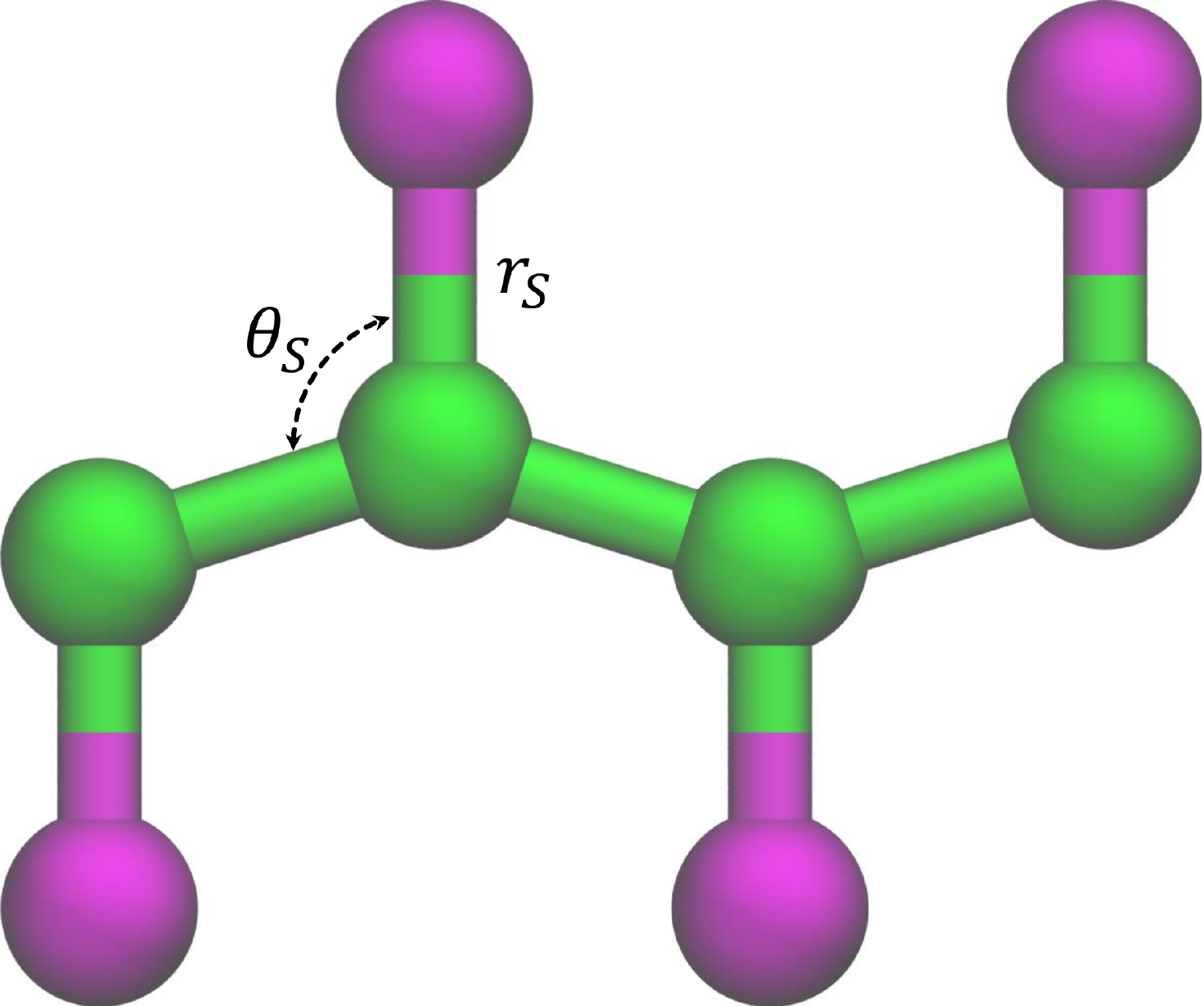"}

\includegraphics[width=2.50in]{"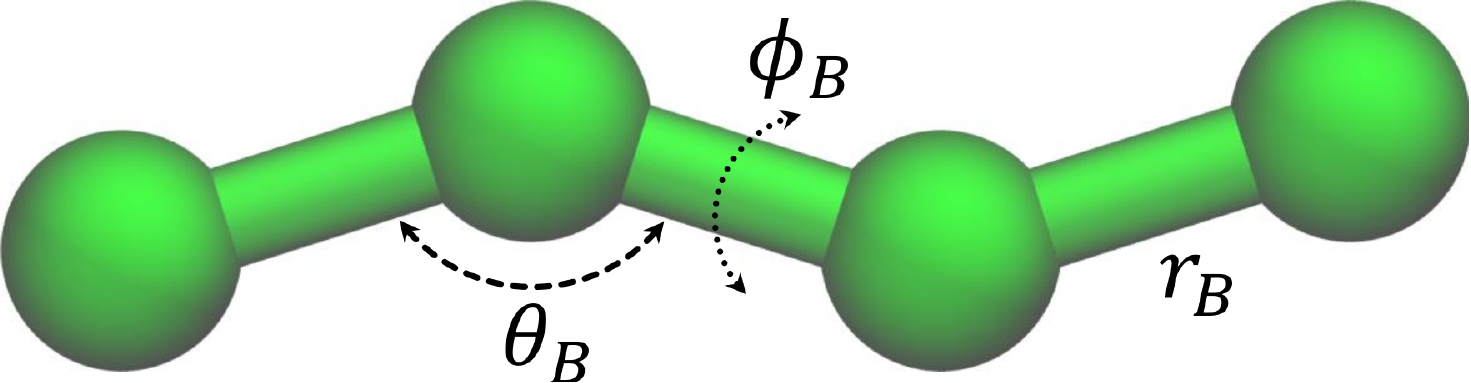"}
\caption{Depiction of the two types of coarse-grained models studied in this work: linear  (bottom) and 1-1 (top). Bond lengths between backbone beads ($r_B$), bond lengths between backbone and side chain beads ($r_S$), bond-bending angles  ($\theta_B$, $\theta_S$), and backbone pseudotorsion angle ($\phi_B$) definitions are labeled.}
\label{fig:cg_tetramer}
\end{figure}

We employ the \texttt{cg\_openmm} framework for building coarse-grained models, which is described in detail in previous work.~\cite{Walker2021}
Here, we study knotting in two types of minimalist coarse-grained models: a linear oligomer consisting of only backbone beads, and a 1-1 model in which each backbone bead has a single side chain bead attached (Fig. \ref{fig:cg_tetramer}). All oligomers in this work are 42 monomers in length.

Isotropic nonbonded interactions are given by a standard Lennard-Jones 12-6 function:
\begin{equation}
    U_{nonbond} = 4\varepsilon\left[ \left( \frac{\sigma}{r} \right) ^{12}- \left( \frac{\sigma}{r} \right) ^{6} \right]
    \label{eqn:LJ_eq}
\end{equation}
Throughout this work, we employ a reduced units system in which the $\sigma$ and $\epsilon$ Lennard-Jones parameters are the base length and energy units, respectively. Backbone and side chain beads are assigned identical Lennard-Jones parameters. Due to the small number of particles in the oligomers, no nonbonded cutoff is used. Nonbonded interactions between 1-2 and 1-3 neighbors are excluded, and nonbonded interactions between 1-4 neighbors are given full weight.
    
Bonds between coarse-grained particles are given by stiff harmonic potentials:
\begin{equation}
  U_{bond}=\frac{k_b}{2}(r-r_o)^2
  \label{eqn:bond_eq}
\end{equation}
In all cases equilibrium bond lengths ($r_{B,o}=r_{S,o}$) are set to 1, and bond stiffness force constants ($k_b$) are set to a dimensionless value of 250, which maintains roughly constant bond lengths throughout the knotting transitions. Note that setting $r_o = \sigma$ leads to a ratio of $r_o/r_{min} = 0.891$, where $r_{min}$ is the equilibrium distance in the Lennard-Jones 12-6 function, which has been found to be favorable for homopolymer knot formation.~\cite{Marenz2016,Majumder2021} 
Bond-bending angles acting on three consecutive beads are given by harmonic potentials:
\begin{equation}
  U_{angle}=\frac{k_a}{2}(\theta-\theta_o)^2
  \label{eqn:angle_eq}
\end{equation}
In equation~\ref{eqn:angle_eq}, $k_{a}$ is the angle stiffness force constant and $\theta_o$ the equilibrium angle value. A stiff angle force constant of 320 $\mathrm{rad}^{-2}$ is applied in all cases, including side chain angles. In the 1-1 models, side chain equilibrium angles are set to 95 degrees, which orients the side chain beads away from the backbone-backbone contacts in the knotted core. This harmonic potential differs from the cosine bending potentials used in previous homo-oligomer knotting studies~\cite{Marenz2016,Majumder2021}, which effectively used an equilibrium value of 180 degrees (fully extended). The models of Marenz and Majumder also included 1-3 nonbonded interactions in contrast to our models, which exclude them.

In certain experiments we applied a pseudotorsion potential of single periodicity which acts on four consecutive backbone beads: 
\begin{equation}
  U_{torsion}=k_t[1+\cos{(\phi-\phi_t)}] \label{eqn:torsion_eq}
\end{equation}
In equation~\ref{eqn:torsion_eq}, $k_t$ is the torsion stiffness force constant, and $\phi_t$ is the torsion phase angle. In all cases where torsion potentials are used, we set $\phi_t$ to 180 degrees, which results in a single minimum at 0 degrees corresponding to the \textit{cis} conformation. Torsions involving side chain beads are not considered, with such 1-4 pairs interacting by nonbonded potentials only. We note that using backbone torsion potentials along with stiff, wide backbone bending angles may cause numerical instability in OpenMM due to the singularity that occurs in the torsion potential calculation for 3 or more co-linear beads. Though we did not encounter such instability for angles up to 155 degrees with $k_a=320$, consideration of higher angles and/or weaker $k_a$ may require an alternative implementation, such as the solution proposed by Howard et. al. for removing the singularity.~\cite{Howard2017}

Force field parameters that remain fixed throughout all experiments are summarized in Table \ref{table:fixed_FF_params} below:

\begin{table}[H]
  \caption{Fixed force field parameters}
  \label{table:fixed_FF_params}
  \begin{tabular}{||c|c||}
    \hline
    Parameter & Value \\
    \hline
    $\sigma_B = \sigma_S$ & 1 \\
    $\varepsilon_B = \varepsilon_S$ & 1 \\
    $r_{B,o} = r_{S,o}$ & 1 \\
    $k_b$ & 250 \\
    $k_a$ & 320 ($\mathrm{rad}^{-2}$)\\
    $\theta_{S,o}$ & 95 ($\mathrm{degrees})$ \\
    $\phi_{B,o}$ & 0 ($\mathrm{degrees})$ \\
    \hline
  \end{tabular}
\end{table}    

Variable parameters include the torsion stiffness force constant ($k_t$), the backbone equilbrium angle ($\theta_{B,o}$), and the presence or absence of side chain beads.

\subsection{Replica Exchange Molecular Dynamics Simulations}~\label{methods_replica_exchange}
Slow folding kinetics and the possibility of numerous metastable intermediate states necessitate the use of enhanced sampling methods for accurately computing thermodynamic properties of molecular knots. Temperature replica exchange molecular dynamics (parallel tempering) simulations, which we use in this work, allow for simulating systems with significant free energy barriers, such as knotting, across a range of temperatures. In parallel tempering MD, a set of replicas run concurrently in separate simulation boxes, each with a specified temperature, is subject to configurational swaps attempted at fixed time intervals, which are either acccepted or rejected based on a Metropolis acceptance criterion. Transitioning from a lower to higher temperature allows the system to escape local energy minima, such as a partially folded knot. Moreover, the trajectory and energy files output from parallel tempering simulations allow for the computation of thermodynamic and structural properties as functions of temperature. Note that other schemes such as Hamiltonian replica exchange allow for sampling in force field parameter space, rather than, or in addition to, temperature space. For example, a 2D replica exchange approach in both temperature and bending stiffness force constant was successfully used by Marenz and Janke~\cite{Marenz2016} to generate phase diagrams of semi-flexible homo-oligomers. 

In this work we use the \texttt{cg\_openmm} Python framework developed by us for setting up temperature replica exchange MD simulations in OpenMM. \texttt{cg\_openmm} itself wraps around the \texttt{openmmtools} multistate framework.~\cite{Chodera2011,Friedrichs2012,Eastman2010a,Eastman2010b} We used the Signac~\cite{Adorf2018} Python workflow manager to streamline the simulation and analysis operations. Following protocol we found to provide effective sampling across folding transitions for helix systems, we first run a preliminary REMD simulation with logarithmically distributed temperatures, and then optimize the temperatures using the constant entropy increase (CEI) method~\cite{Sabo2008} for a much longer production run. Logarithmic temperature spacing is effective for systems with constant heat capacity,~\cite{Kofke2002} but may lead to poor sampling near transitions with large free energy barriers. 

All REMD simulations are run in the canonical ensemble and use a Langevin integrator with time step of 5 fs and collision frequency of 5 $\mathrm{ps}^{-1}$. Replica exchange moves are attempted every 1 ps. Ideally, exchange moves should be attempted as frequently as possible to maximize sampling efficiency across all states.~\cite{Sindhikara2008} However, we previously~\cite{Walker2021} found these parameters to provide a satisfactory balance between computational speed and sampling efficiency for simulations of helical foldamers.

Preliminary simulations with logarithmic temperature sets and 12 replicas are run for 100-200 ns, with the first 50-150 ns discarded as an equilibration period. Replica energies are then decorrelated using the pyMBAR \texttt{timeseries} module.~\cite{shirts_statistically_2008} Using heat capacity versus temperature data computed from the decorrelated replica energies via multistate Bennett acceptance ratio (MBAR) reweighting~\cite{shirts_statistically_2008}, we optimize the set of replica temperatures such that the entropy increase between all adjacent temperatures is equal. The optimal temperatures are obtained by solving the following system of integral equations, with the constraint that the lower and upper temperature bounds are maintained from the original set:

\begin{equation}
    {\int_{T_{i}}^{T_{i+1}} \frac{C_{V}(T)}{T} \,dT} = \frac{\Delta S}{N}
    \label{eqn:CEI_equation}
\end{equation}

In equation \ref{eqn:CEI_equation}, $C_{V}$ is constant volume heat capacity, $N$ is the number of replicas, and $\Delta S$ is the constant entropy increase between any two adjacent temperature pairs. Using the CEI-optimized set of temperatures, a production simulation is run for 1.2 $\mu s$ per replica, totaling 14.4 $\mu s$. The first 200 ns of this longer simulation are discarded as an equilibration period in all cases. Each production run, including all 12 replicas, takes approximately 1 week of wall clock time on a single NVIDIA V100 GPU.

\subsection{Heat capacity decomposition}~\label{methods_heat_capacity}
Heat capacity as a function of temperature is a useful indicator of a folding transition, which appears as a peak due to the presence of both folded and unfolded populations in the transition region. Applying the MBAR reweighting algorithm~\cite{shirts_statistically_2008} to the decorrelated replica energy time series, heat capacity may be computed as a  nearly continuous function of temperature, including at arbitrary intermediate temperatures not explicitly sampled. Specifically, heat capacity is computed in \texttt{cg\_openmm} from the finite difference derivative of potential energy with respect to temperature.

From the heat capacity curve as a function of temperature, information regarding both the (thermal) stability of a foldamer and the cooperativity of the folding transition can be extracted. Thermal stability is proportional to the melting point, which we take as the temperature at which the heat capacity is at a maximum. In systems with 2-state folding behavior, cooperativity is related to the characteristic width of the folding transition,~\cite{Klimov1998,Noel2012} which we define as the full-width half-maximum of the heat capacity peak.

\begin{figure}[H]
\includegraphics[trim=0in 0in 0in 0in, clip, width=6.5in]{"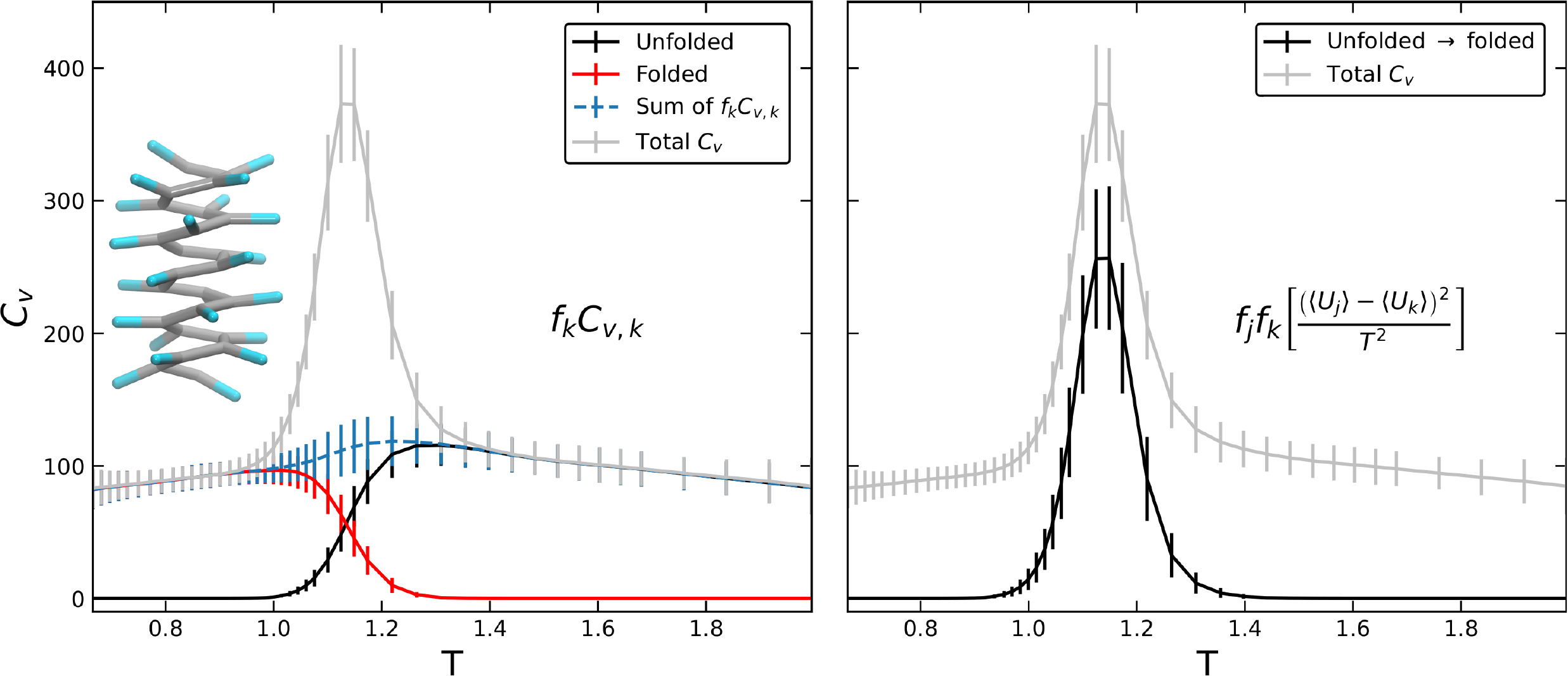"}
\caption{Example heat capacity decomposition for a cooperative helix foldamer simulated in previous work\cite{Walker2021}: the sum of weighted $C_{v,k}$ for unfolded and folded conformational states is flat within uncertainty (left), and virtually all of the total $C_v$ peak arises from the unfolded-folded transition term (right). Total $C_v$ is shown in gray for reference. Uncertainties are the standard deviation from bootstrapping the energies.}
\label{fig:cv_decomp_helix}
\end{figure}

For systems with more than two conformational states (i.e., where multiple folded states and/or intermediates exist), heat capacity versus temperature curves may contain multiple peaks corresponding to multiple cooperative transitions. Likewise, what may appear as a smooth, single peak in heat capacity may actually arise from the sum of multiple different transitions that occur over similar temperature ranges.~\cite{Zhou1999} We use the standard statistical mechanical definition of heat capacity ($C_v=\frac{d\langle U\rangle}{dT}$) to deconstruct the total heat capacity versus temperature curve into contributions from each conformational state and each transition between states (equation~\ref{eqn:cv_decomp_equation}). This analysis allows for a clearer picture of the cooperativity of each individual state transition. In equation~\ref{eqn:cv_decomp_equation}, $f_k$ is the fraction of structures of the total population belonging to conformational state $k$, $C_{v,k}$ is heat capacity computed using only the samples belonging to state $k$, and $\langle U\rangle_k$ is the potential energy expectation for state $k$.
For a detailed derivation of the heat capacity decomposition, refer to Appendix A.
\begin{eqnarray}
\frac{d\langle U\rangle}{dT} &=& \sum_k {f_k C_{v,k}} + \frac{1}{k_B T^2}\left(\sum_k\sum_{j > k} f_k f_j \left(\langle U\rangle_j -\langle U\rangle_k\right)^2\right)  
\label{eqn:cv_decomp_equation}
\end{eqnarray}
The first term in equation~\ref{eqn:cv_decomp_equation} is essentially a weighted sum of the `baseline' heat capacites $C_{v,k}$ of each individual configurational state $k$. The second term represents the sum of the energy changes associated with transitions between each possible configurational state pair. Previous derivations have divided heat capacity into similar baseline and transition components,~\cite{Shirley1995,Zhou1999} but to our knowledge the succinct form of the state-to-state transition terms in equation~\ref{eqn:cv_decomp_equation} is a new result. It is important to note that this decomposition holds true for \textit{any} definition of configurational states. Well-defined configurational states, distinguishable by one or more order parameters, should ideally have constant or nearly constant $C_{v,k}$ as a function of temperature, and the total observed $C_v$ peak should arise solely from the transition terms. Poorly defined states, which do not capture existing well-defined transitions, will instead have large changes within individual $C_{v,k}$ terms and small transition terms. As an example, we computed the heat capacity decomposition terms for a highly cooperative 1-1 helix foldamer model simulated in a previous study~\cite{Walker2021} (Figure~\ref{fig:cv_decomp_helix}). The sum of the weighted folded and unfolded state heat capacities ($f_kC_{v,k}$) amounts to a flat baseline heat capacity of the system (at least over the range of $T$ where folded/unfolded configurations are present) while the unfolded-folded transition term comprises nearly the entirety of the observed heat capacity peak.

\subsection{DBSCAN Clustering and state classification}~\label{methods_clustering_classification}
\subsubsection{DBSCAN Clustering}~\label{methods_clustering}
Stable folded states are first identified by clustering the replica exchange trajectories based on the RMSD's among all structures. This allows for identifying groups of frequently visited structures. As in our previous study on simple helix models, we use the DBSCAN~\cite{Ester1996,Schubert2017} clustering algorithm, as implemented in the \texttt{scikit-learn} Python package~\cite{Pedregosa2011}. The MDTraj~\cite{McGibbon2015} Python package is used to compute the RMSD's and perform other structural analyses. As a density-based clustering method, DBSCAN discards data points with low density of neighbors as `noise', leaving only the low-temperature higher density data points to be grouped into clusters. We also apply a pre-clustering filtering step, in which a user-specified fraction of data to filter out is met by optimizing a distance cutoff parameter and number of near neighbors within said cutoff (in RMSD space). This effectively discards a portion of the noise data points which have no or few similar structures.

We found that in all cases, a negligible amount of structures from the 6 highest temperature states were assigned to a cluster, and so we only input the lowest 6 state trajectories into the DBSCAN algorithm. Due to the large size of the trajectories (1 million samples/replica), we selected every $200^{th}$ frame for clustering, effectively sampling every 200 ps. We considered end-to-end symmetry when computing the RMSD's for clustering, such that structures differing only in particle index labeling are treated as identical. However, we treated right and left-handed enantiomers as distinct to confirm the adequacy of sampling in our replica exchange simulations.

In all but one case studied here, more than one stable folded state was observed in the replica trajectories, and in some cases, knotted states with nearly identical contact patterns but different crossing patterns were identified. In addition, both right and left-handed enantiomers of chiral molecular knots are sampled due to the symmetric torsion potential used. Great care must therefore be taken in selecting appropriate DBSCAN clustering parameters. We established the following protocol to identify the different knotted states.

We first filtered out 50\% of the data with fewest near neighbors in RMSD space, and fixed the value of the DBSCAN core point minimum number of samples parameter at 50. Then, we tuned the DBSCAN $\epsilon$ radius parameter to maximize the number of topologically distinct clusters without any duplicate cluster medoids containing the same crossing patterns but differing in minor structural aspects. Such duplicate clusters were determined by comparing their knot types (by both direct closure and the distributions from stochastic closure) and by visual inspection of the medoid structures. Too small an $\epsilon$ parameter can lead to such duplicates, while too large an $\epsilon$ parameter may fail to distinguish between knots with different crossing patterns. A complication for systems with intermediate states is that they may exist predominantly at intermediate temperatures, which inherently will have higher RMSD due to thermal fluctuations than other folded states existing mainly at low temperatures. In particularly challenging cases where clusters found using the above procedure still contained samples with negative silhouette scores, indicating the sample is more similar to a different cluster, or when known stable structures observed in the replica trajectories were not identified, we also adjusted the minimum number of samples parameter and filtering percentage. A table listing the DBSCAN parameters used for each simulation is provided in the supporting information (Table \ref{table:cluster_params}).

Once clusters have been identified, medoids, the structures most representative of each cluster, are selected as the structure whose sum of RMSD's to other members in the cluster is smallest.~\cite{Walker2021} All medoids are then energy-minimized in OpenMM, to better represent the true native configurations.

\subsubsection{Conformational state classification}~\label{methods_classification}
Previously, for 2-state helix/unfolded systems studied with the \texttt{cg\_openmm} framework, a native structure was identified based on the helical backbone native contacts, and structures were classified as `folded' or `unfolded' based on whether the native contact fraction was greater or less than the value at the center of a sigmoid function fit to the average native contact fraction versus temperature curve.~\cite{Walker2021} For systems with more than two conformational states, fitting a single native contact fraction expectation curve is not sufficient to characterize the unfolded to folded transition(s). Nor is it sufficient to characterize the folded-to-unfolded transitions by knot classification alone, given that multiple structural variants of the same knot type may exist, and knot closures often result in ambiguous solutions, especially when applied to random globules.

First, reference medoid structures are identified from clustering and then energy-minimized. Since right and left-handed forms have equivalent contact patterns, we selected for each folded state the medoid whose cluster is the larger of the two, with the reasoning that larger clusters will produce medoids closer to the `true' native state. Contact pairs are then defined within each reference medoid, based on a common fixed distance cutoff of $2\sigma$. Contact maps for all medoids are provided in the supporting information, Section~\ref{SI_clustering}. Contact fractions are then computed for all structures in the REMD trajectories with respect to each reference medoid. We define $Q_k$ as the contact fraction with respect to the reference medoid for conformational state $k$. Then, the average native contact fraction versus T curves are computed (using the entire dataset) for each conformational state reference using MBAR. These curves provide reference values of an `unfolded' contact fraction with respect to each medoid, taken as the value of $Q_k$ at the highest temperature simulated. For side chain models, only backbone beads are included in the native contact fraction. A multiplicative tolerance factor of 1.3 was applied when determining whether the `native' pairs are close enough to their native distances to be considered contacting. Previously, this tolerance factor was found to provide a maximal distinction between Q in the unfolded and folded states for simple 1-1 helix models. Here we also considered tolerance factors of 1.1 and 1.5, as well as a stricter $1.25\sigma$ contact distance cutoff for the 1.3 tolerance factor. The effect of these contact parameters on population distributions and heat capacity decompositions are compared in the supporting information for selected experiments in Section~\ref{SI_contact_tolerance}. Generally, changes to the contact definitions transfer populations of the unfolded globule state to/from the folded states that exist at intermediate temperatures, and have a negligible effect on the relative populations of the knotted states at low temperature. However, the lack of a clear dividing order parameter between unfolded globule and folded/intermediate states in itself points to the relatively weak cooperativity of the homopolymer knots compared to the helices---i.e., that there is a significant population of partially folded states preventing clear configurational state classification.

We found that in some cases, folded structures with different topologies had nearly identical sets of contacts (see for example, supporting information Section~\ref{SI_medoids_theta155_kt04}). Instead of assigning a folded state based on the contact fraction with respect to each reference medoid, we used the RMSD with respect to each medoid reference. Recall that these medoids were determined from RMSD clustering in the first place. In computing RMSD, both end-to-end (particle index labeling) symmetry and mirror image symmetry are taken into account, in which the smallest of the 4 RMSD's is used.

Next, all production frames in the replica exchange trajectory set are classified into configurational states as follows: a structure is classified into folded state $k$ if the RMSD to the reference medoid for state $k$ is the smallest among all folded states, and $Q_k$ is greater than or equal to a critical contact value $Q_{k,o}=\frac{1+Q_{k,u}}{2}$, where $Q_{k,u}$ is the unfolded reference value for state $k$. In other words, a structure is folded if $Q_k$ is closer to 1 than to the unfolded reference value. If this critical contact value is not met, the structure is classified as unfolded.

The unfolded state in all systems studied here consists of both compact globule structures with low contact fractions, and extended coil structures with low contact fractions. We therefore subdivided the unfolded configurations into `unfolded coil' and `unfolded globule' states based on a radius of gyration threshold ($R_{g,o}$) set as the midpoint of a sigmoidal fit to the mean radius of gyration versus T data within the unfolded set of configurations (for further details, see supporting information, Section~\ref{SI_Rg_vs_T}). The conformational state classification scheme for any arbitrary amount of folded states is summarized in Table \ref{table:state_classification}.

\begin{table}[H]
  \caption{Configurational state classification scheme}
  \label{table:state_classification}
  \begin{tabular}{||m{3.5cm}|m{4cm}|m{7.5cm}||}
    \hline
    \textbf{Conformational state} & \textbf{Criteria} & \textbf{Explanation}\\
    \hline
    Folded state $k$ & \makecell{$RMSD_k > RMSD_{i\neq k}$ \\ $Q_k \geq Q_{k,o}$} & RMSD to state $k$ reference medoid is smallest, contact fraction $Q_k$ at or above folded cutoff\\
    \hline
    Unfolded globule & \makecell{$RMSD_k > RMSD_{i\neq k}$ \\ $Q_k < Q_{k,o}$ \\ $R_g \leq R_{g,o}$} & RMSD to state $k$ reference medoid is smallest, contact fraction $Q_k$ below folded cutoff, Rg at or below value at sigmoid center\\
    \hline
    Unfolded coil & \makecell{$RMSD_k > RMSD_{i\neq k}$ \\ $Q_k < Q_{k,o}$ \\ $R_g > R_{g,o}$} & RMSD to state $k$ reference medoid is smallest, contact fraction $Q_k$ below folded cutoff, Rg above value at sigmoid center\\
    \hline
  \end{tabular}
\end{table}

Once all structures are classified, it is then possible to compute population distributions as a function of temperature, the complete heat capacity decomposition, and the free energy, entropy, and enthalpy changes for all configurational state transitions.

To verify that each of the knots identified are indeed stable, distinct states, we visualized a superposition of member structures of each state, shown in the supporting information, Section~\ref{SI_conf_state_ensembles}, which shows clearly defined ensembles across all parameter sets, with the possible exception of $\theta_{B,o}=140^\circ$. More challenging is the validation of the native contact cutoff criteria separating folded versus unfolded structures. For the 1-1 helix foldamers, we optimized a multiplicative native contact distance tolerance factor to maximize the difference in native contact fraction in the folded and unfolded states.~\cite{Walker2021} For systems with 3 or more configurational states, a different strategy is needed, since contact fraction expectations computed for the entire population with respect to a single native reference structure is insufficient. One could in principle compute the Q versus T expectation curves for each state transition, using only the structures involved in that transition, and where Q is computed with respect to the folded or intermediate state of interest. This, however, would require knowledge of the specific knotting pathway, such as by a series of kinetics experiments at relevant transition temperatures.

\subsection{Knot closure and classification}~\label{methods_knot_classification}
We use the Python package Topoly~\cite{Dabrowski-Tumanski2021} to identify the Alexander-Briggs knot type of all medoid structures based on the HOMFLY polynomial, also referred to as HOMFLY-PT~\cite{Freyd1985,Przytycki1987}. Importantly, the HOMFLY knot invariant distinguishes between different chiral enantiomers of all knot types encountered in this work, and is unique for all knots with 8 crossings or fewer.~\cite{Ramadevi1994} Several different closure strategies have been devised for molecular knots, and the resultant knot type can vary depending on this choice.~\cite{Virnau2010,Perego2019,Millett2013}. We computed knot types using direct closure and also by a stochastic closure method in which the end beads are connected to separate random points on an outer sphere, which are then joined by an arc along the sphere. For further discussion of knot closure, refer to the supporting information, Section~\ref{SI_knot_closure}, and the Topoly~\cite{Dabrowski-Tumanski2021} online documentation. In most cases, the knot types from both closure methods are in agreement. While we present the direct closure classifications in Section~\ref{methods}, results for the stochastic closures for all medoids are shown in supporting information, Section~\ref{SI_clustering}.

We reiterate that Alexander-Briggs knot classification is not a sufficient criteria, or perhaps even a necessary one, for determining whether a structure belongs to a specific `folded' state. Molecular knots of the same knot type, but differing in conformation, can rearrange to a singular structure by a series of Reidemeister moves, but the contact patterns and energies of the original structures may differ significantly. Here we simply use the knot type to characterize the medoids from RMSD clustering, rather than as a means for sorting structures into conformational states. Our approach to knotted structure characterization is in contrast to that in references~\citenum{Marenz2016} and~\citenum{Majumder2021}, for example, where a knot order parameter derived from the Alexander polynomial was used to identify phases encompassing ensembles of conformations, rather than precise 3-dimensional structures.

\vspace{12pt}
\section{Results and discussion}~\label{results}

\subsection{Effect of equilibrium angles on knot topology}~\label{results_vary_theta}
Small changes to the equilibrium backbone bond-bending angle ($\theta_{B,o})$ create a diverse set of stable knot topologies, including $3_1$, $5_1$, $7_1$, $8_{19}$ knots and the unknot, $0_1$. In order to find these knots, we ran a series of REMD simulations for linear homo-oligomers containing 42 monomers and with backbone torsion potentials off ($k_t=0$) for $\theta_{B,o}$ = 140, 145, 150, and 155 degrees, with all other force field parameters fixed as listed in Table \ref{table:fixed_FF_params}. Using these parameters, angles above and below the specified range did not yield stable knots: for $\theta_{B,o}=160^\circ$, the extended conformation is favored over the knotted form; $\theta_{B,o}=135^\circ$ did not result in any stable folded structures. Manipulation of $k_a$ could likely produce stable knots for those $\theta_{B,o}$, but due to the already complex analysis of the knotted systems, we did not attempt to produce a complete phase diagram in $k_a$ and $\theta_{B,o}$ space.

The knots discovered from clustering the REMD trajectories for varying $\theta_{B,o}$ are summarized in Figure \ref{fig:medoids_all_theta} below, along with their knot types computed from direct closure and HOMFLY invariants using the Topoly~\cite{Dabrowski-Tumanski2021} package. For simplicity, only right-handed enantiomers are displayed, though we also observed left-handed enantiomers in all cases. We observed roughly equal populations of right and left-handed enantiomers, indicating satisfactory sampling of the various knotted states. This is demonstrated by the population versus temperature distributions of right and left-handed clusters, shown in the supporting information Section~\ref{SI_cluster_populations}. Note that the folded state numbers for each $\theta_{B,o}$ are separate designations. Complete charts of both right and left-handed enantiomers, along with the contact maps of each, are shown in the supporting information, Section~\ref{SI_clustering}. In addition to the simple unknots ($0_1$) and trefoils ($3_1$) found for $\theta_{B,o}=155^\circ$, we discovered distinct trefoils for $\theta_{B,o}$ = 140, 145, and 150 degree systems, as well as more complex $5_1$, $7_1$, and $8_{19}$ knots. Interestingly, some of the $\theta_{B,o}=140^\circ$ knotted structures are asymmetric, which is in contrast to most experimentally synthesized molecular knots.~\cite{Lim2015,Fielden2017} Subtle differences in crossing patterns characterize some state divisions, such as states 2 versus 3 for $\theta_{B,o}=140^\circ$ or states 2 versus 4 for $\theta_{B,o}=155^\circ$.

The knot types discovered here are consistent with those reported by Polles et al.~\cite{Polles2015} from simulations of the self-assembly of rigid helical fragments with sticky ends, further supporting the notion that $3_1$ and $8_{19}$ torus knots are inherently more `designable' than other knots. The rigid helical fragment systems bear some similarity to our stiff harmonic angle oligomers, and also suggest that knots may form for other equilibrium torsion angles. Such torsion potentials with non-zero equilibrium angles could also be used to favor one enantiomer over the other.

\begin{figure}[H]
\includegraphics[trim=0in 2.5in 2.5in 0in, clip, height=5in]{"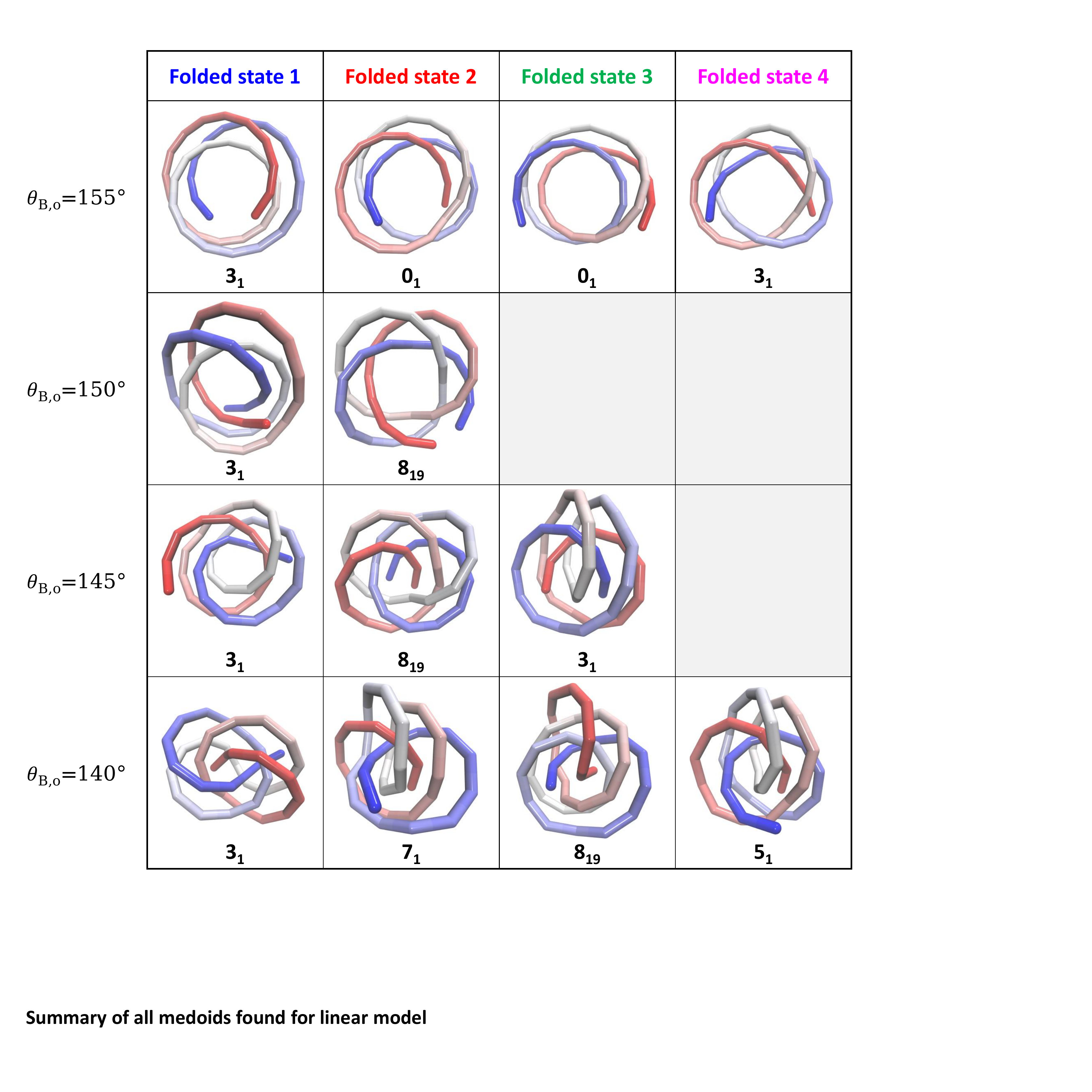"}
\caption{The topology of knotted states identified from RMSD clustering varies significantly with small changes to $\theta_{B,o}$, with both knot complexity and diversity increasing as $\theta_{B,o}$ is lowered. Structures are colored by residue index. Knot types shown are from direct closure and the HOMFLY polynomial invariant. Only right-handed enantiomers are depicted, though left-handed enantiomers are also sampled.}
\label{fig:medoids_all_theta}
\end{figure}

\begin{figure}[H]
\includegraphics[trim=0in 0in 0in 0in, clip, width=6.5in]{"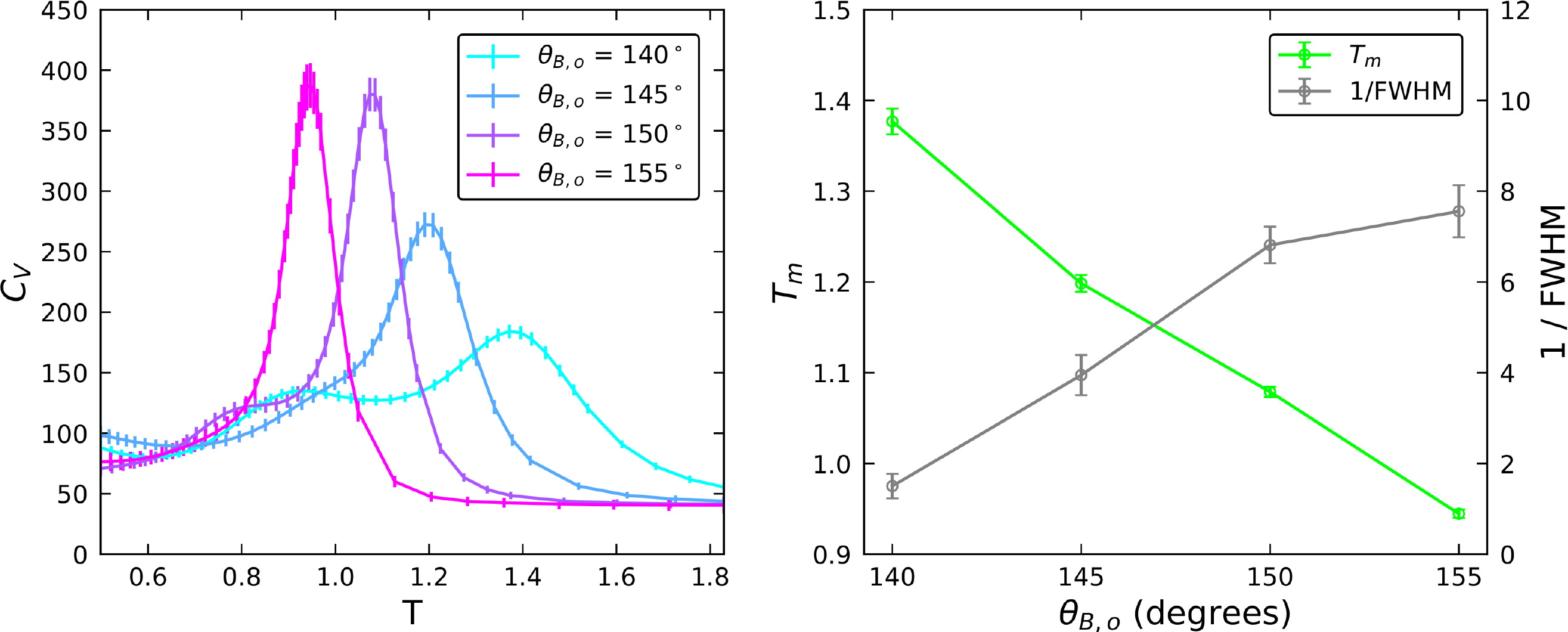"}
\caption{Total heat capacity versus temperature curves (left) and the corresponding melting point and inverse full-width half-maximum (right) indicate that knots with 150 and 155 degree equilibrium angles undergo a more cooperative overall transition from unfolded coil to knotted states than the 140 and 145 degree systems. The 150 and 155 degree systems are also less stable, manifested by lower melting points, likely a consequence of having fewer knot crossings, and thereby fewer contact pairs, than the 140 and 145 degree systems. Uncertainties are the standard deviation from bootstrapping the energies.}
\label{fig:cv_full_vary_theta}
\end{figure}

Overall heat capacity versus temperature curves for each of the backbone equilibrium angles studied (Figure~\ref{fig:cv_full_vary_theta}) at first suggest highly cooperative transitions from unfolded coil to knotted states for the wider $\theta_{B,o}=150,155$ degree angles, and weakly cooperative transitions for the smaller $\theta_{B,o}=140,145$ degree angles. Melting points, taken as the temperature at which heat capacity is maximal, imply that thermal stability of the knots increases with decreasing backbone angle. However, as previous theoretical and simulation studies have demonstrated, a single, smooth heat capacity peak does not necessarily translate to 2-state folding behavior.~\cite{Zhou1999,Bakk2004} Indeed, as seen from the multiple folded states for each $\theta_{B,o}$ in Figure~\ref{fig:medoids_all_theta}, this seemingly simple heat capacity behavior warrants a more detailed analysis through the heat capacity decomposition.

\begin{figure}[H]
\includegraphics[trim=0in 0in 0in 0in, clip, width=6.5in]{"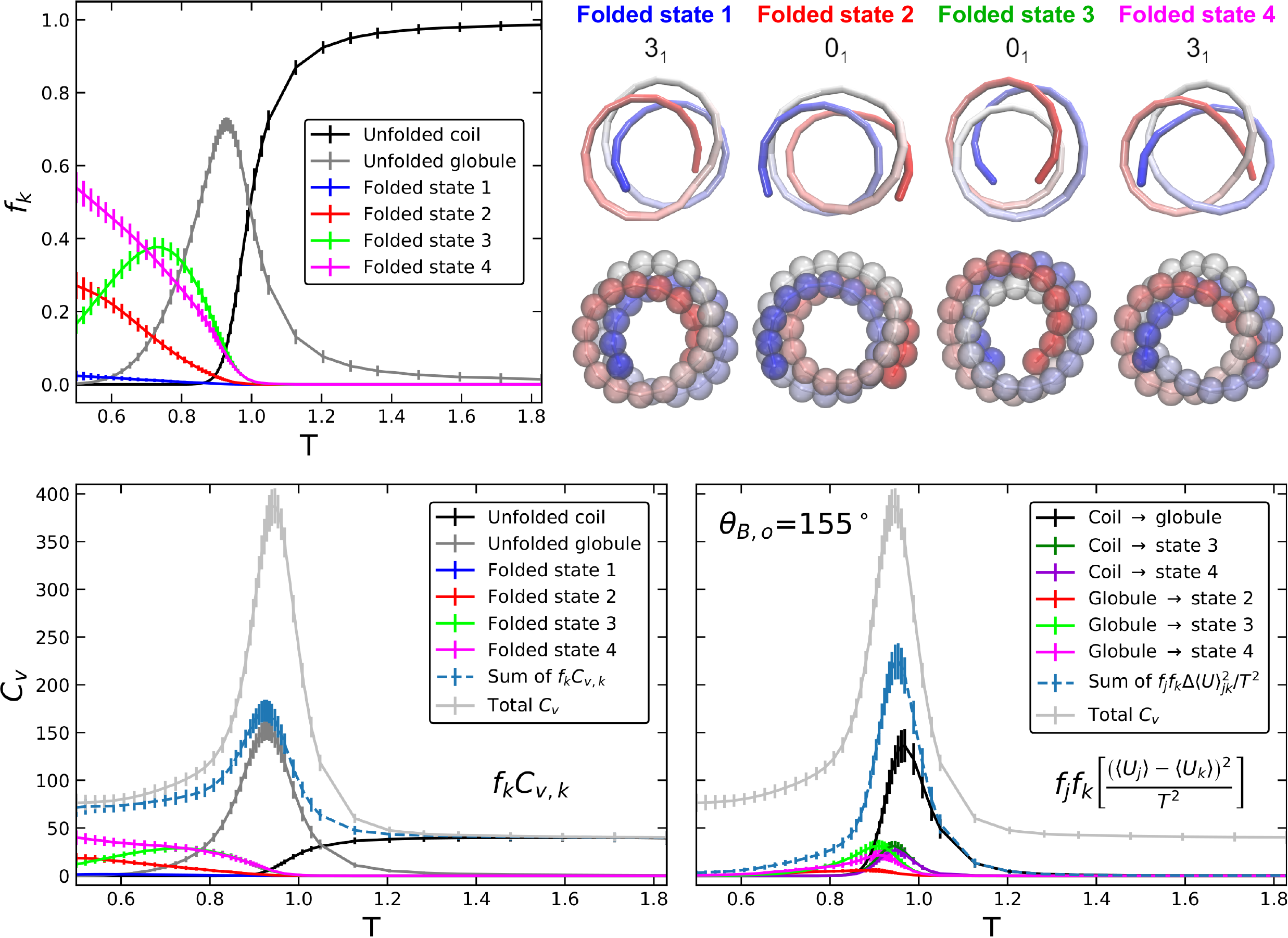"}
\caption{Heat capacity decomposition (bottom) for $\theta_{B,o}=155^\circ$ and $k_t=0$ shows dominant contributions from the random globule state and coil-to-globule transition to total heat capacity (light gray). The sums of each set of terms in equation \ref{eqn:cv_decomp_equation} are shown in dashed blue. Population fraction versus temperature distributions (top left) show no clear native state at low temperature, with all 4 folded states in coexistence. All error bars are the standard deviation propagated from bootstrapping the energies. For clarity, only transition terms with peak values > 5 are displayed.}
\label{fig:theta155_summary}
\end{figure}

\begin{figure}[H]
\includegraphics[trim=0in 0in 0in 0in, clip, width=6.5in]{"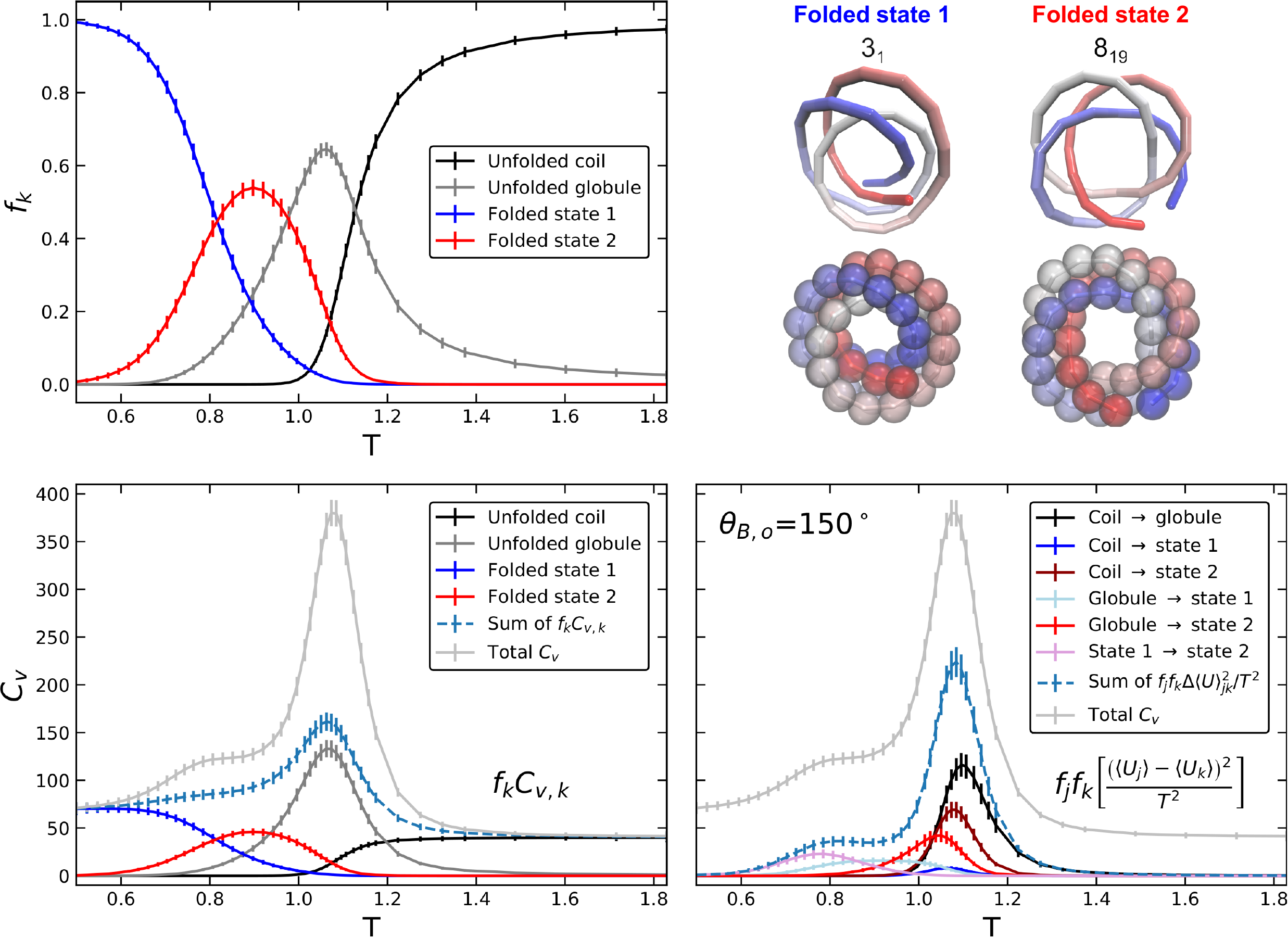"}
\caption{Population fraction versus temperature distributions for each configurational state for $\theta_{B,o}=150^\circ$ indicate a native $3_1$ knot (folded state 1) is dominant at low temperature. As temperature is increased, the $3_1$ knot gives way to a $8_{19}$ knot (folded state 2). In the heat capacity decomposition, the globule state and coil-to-globule transition terms are dominant, but the globule/state 2 and coil/state 2 coexistence terms also contribute significantly to the primary total heat capacity peak. A secondary heat capacity peak (shoulder) centered at $T\sim0.8$ arises from the state 1/state 2 and globule/state 1 transition terms, suggesting a mildly cooperative transition between the $3_1$ and $8_{19}$ knots. All error bars are the standard deviation propagated from bootstrapping the energies.}
\label{fig:theta150_summary}
\end{figure}

\begin{figure}[H]
\includegraphics[trim=0in 0in 0in 0in, clip, width=6.5in]{"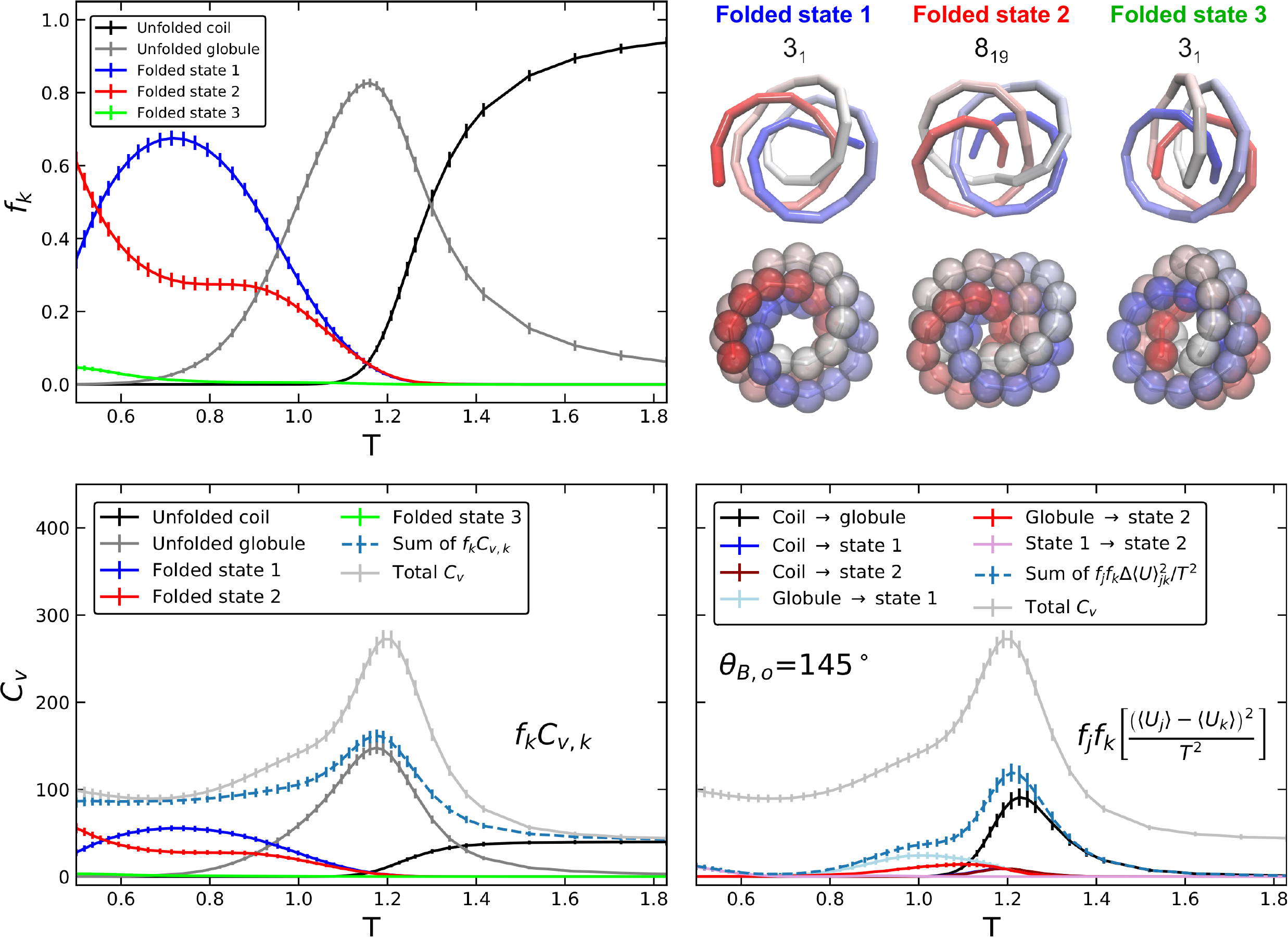"}
\caption{For the $\theta_{B,o}=145^\circ$ system, with decreasing temperature the population of the $8_{19}$ knotted state increases, suggesting it is likely the native state. Two distinct $3_1$ knots are identified, but folded state 1 is heavily favored over state 3. The heat capacity decomposition again shows the dominance of the individual globule state and coil-to-globule transition terms. The shoulder below the primary $C_v$ peak is found to originate from two globule-to-knot transition terms. For clarity, only transition terms with peak values > 5 are displayed.}
\label{fig:theta145_summary}
\end{figure}

\begin{figure}[H]
\includegraphics[trim=0in 0in 0in 0in, clip, width=6.5in]{"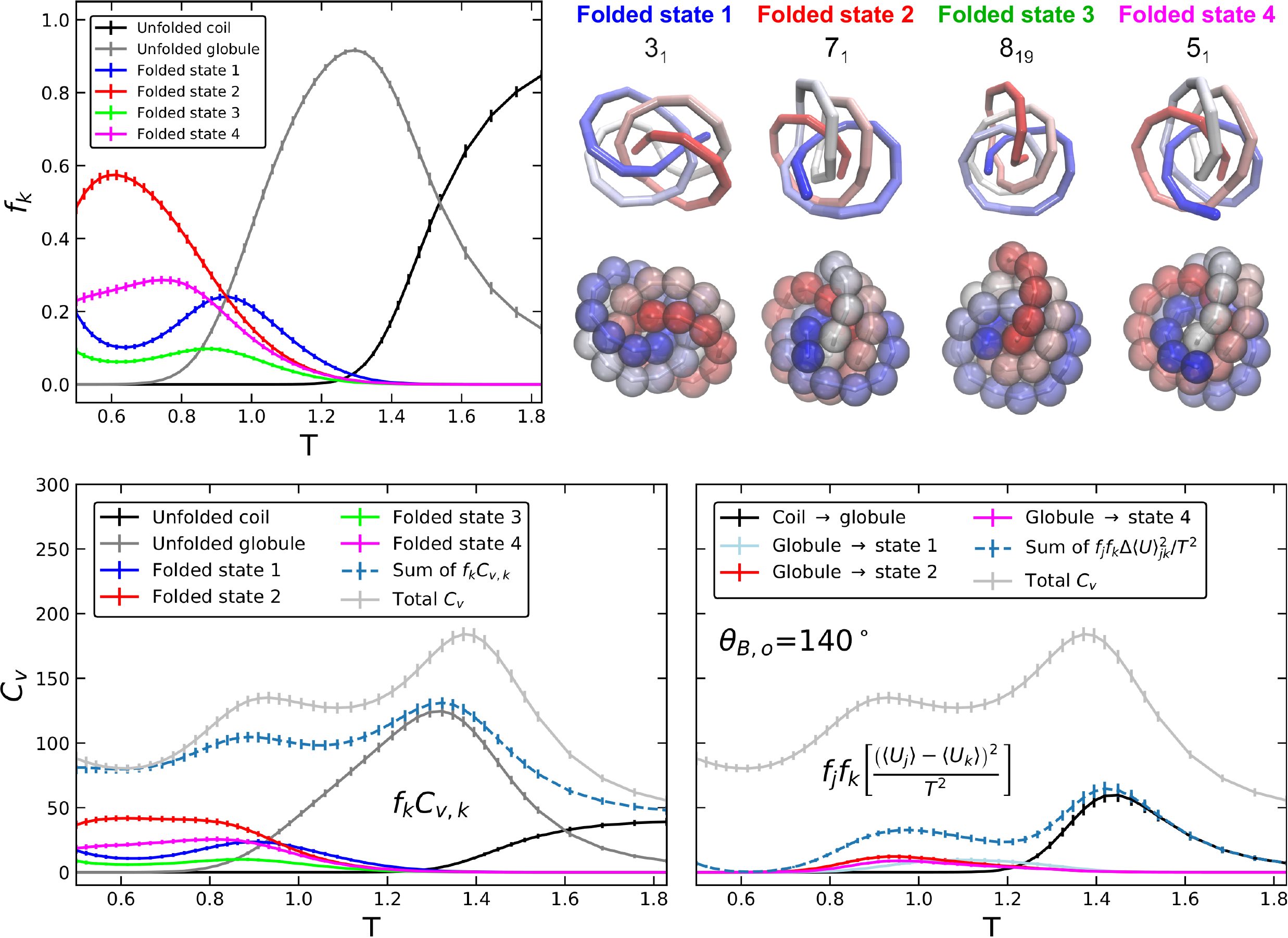"}
\caption{Population fraction versus temperature curves for $\theta_{B,o}=140^\circ$ show coexistence of $3_1, 5_1, 7_1$, and $8_{19}$ knots across low and intermediate temperatures. Due to similar energies and small $f_jf_k$ scaling factors among the knotted states, only the coil-to-globule term contributes significantly to the coexistence contributions. The very broad, shallow contributions of the globule/knot transition terms suggest marginal or no cooperativity. For clarity, only transition terms with peak values > 5 are displayed.}
\label{fig:theta140_summary}
\end{figure}

The heat capacity decompositions for each $\theta_{B,o}$ (Figures~\ref{fig:theta155_summary}-\ref{fig:theta140_summary}) reveal that in all cases the coil-to-globule transition term is the dominant contribution to total heat capacity, rather than highly cooperative knotting steps. Left-hand-side plots contain the individual configurational state $f_k C_{v,k}$ terms, and right-hand-side plots contain the coexistence/transition $f_k f_j \left(\langle U\rangle_j -\langle U\rangle_k\right)^2 $ terms. Unweighted heat capacities ($C_{v,k}$) for all states in each system are provided in the supporting information, Sections \ref{fig:SI_unweighted_Cv_vary_kt} and \ref{fig:SI_unweighted_Cv_vary_theta}. These unweighted heat capacities are approximately constant with temperature, with the exception of the random globule state, confirming the correct classification of folded structures into their configurational states.

Accompanying population fraction ($f_k$) distributions in Figures~\ref{fig:theta155_summary}-\ref{fig:theta140_summary} demonstrate complex dynamics in the relative amounts of the knotted and unfolded globule structures with changing temperature. In the case of $\theta_{B,o}=150^\circ$, a clear native structure ($3_1$ knot) is identified, with a $8_{19}$ knot existing only at intermediate temperatures. In the other 3 angle systems, for which we identified 3-4 knotted states, it is less clear which knots are the native states, and which are intermediates. Extending the replica temperature range to lower values may reveal a single native state, but this would further exacerbate already challenging sampling of the knotted structures.

\begin{figure}[H]
\includegraphics[trim=0.25in 0.1in 0.25in 0.25in, clip, width=3.25in]{"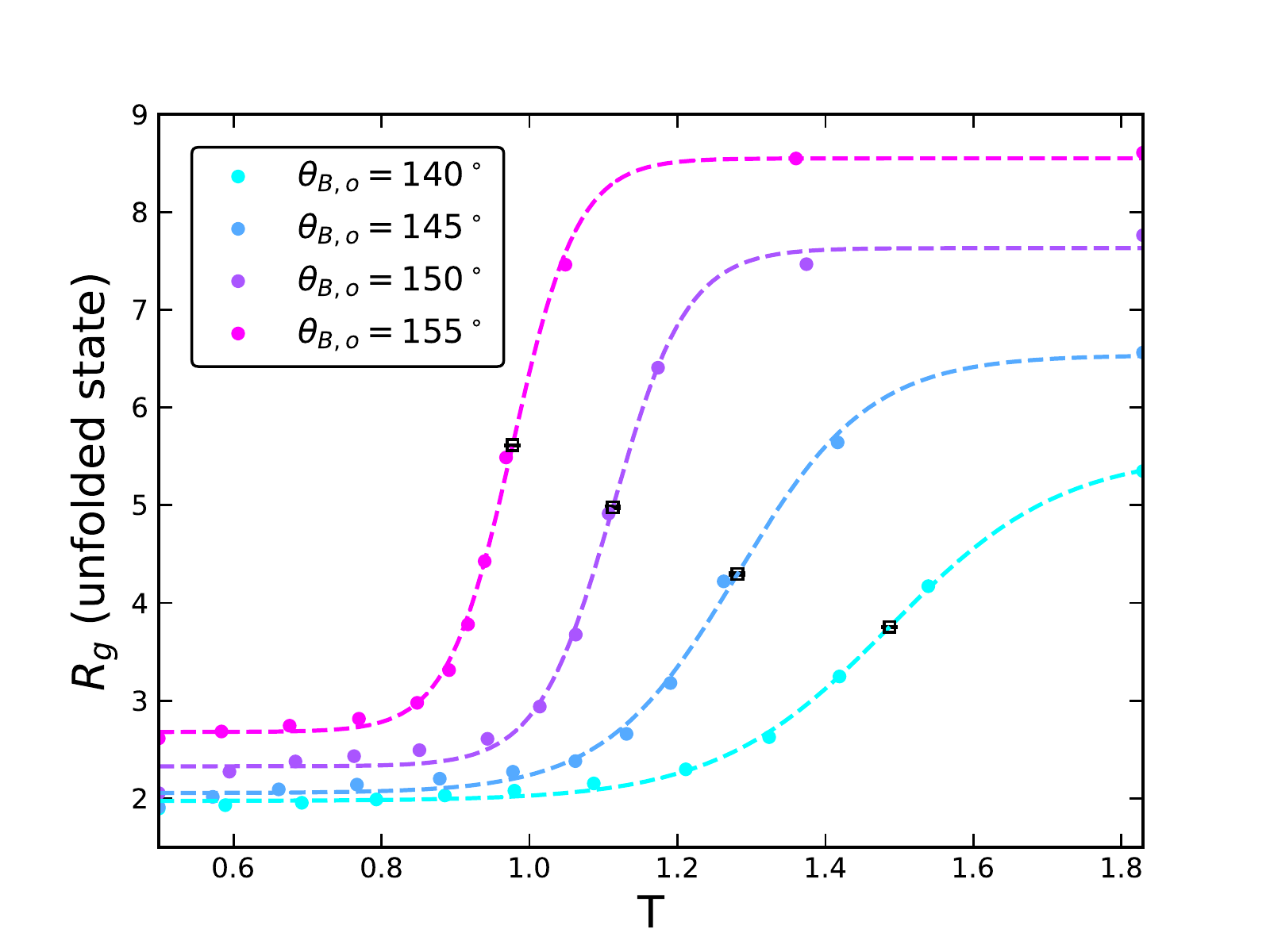"}
\caption{Mean radius of gyration versus temperature curves and hyperbolic fits (dashed lines) for varying backbone equilibrium angles indicate a significant weakening of coil-to-globule transition cooperativity with decreasing angle. Black squares mark the coil-to-globule transition centers. Standard errors of the mean for $R_g$ are smaller than the symbol size and are thus omitted.}
\label{fig:rg_vary_angles}
\end{figure}

The sharpness of the coil-to-globule transition is an another measure of the cooperativity of that transition. With analogy to the native contact fraction versus T curves in the 1-1 helix models characterizing helix-to-coil transitions, cooperativity of the coil-to-globule transition is inversely related to the temperature range over which the transition occurs. We take this characteristic width of the coil-to-globule transition as the width parameter in a sigmoidal switching function (equation \ref{eqn:rg_sigmoid_fit}). As $\theta_{B,o}$ is decreased, the coil-to-globule transition occurs over a much broader temperature range as seen from the radius of gyration versus temperature curves in the unfolded states (Figure~\ref{fig:rg_vary_angles}). This is in line with the broadening of the coil-to-globule transition contribution to heat capacity with decreasing $\theta_{B,o}$. Given the stiff angle potentials, this lower cooperativity of the coil-to-globule transition is likely a consequence of stronger nonbonded interactions in the unfolded coil state for low backbone angles, relative to high backbone angles which adopt more extended conformations. A smaller energy difference between the unfolded coil and unfolded globule states results in a weaker driving force for collapse. Hyperbolic fitting parameters are listed in the supporting information, Table~\ref{table:rg_fitting_params}.

\subsection{Tuning knot populations via torsion stiffness}~\label{results_vary_kt}
Tuning backbone torsion stiffness provides some level of control over the populations of the different knotted states as a function of temperature. Backbone torsions stabilize some knotted states observed in Section \ref{results_vary_theta}, and destabilize others, leading to different native knots. Torsion stiffness is also found to be a key parameter for tuning cooperativity, in alignment with findings in our prior study on 1-1 helix foldamer models.~\cite{Walker2021} Here we focus on the $\theta_{B,o}=155^\circ$ system as an example. With varying $k_t$, all other parameters are fixed as specified in Table~\ref{table:fixed_FF_params}. Recall that the backbone torsion potentials favor the \textit{cis} conformation. We chose $\theta_{B,o}=155^\circ$ out of the range of angles studied in section~\ref{results_vary_theta} because it yielded the sharpest heat capacity peak with torsions turned off, and the folded conformational states are relatively simple (two types each of trefoils and unknots).

Applying the RMSD clustering approach outlined in Section~\ref{methods_clustering}, we identified the same 4 stable folded conformations (right/left-handed pairs) for $k_t=0.4$ as for $k_t=0$ (cf. Figure~\ref{fig:medoids_all_theta}). For $k_t=0.8$ and higher, we detected folded states 1-3, and no additional states, but did not detect folded state 4, a trefoil knot with distant chain ends. The medoids for the 4 folded states encountered did not change noticeably across the different $k_t$. Further details on clustering results for all $k_t$ are shown in the supporting information, Section \ref{SI_clustering}.

\begin{figure}[H]
\includegraphics[trim=0in 0.0in 0in 0.0in, clip, width=6.5in]{"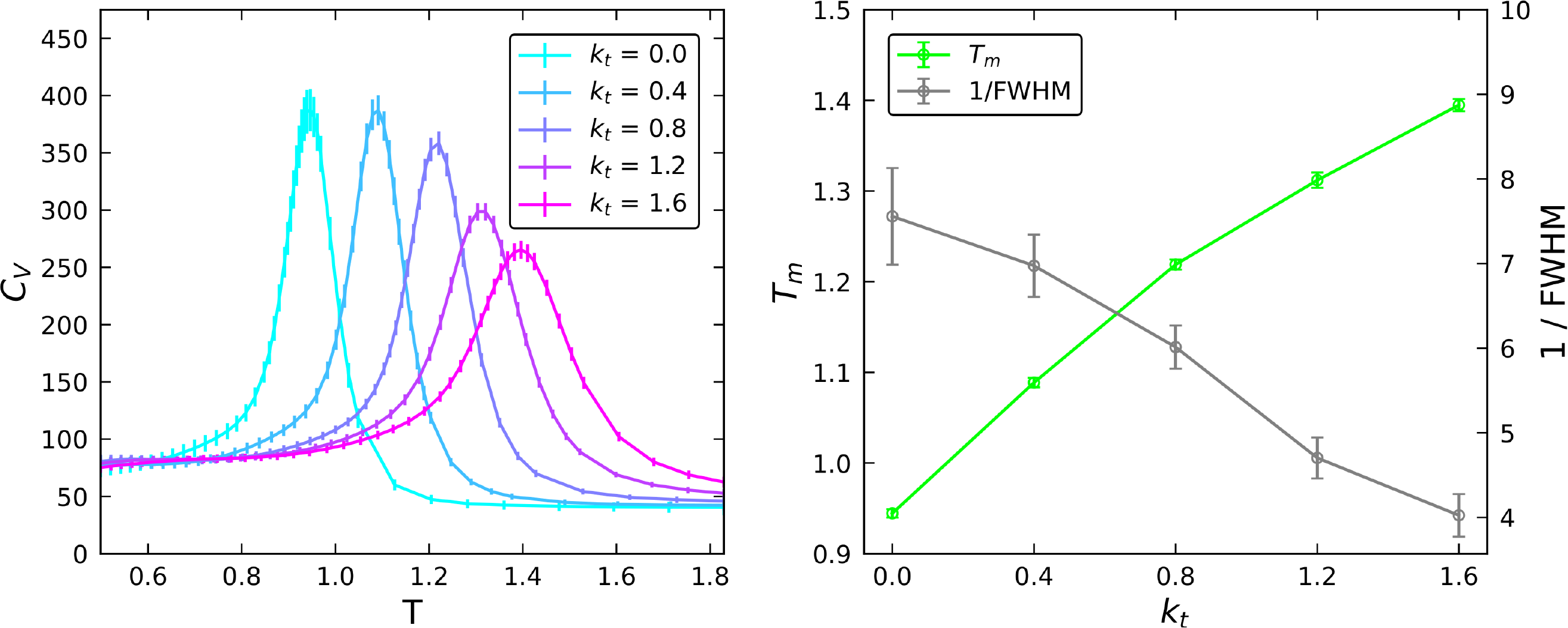"}
\caption{Total heat capacity versus temperature curves for varying $k_t$ and fixed value of $\theta_{B,o}=155^\circ$ in linear 42-monomer models indicate that with increasing torsion stiffness, thermal stability (melting point) increases, and overall cooperativity (1/FWHM) decreases. However, this seemingly simple overall heat capacity behavior hides the complexity of the multiple conformational state transitions contributing to the $C_v$ peak, which are explored in detail in the heat capacity decomposition (Fig.~\ref{fig:Cv_decomp_vary_kt_1_2_3_4}).}
\label{fig:Cv_full_vary_kt}
\end{figure}

The overall cooperativity of the $\theta_{B,o}=155^\circ$ transition from unfolded coil to knots/unknots, as measured by the inverse of the full-width half-maximum of the total heat capacity curve, increases with decreasing torsion stiffness (Figure~\ref{fig:Cv_full_vary_kt}). This behavior is consistent with results from a scan of backbone torsion stiffness in the 1-1 helix models.\cite{Walker2021} Cooperativity of the 1-1 helices was found to be maximal at a critical stiffness below which the helix was destabilized. As we saw in Section~\ref{results_vary_theta}, stable knots form in the absence of backbone torsion potentials, a notable difference between the helices and knots.

Thermal stability of the knots/unknots, as measured by the melting point or temperature at which total $C_v$ is maximal, increased with torsion stiffness. This again mirrors behavior of the 1-1 helices. Stronger backbone torsions serve to lock in the native state, reducing the chance of destabilization by thermal fluctuations. However, there is a trade-off between high stability and high cooperativity: stiff torsions also decrease the entropy of the unfolded states, diminishing the entropic driving force for folding.

\begin{figure}[H]
\includegraphics[trim=0in 0in 0in 0in, clip, width=6.5in]{"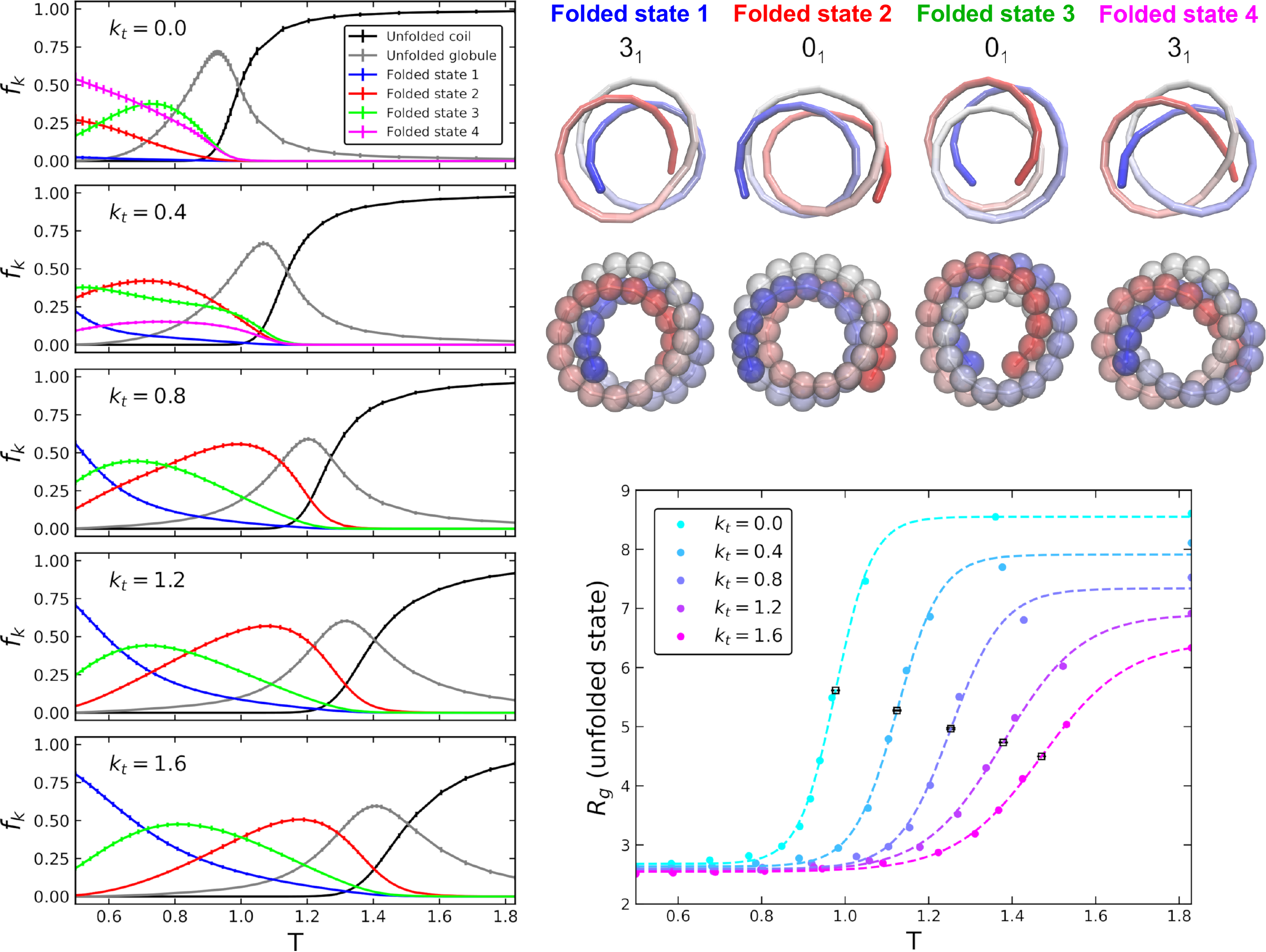"}
\caption{Left: Population distributions of unfolded coil, unfolded globule, and folded states as functions of temperature for the linear 42-monomer systems with $\theta_{B,o}=155^\circ$ reveal precisely how the populations shift with torsion stiffness ($k_t$). With increasing $k_t$, folded state 1, scarcely present for $k_t=0$, becomes increasingly the dominant configurational state at low temperature, while folded state 4 is destabilized by stiffer torsions. Folded state medoids shown are the energy-minimized right-handed enantiomers from the $k_t=0.4$ simulation - these structures did not change from one $k_t$ to the next. Knot types listed are from direct closure and HOMFLY polynomial invariants. We note that for folded state 3, weak torsion force constants ($k_t$=0.0-0.4) yielded a $0_1$ (unknot) classification, while the stronger torsion force constants ($k_t$=0.8-1.6) yielded a $3_1$ classification. Bottom right: mean radius of gyration among unfolded state conformations as a function of $k_t$, fit to a sigmoidal switching function (equation \ref{eqn:rg_sigmoid_fit}, dashed lines), reveals that the coil-to-globule transition occurs over a narrower temperature range as $k_t$ is decreased to 0. The coil-to-globule transition centers, marked by black squares, indicate that the coil-to-globule transition temperature increases with $k_t$, and $R_g$ at the transition decreases with increasing $k_t$, as expected. Standard errors of the mean for $R_g$ are smaller than the symbol size and are thus omitted.}
\label{fig:populations_all_kt}
\end{figure}

Conformational state populations as a function of temperature (Figure~\ref{fig:populations_all_kt}) signal a change in the native knot with the addition of torsion potentials: with no torsion potential, folded state 4 is the dominant state at the lowest temperature simulated, and folded state 1 is present in low amounts. As $k_t$ is increased, folded state 4 is destabilized to the point where it is no longer detectable in our clustering algorithm by $k_t=0.8$, and folded state 1 increasingly becomes favored as the native state over the intermediate states 2 and 3 at the lowest temperature. For $k_t=0.4$, there is no clear native state based on the range of temperatures simulated. For $k_t=0.8-1.6$, increasing $k_t$ effectively shifts all of the population distributions towards higher temperatures. Due to the coexistence of 3 or more states at many temperatures, we cannot pinpoint the knotting mechanism from analysis of these replica exchange simulations, though it is reasonable to hypothesize for $k_t=0.8-1.6$ that an initial coil-to-globule collapse precedes the formation of the native trefoil knot (folded state 1). In future work, kinetics simulations conducted at a series of single temperatures could be used to elucidate the pathway(s) by which the intermediates and knots are related, as in reference~\citenum{Wu2018}.

The mean radius of gyration versus temperature curves for varying $k_t$ (Figure~\ref{fig:populations_all_kt}) again provide an alternative metric of the cooperativity of the coil-to-globule transition. With increasing $k_t$, the width of the coil-to-globule transition increases, and the transition temperature increases. The radius of gyration in the unfolded state decreases with increasing torsion stiffness, which is consistent with the hypothesis that lower entropy in the unfolded state with stiffer $k_t$ is a cause of weaker cooperativity.

\begin{figure}[H]
\includegraphics[trim=0in 0.0in 0in 0.0in, clip, height=7in]{"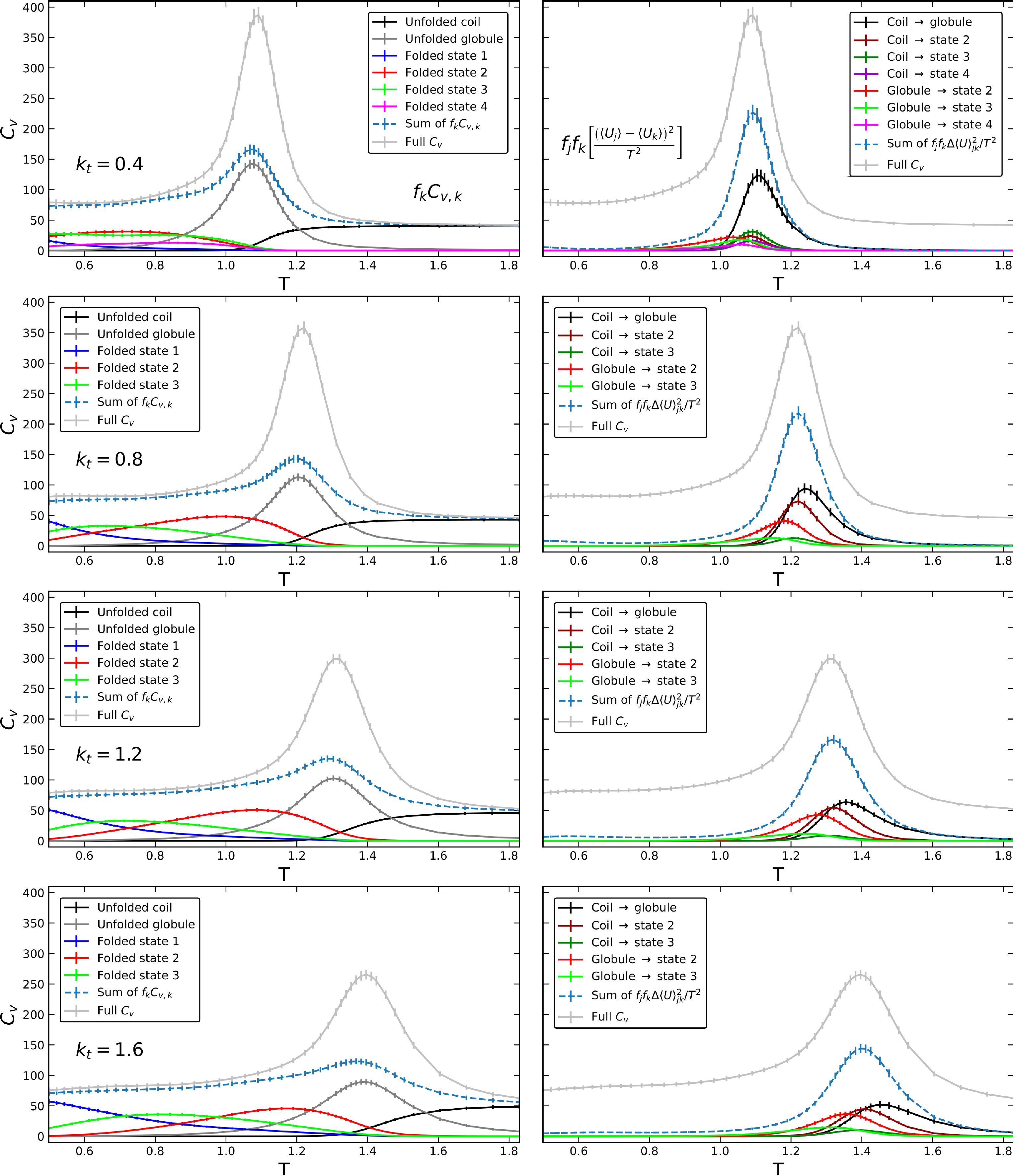"}
\caption{Heat capacity decompositions for $\theta_{B,o}=155^\circ$ and varying torsion stiffness allow for a detailed analysis of how the cooperativity of each individual transition is affected. For the $k_t=0$ heat capacity decomposition, see Figure~\ref{fig:theta155_summary}. The left-hand-side contains the heat capacity contributions by configurational state weighted by population fraction.  The right-hand-side contains the contributions to heat capacity from each state-to-state transition. For clarity, only transition terms with maxima > 5 are displayed, a criteria which no folded state to folded state combinations meet. All error bars are estimated as the standard deviation from bootstrapping the energies.}
\label{fig:Cv_decomp_vary_kt_1_2_3_4}
\end{figure}

The heat capacity decompositions (Figure \ref{fig:Cv_decomp_vary_kt_1_2_3_4}) demonstrate weakening cooperativity of the coil-to-globule transition with increasing $k_t$, as shown by decreasing height and increasing width of the coil-to-globule peaks. In addition, comparing $k_t=0.8$--$1.6$, which contain the same 3 folded states, the globule/knot and coil/knot coexistence terms also broaden with increasing $k_t$. Whereas the coil-to-globule transition dominates for $k_t=0$--$0.4$, increasing $k_t$ over the range of 0.8--1.6, the height and width of the coil/state 2 and globule/state 2 transition terms approach that of the coil-to-globule transition term. Note that the case of $k_t=0$ is shown above in Figure~\ref{fig:theta155_summary}. In no case did a folded state to folded state transition meet the threshold of 5 $C_v$ units, due to similar energies between the different knot and unknot forms.

The sums of the weighted individual state heat capacities ($f_{k}C_{v,k}$), shown on the left-hand-side in dashed blue, and in particular the contribution from the random globule state, also become flatter with increasing $k_t$. The peaks in unweighted heat capacities for the unfolded globule state also decrease with increasing $k_t$ (Figure \ref{fig:SI_unweighted_Cv_vary_kt}), indicating this is not due to shifting population fractions. The flattening of the random globule term is likely due to a decrease in the microstates sampled in that state as increasing torsion stiffness restricts conformational space. The unweighted heat capacities of all folded states are roughly constant with temperature, supporting the classification criteria used to define the folded states. As such, the weighted heat capacities of the folded states mirror their population distributions (Figure \ref{fig:populations_all_kt}).

\subsection{Effect of side chains on knot populations and cooperativity}~\label{results_side chains}
The addition of side chain beads enhances globule-to-knot cooperativity for the $\theta_{B,o}=155^\circ$ system compared to the linear model. The additional excluded volume of the side chains also decreases the number of stable knotted states. We repeated the replica exchange simulations and thermodynamic analysis on 42-monomer 1-1 models containing one backbone bead and one side chain bead per residue, with $\theta_{B,o}=140,145,150,155^\circ$ and backbone torsion force constants of $k_t=0.4$. Thus, direct comparisons between the 1-1 and linear models can be made for $\theta_{B,o}=155^\circ$. We did not observe stable knots for 140 and 145 degree angles in this 1-1 model, likely due to excessive steric repulsion introduced by the side chain beads, so here we focus on the $\theta_{B,o}$ = 150 and 155 degree systems. Force field parameters for the 1-1 systems are described in Section~\ref{methods_cgmodel} and summarized in Table~\ref{table:fixed_FF_params}. Note that the side chain beads were assigned identical Lennard-Jones parameters as the backbone beads.

\begin{figure}[H]
\includegraphics[trim=0in 0in 0in 0in, clip, width=6.5in]{"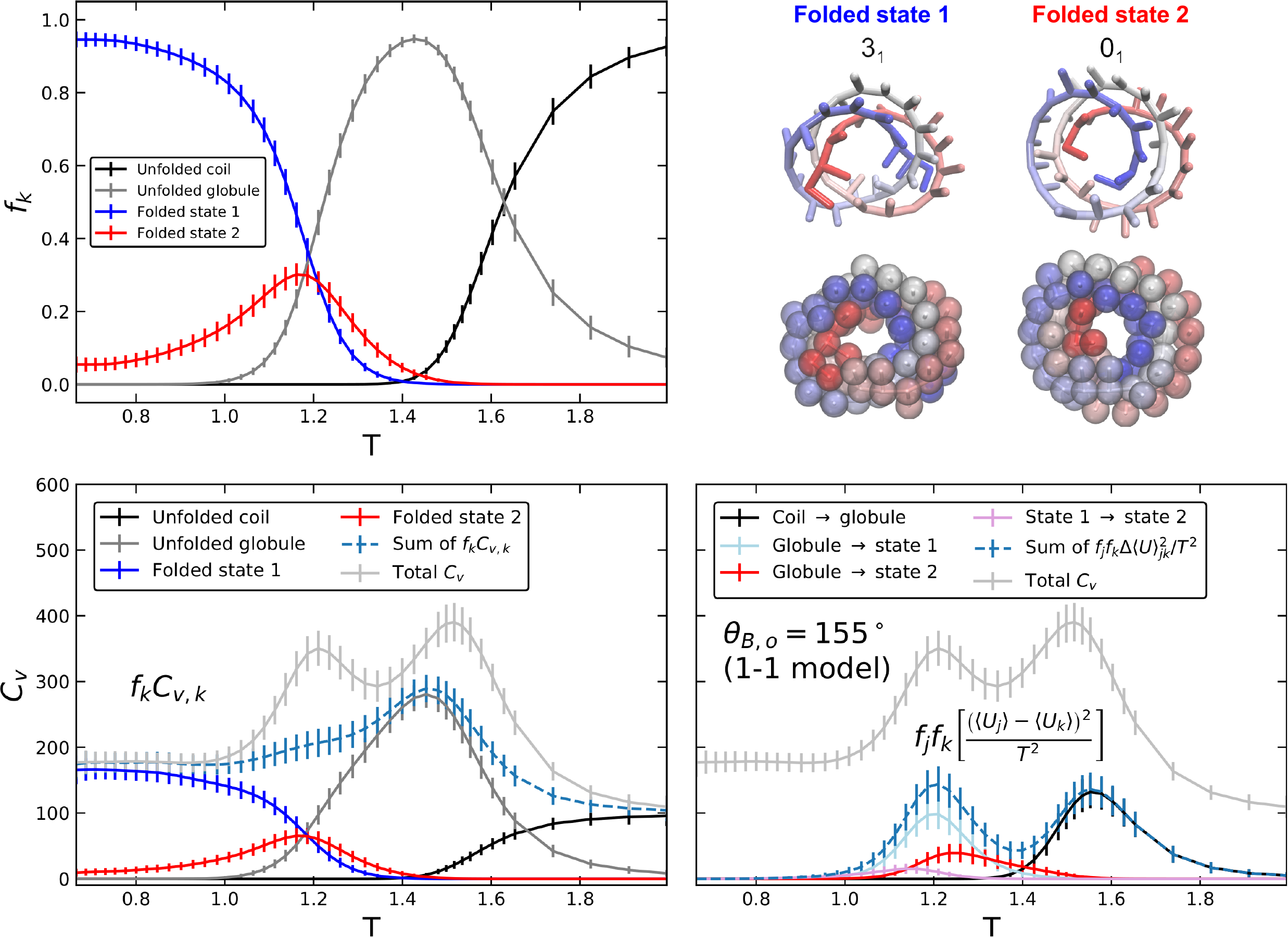"}
\caption{Population fraction versus temperature curves for the $\theta_{B,o}=155^\circ$ 1-1 model show that a native $3_1$ knot (folded state 1) is dominant at low temperature, and a single intermediate (folded state 2, an unknot) exists in relatively low amounts. The heat capacity decomposition indicates that adding side chains significantly enhances cooperativity of the globule-to-knot transitions, in part due to a reduction in the number of stable folded states. The globule to native knot transition term rivals the coil-to-globule transition term in height and width, and unlike in the corresponding linear model, these transitions occur at significantly different temperatures.}
\label{fig:theta155_SC_summary}
\end{figure}

The heat capacity decomposition for the $\theta_{B,o}=155^\circ$ 1-1 model (Figure \ref{fig:theta155_SC_summary}) reveals moderately cooperative globule-to-knot transitions that rival the coil-to-globule transition peak. Unlike the linear model for the same backbone angle and torsion stiffness parameters, only 2 folded states were identified from DBSCAN clustering. These folded states 1 and 2 in the 1-1 model have the same backbone crossing patterns of folded states 4 and 2 in the linear model, respectively (cf. Figure \ref{fig:populations_all_kt}). The inclusion of side chains therefore disfavors folded states 1 and 3 in the linear model, the former of which appears to be its native state. The population distribution for the $\theta_{B,o}=155^\circ$ side chain model closely resembles that of the $\theta_{B,o}=150^\circ$ linear model: a native knot (in this case folded state 1, a $3_1$) is highly dominant at low temperature, and an intermediate (folded state 2, an unknot) exists at moderate temperatures. The globule-to-knot and coil-to-globule transition peaks in the 1-1 model are also characterized by larger temperature separation than those in the linear model. In the linear model, poor overlap between globule and folded state 1 populations led to a negligible transition term for that pair, and other globule-to-knot transitions occurred at roughly the same temperature as the coil-to-globule transition (cf. Figure~\ref{fig:Cv_decomp_vary_kt_1_2_3_4}). By tuning the side chain angle or side chain nonbonded interaction parameters, one might be able to selectively destabilize folded state 2, even further enhancing cooperativity.

\begin{figure}[H]
\includegraphics[trim=0in 0in 0in 0in, clip, width=6.5in]{"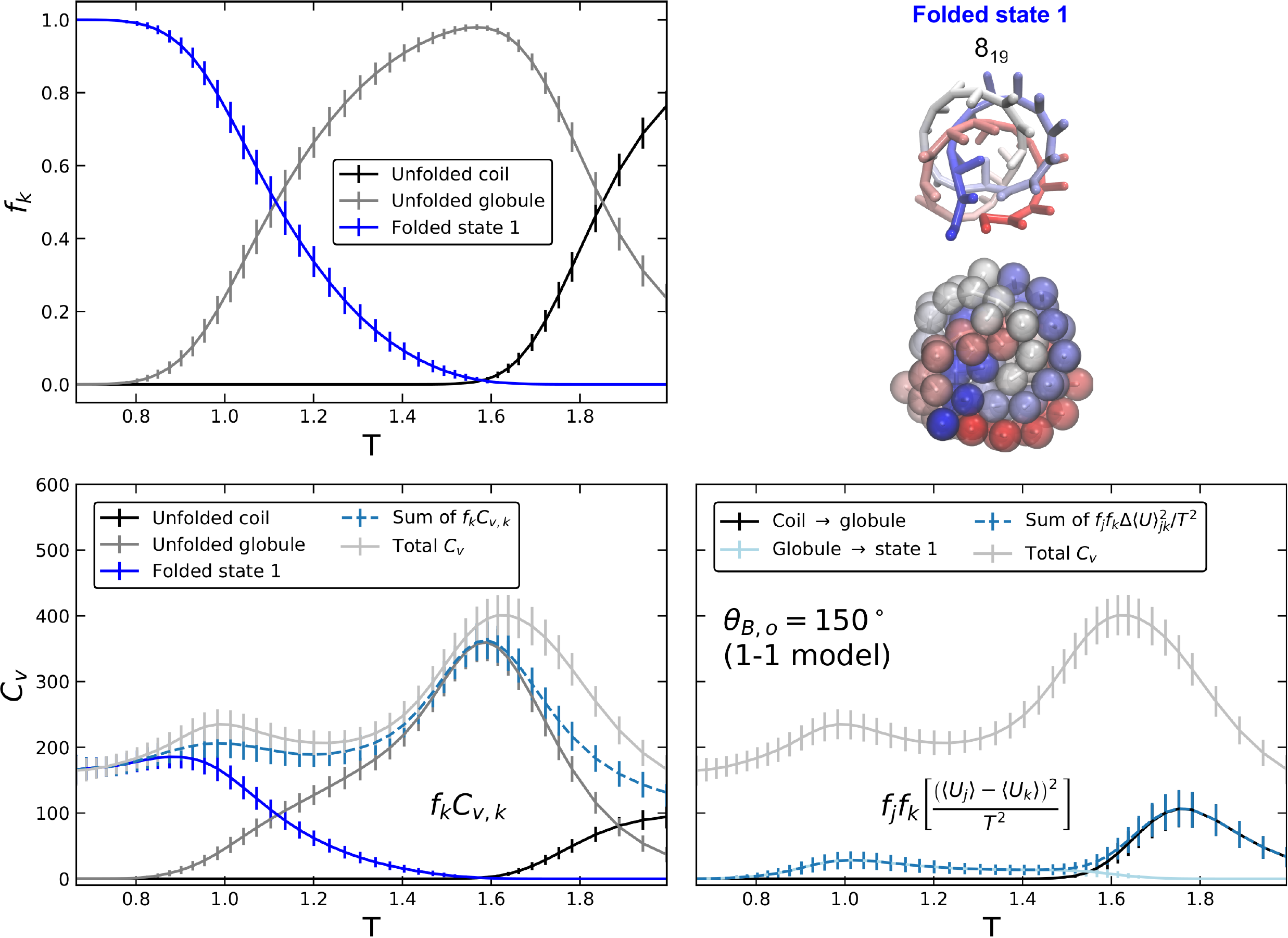"}
\caption{For the $\theta_{B,o}=150^\circ$ 1-1 model with side chains, only 1 folded state was identified, along with the unfolded globule and coil states. Yet, cooperativity of knotting (globule to state 1) is poor, as evidenced by a broad, shallow transition peak in the heat capacity decomposition.}
\label{fig:theta150_SC_summary}
\end{figure}

We achieved a stable $8_{19}$ knot for the 1-1 model with $\theta_{B,o}=150^\circ$, and identified no other stable folded states using the DBSCAN clustering protocol outlined in Section~\ref{methods_clustering}. Despite only three configurational states (knot, globule, coil), cooperativity of the globule-to-knot transition assessed from the heat capacity decomposition appears quite poor (Figure~\ref{fig:theta150_SC_summary}). Moreover, the cooperativity of the coil-to-globule transition is among the weakest out of all systems studied in this work, in terms of the sigmoid width parameter of the fit to radius of gyration versus temperature of the unfolded states (cf. Table~\ref{table:rg_fitting_params}). The $8_{19}$ knot with side chains most closely resembles folded state 2 in the linear model with $\theta_{B,o}=150^\circ$, again signaling a shift in the native state backbone topology in the linear versus 1-1 models. Note that a direct comparison of heat capacities cannot be made for this case due to differences in the backbone torsion stiffness.

\section{Conclusions}~\label{conclusions}

In this study we applied a heat capacity decomposition analysis to coarse-grained linear and 1-1 homo-oligomer knot systems to study their thermodynamics in great detail. The heat capacity decomposition comprises two groups of terms: a sum of weighted heat capacities computed within each configurational state, and a sum of state-to-state transition terms proportional to the difference in average potential energies of the two states. The heat capacity decomposition holds for any arbitrary subdivision of configurational space---however, a successful classification of conformational states will ideally lead to constant $C_{v,k}$ within each individual state, and peaks only in the state-to-state transition terms. This detailed heat capacity analysis allows for an assessment of the cooperativity of each state-to-state transition in complex systems, and the adequacy of the state definitions themselves. Smooth population distributions of each configurational state computed as functions of temperature, which are a critical component of the heat capacity decomposition, reveal complex behavior and hint at the possibility of designing globule-to-knot and knot-to-knot transitions induced by changes in temperature.

Knot topology and diversity varied dramatically with small 5 degree changes in backbone equilibrium angle values. Through a series of REMD simulations with RMSD clustering used to identify and classify folded states, we showed that thermodynamically stable physical knots form for linear homo-oligomers with 42 monomers and stiff harmonic angle potentials with backbone angles in the range of approximately 140 to 155 degrees. With decreasing equilibrium backbone angle, knot topology became more complex and varied, ranging from $3_1$ and $0_1$ types for $\theta_{B,o}=155^\circ$ to $3_1$, $5_1$, $7_1$, and $8_{19}$ knots all in coexistence for $\theta_{B,o}=140^\circ$. Stable $8_{19}$ and $3_1$ knots also formed for 1-1 homo-oligomers with 150 and 155 degree angles, respectively. To the best of our knowledge, these are the first such knots reported for 1-1 homo-oligomer models.

Applying the heat capacity decomposition, we revealed that the sharp peaks in total heat capacity observed for the homo-oligomer knots are primarily due to the coil-to-globule transition. However, some parameter sets did produce globule-to-knot transition peaks comparable in height and width to the coil-to-globule peak, suggesting moderate cooperativity of the knotting step. These more cooperative knots included the 1-1 model with $\theta_{B,o}=155^\circ$ and linear models with $\theta_{B,o}=150^\circ$ (no torsions) and $\theta_{B,o}=155^\circ$ (stiff torsions). All knot systems were characterized by a coil-to-globule transition at high temperature, with low contact fractions throughout the transition, and one or more stable knotted states existing at low temperatures. While all knotted states had constant unweighted heat capacities ($C_{v,k}(T)$) within uncertainty, the unfolded globule state consistently showed a peak within the state due to the relatively high conformational diversity. This peak could be suppressed by introducing stiffness via backbone torsion potentials.

In addition to topological changes due to varying backbone angles, we found that the native knot can change by manipulating torsion stiffness (Section~\ref{results_vary_kt}), and by adding side chain beads to linear models (Section~\ref{results_side chains}). Torsion stiffness also shifted the population fraction curves of all configurational states.

It is important to note that the existence of a peak in the $f_jf_k(\langle U_j\rangle-\langle U_k\rangle)^2$ term does not necessarily mean the knotting pathway includes that direct transition, such as the transitions from unfolded coil directly to a knotted state. The transition peaks signify the coexistence of the pairs of states differing in average energies. To discern the knotting pathway(s) and mechanisms for each of the systems studied, kinetic studies would be needed.

The overall trends in heat capacity decompositions were not sensitive to the precise definition of the contact fraction criteria for dividing unfolded versus knotted states. However, the presence of a significant population of structures with intermediate contact fractions, and lack of a better order parameter for the globule-to-knot transitions given the ambiguity of knot identification from closure schemes, make the choice of an unfolded versus folded/knotted cutoff somewhat arbitrary. This lack of a definitive order parameter itself points to the weak cooperativity of the homo-oligomer knots, on the basis that highly cooperative transitions have a large free energy barrier between the states at which few intermediates will be found.~\cite{Munoz2016} Nonetheless, we found that RMSD to reference medoids does adequately separate one knotted state from another.

If we can achieve configurationally well-defined knotted structures simply by way of stiff angle potentials in homo-oligomers with isotropic nonbonded interactions, it is very likely that their cooperativity can be significantly enhanced by manipulating sequences and local bending stiffness in hetero-oligomers with 2 or more bead types. This direction towards designing precise knotted hetero-oligomers was pursued by Cardelli and coworkers~\cite{Cardelli2018} in simulations of chains of patchy particles. Notably, in that study a designed $5_2$ twist knot was successfully achieved through manipulation of sequence in a 3-monomer alphabet. With refinement, the knots studied here can be further developed into secondary structural elements to potentially be used for self-assembly or other applications, provided that real chemical analogs can be found.

\vspace{12pt}
\section{Appendix A: Derivation of Heat Capacity Decomposition}~\label{A}
\setcounter{equation}{0}
\renewcommand{\theequation}{A.\arabic{equation}}

We first partition configurational space into $n$ nonoverlapping parts, with these configurationally defined states labeled $0$, $1$, ..., $n$. We then define the following partition functions:
\begin{eqnarray}
Z &=& \sum_{i\in 0+1+...+n} e^{-\beta U_i} \\
Z_{k} &=& \sum_{i\in k} e^{-\beta U_i}
\end{eqnarray}
We use $Z$, usually used as the configuration integral, to represent the partition function, as $Q$ represents native contact fraction in this paper. Functionally, the two are equivalent, since all contributions from the kinetic energy portion of the partition function cancel out between folded and unfolded states. In the partition functions, $U_i$ are the energies of each microstate $i$, and $k$ designates the conformational state index. Noting that $Z = Z_{0} + Z_{1} + ... + Z_{n}$, the ensemble averages for potential energy of the system can be expressed as:
\begin{equation}
\langle U \rangle = \sum_{k} {\left[ \frac{Z_{i}}{Z_{0}+Z_{1}+...+Z_{n}}\langle U \rangle_{k}\right]}
\label{eqn:U_Q_summation}
\end{equation}
We introduce the following notation representing the population fraction of each configurational state:
\begin{equation}
f_{k} = \frac{Z_{k}}{Z_{0}+Z_{1}+...+Z_{n}} = \frac{Z_{k}}{Z}
\label{eqn:f_definition}
\end{equation}
Thus, another representation of $\langle U\rangle$ is:
\begin{equation}
\langle U\rangle = \sum_{k} {f_k \langle U_k\rangle}
\label{eqn:U_f_summation}
\end{equation}
With some rearrangement of equation \ref{eqn:f_definition} we get, for example:
\begin{equation}
f_{0} = \frac{1}{1+\frac{Z_{1}}{Z_{0}} + ... + \frac{Z_{n}}{Z_{0}}}
\end{equation}
Hence, the fraction of each conformational state can be computed from the sum of all pairwise free energy changes:
\begin{equation}
f_{0} = \frac{1}{1+e^{-\beta (G_{1}-G_{0})} + ... + e^{-\beta (G_{n}-G_{0})}}
\label{eqn:f_from_free_energy}
\end{equation}
The free energy changes are computed from ratio of the probability expectations of the pair of configurational states:
\begin{equation}
    \Delta G_{ij}(T) = -k_BT\ln{\left( \frac{\langle P_i\rangle}{\langle P_j\rangle}\right)}
\end{equation}
The fractions $f_k$ by definition sum to unity. Applying the expression for $\langle U\rangle$ in equation \ref{eqn:U_f_summation} to the definition of heat capacity, we get the following relation:
\begin{equation}
\frac{d\langle U\rangle}{dT} = \sum_{k} {\left[ \frac{df_{k}}{dT} \langle U \rangle_{k} + f_{k} \frac{d\langle U \rangle_{k}}{dT}\right]}
\label{eqn:cv_decomp_before_derivatives}
\end{equation}
Taking the temperature derivative of the population fraction (equation \ref{eqn:f_definition}):
\begin{eqnarray}
\frac{df_k}{dT} &=&  \frac{d}{dT}\left(\frac{Z_k}{Z}\right) \nonumber \\ 
&=& \frac{dZ_k}{dT}(Z)^{-1} -Z_k(Z)^{-2}\frac{dZ}{dT} \nonumber \\
&=& \frac{Z_k \langle  U\rangle_k}{k_BT^2}Z^{-1} -Z_k Z^{-2}\frac{Z \langle U\rangle}{k_BT^2} \nonumber \\
&=& \frac{Z_k}{Z} \left(\frac{\langle  U\rangle_U}{k_BT^2} - \frac{\langle U\rangle}{k_BT^2} \right) \nonumber \\
\frac{df_k}{dT} &=& f_{k}\left(\frac{\langle  U\rangle_{k}}{k_BT^2} - \frac{\langle U\rangle}{k_BT^2} \right)
\label{eqn:df_dT_expression}
\end{eqnarray}
We now define an intra-state heat capacity for configurational state $k$ as $C_{v,k} = d\langle U \rangle_{k}/dT$. Applying this definiton along with the result in equation~\ref{eqn:df_dT_expression}, the total heat capacity decomposition (equation~\ref{eqn:cv_decomp_before_derivatives}) becomes:
\begin{eqnarray}
\frac{d\langle U\rangle}{dT} = f_0 C_{v,0} + f_1 C_{v,1} + ... + f_n C_{v,n} + 
f_0 \langle U \rangle_0 \left(\frac{\langle U \rangle_0}{k_BT^2} - \frac{\langle U \rangle}{k_BT^2}\right) + \\ \nonumber
f_1 \langle U \rangle_1 \left(\frac{\langle U \rangle_1}{k_BT^2} - \frac{\langle U \rangle}{k_BT^2}\right) + ... + f_n \langle U \rangle_n \left(\frac{\langle U \rangle_n}{k_BT^2} - \frac{\langle U \rangle}{k_BT^2}\right)
\end{eqnarray}
Applying the definition of $\langle U\rangle$ in equation \ref{eqn:U_f_summation} and collecting the $f_k \langle U\rangle_k$ prefactors:
\begin{eqnarray}
\frac{d\langle U\rangle}{dT} &=& \sum_k {f_k C_{v,k}} + \sum_k
{f_k \frac{\langle U\rangle_k^2}{k_BT^2}} - \frac{\langle U \rangle}{k_BT^2} \left( f_0 \langle U\rangle_0 + f_1 \langle U\rangle_1 + ... + f_n \langle U\rangle_n \right) \nonumber \\ 
&=& \sum_k {f_k C_{v,k}} + \sum_k
{f_k \frac{\langle U\rangle_k^2}{k_BT^2}} - \frac{\langle U \rangle^2}{k_BT^2}
\end{eqnarray}
A more useful result can be obtained by rearranging the above expression into contributions from each conformational state transition, rather than as a deviation of the sum of $f_k\langle U\rangle_k^2$ terms from the total potential energy of the mixture.

We note that: $f_k  - f^2_k = f_k(1-f_k) = f_k(\sum_{j\ne k)} f_j)$, so that: 
\begin{eqnarray}
\frac{d\langle U\rangle}{dT} &=& \sum_k {f_k C_{v,k}} + \frac{1}{k_B T^2}\left(\sum_k \left(f_k \langle U\rangle_k^2 - f_k^2 \langle U\rangle_k^2 - \sum_{j > k} 2 f_k f_j \langle U\rangle_k \langle U\rangle_j\right)\right) \\ \nonumber 
&=& \sum_k {f_k C_{v,k}} + \frac{1}{k_B T^2}\left(\sum_k \left(f_k(1-f_k) \langle U\rangle_k^2 - \sum_{j > k} 2 f_k f_j \langle U\rangle_k \langle U\rangle_j\right)\right) \\ \nonumber 
&=& \sum_k {f_k C_{v,k}} + \frac{1}{k_B T^2}\left(\sum_k \left(\left(f_k \sum_{j\ne k} f_j\right)\langle U\rangle_k^2 - \sum_{j > k} 2 f_k f_j \langle U\rangle_k \langle U\rangle_j\right)\right) \\ \nonumber 
&=& \sum_k {f_k C_{v,k}} + \frac{1}{k_B T^2}\left(\sum_k\left( \sum_{j \ne k} f_k f_j \langle U\rangle_k^2 -\sum_{j > k} 2 f_k f_j \langle U\rangle_k \langle U\rangle_j\right) \right) \\  \nonumber 
&=& \sum_k {f_k C_{v,k}} + \frac{1}{k_B T^2}\left(\sum_k\sum_{j > k} f_k f_j \langle U\rangle_j^2 + f_k f_j \langle U\rangle_k^2 - 2 f_k f_j \langle U\rangle_k \langle U\rangle_j\right) \\
\frac{d\langle U\rangle}{dT} &=& \sum_k {f_k C_{v,k}} + \frac{1}{k_B T^2}\left(\sum_k\sum_{j > k} f_k f_j \left(\langle U\rangle_j -\langle U\rangle_k\right)^2\right)  
\label{eqn:cv_decomp_equation_appendix}
\end{eqnarray}


\section{Author Contributions}
\vspace{18pt}
C.C.W. and M.R.S. primarily conceptualized the project and designed the methodology, with additional contributions from T.L.F.; M.R.S. primarily derived the heat capacity decomposition, with additional contributions from C.C.W.; all experiments were conducted by C.C.W.; experiments were analyzed and validated by C.C.W. and M.R.S.; C.C.W. primarily wrote the original manuscript draft with contributions from M.R.S.; editing and review of the manuscript was done by M.R.S. and T.L.F.; M.R.S. supervised the project and obtained resources.

\begin{acknowledgement}

This material is based upon work supported by the U.S. Department of Energy, Office of Science, Office of Basic Energy Sciences, Materials Sciences and Engineering (MSE) Division, under Award Number DE-SC0018651. This work utilized computational resources from the University of Colorado Boulder Research Computing Group, which is supported by the National Science Foundation (awards ACI-1532235 and ACI-1532236), the University of Colorado Boulder, and Colorado State University. This work also used the Extreme Science and Engineering Discovery Environment (XSEDE), which is supported by National Science Foundation grant number ACI-1548562. Specifically, it used the Bridges-2 system, which is supported by NSF  ACI-1928147, located at the Pittsburgh Supercomputing Center (PSC). T.L.F. was in part supported by the U.S. Department of Education's Graduate Assistance in Areas of National Need (GAANN) fellowship program. 

\end{acknowledgement}

\begin{suppinfo}

The following supporting information file is available free of charge:
\begin{itemize}
  \item Additional details on knot classification and DBSCAN clustering protocol; detailed clustering results, including both right and left-handed medoid structures, knot types from direct and stochastic closure, population versus temperature distributions, and contact maps; structural visualization of knotted state ensembles; comparisons of the heat capacity decomposition results for different contact definitions; radius of gyration versus temperature fitting parameters; and unweighted configurational state heat capacities. (PDF)
\end{itemize}

\end{suppinfo}

\bibliography{references}
\end{singlespace}
\end{document}


\newpage
\begin{singlespace}
\section{Knot closures}~\label{SI_knot_closure}

To classify open molecular knots by their Alexander-Briggs type, a closure method must first be selected to generate a closed chain. The simplest method is direct closure, whereby the end beads are connected together by a virtual bond to form a closed knot. In cases where the chain ends are in close proximity, this is a logical choice. However, when chain ends are on opposite sides of a molecule, small deviations in molecular structure can lead to different knot classifications. In contrast to the direct method, stochastic closure methods generate a distribution of knot types by connecting chain ends to different random points on an imaginary sphere surrounding the molecule; the knot type with the highest probability is then selected. 

Various knot detecting software packages and web servers are available which read in a molecular coordinate file, apply closure methods, and use knot invariants to deduce the knot type, including KymoKnot~\cite{Tubiana2018}, pKNOT~\cite{Lai2007}, KNOTS web server~\cite{Kolesov2007}, and knot\_pull~\cite{Jarmolinska2020}, to name a few. To date, more than 2000 knotted and slipknotted proteins have been identified and added to the knot\_prot 2.0 database \cite{Dabrowski-Tumanski2019}. In this study we use the HOMFLY polynomial invariant\cite{Freyd1985,Przytycki1987} in the Topoly Python package\cite{Dabrowski-Tumanski2021} to characterize the knotted structures. We chose the HOMFLY polynomial over other invariants because it can distinguish between chiral entantiomers, and the HOMFLY polynomials are unique for all right and left-handed knots of 8 crossings or fewer.~\cite{Ramadevi1994}

For proteins, it is common to apply a structure reduction algorithm to make knot identification simpler. Though our systems are not quite as complex, the structures are reduced in Topoly using the KMT reduction algorithm of Koniaris and Muthukumar with later modification by Taylor.\cite{Koniaris1991,W.R.Taylor2000}.

We use two different closure methods for comparison: direct closure, whereby chain ends are joined by a straight line, and stochastic closure, whereby chain ends are each connected to separate random points on a large sphere encapsulating the knot, and those points are subsequently joined by an arc. Specifically, Topoly closure type 0 is used for direct closure, and closure type 2 is used for stochastic closure with 20,000 iterations. Further algorithmic details are provided in the Topoly online documentation. Unsurprisingly, some deviations between the two closures occur for structures with chain ends at opposite sides. The distributions of knot types found using the stochastic closure method for each medoid are shown below in section~\ref{SI_clustering}. These distributions are also useful for distinguishing among structures with the same direct closure knot type but differing in other conformational aspects.

\vspace{12pt}
\section{DBSCAN Clustering}~\label{SI_clustering}

DBSCAN clustering by particle coordinate RMSD is performed using the \\ \texttt{analyze\_foldamers}~\cite{analyzefoldamers2022} Python package developed by us, which in turn uses the \texttt{scikit-learn}~\cite{Pedregosa2011} DBSCAN function and MDTraj~\cite{McGibbon2015} RMSD function. The distance parameter $\epsilon$, number of minimum samples $N_{min}$ defining core points, and fraction of low-density data filtered out in a pre-clustering screening step are listed in Table~\ref{table:cluster_params} for each model studied. Table entries with the `sc' designation refer to the 1-1 model with side chains. It is important to note that only the lowest 6 temperature state trajectories were fed into the clustering algorithm, as few folded structures were identified for the higher temperatures, and the percent filtered pertains to this subset. A stride of 200 was used to sample these 6 state trajectories containing 1 million production frames each. As explained in Section~\ref{methods_clustering}, we adopted a general strategy of fixing $N_{min}$ and the pre-clustering filtering percentage to 50 and 50\%, respectively, and selected an $\epsilon$ (by trial and error) just above the value where topologically duplicate cluster medoids start to appear. When known stable structures observed in the replica trajectories were not identified with these guidelines, only then did we also vary $N_{min}$ and the filtering percentage. Decreasing the percent of data filtered out prior to clustering from 50\% to 40\% effectively discards less data from intermediate temperatures, which is important for identifying knotted states occurring predominantly in that temperature range.

\begin{table}[H]
  \caption{DBSCAN clustering parameters}
  \label{table:cluster_params}
  \begin{tabular}{||cc||ccc||}
    \hline
    $\mathbf{\theta_B}$ \textbf{(degrees)} & $\mathbf{k_t}$ & $\mathbf{\epsilon (\sigma)}$ & $\mathbf{N_{min}}$ & \textbf{\% filtered} \\
    \hline
    155 & 0.0 & 0.34 & 50 & 40 \\
    \hline
    155 & 0.4 & 0.30 & 100 & 50 \\
    \hline
    155 (sc) & 0.4 & 0.78 & 50 & 50 \\
    \hline
    155 & 0.8 & 0.272 & 100 & 50 \\
    \hline
    155 & 1.2 & 0.23 & 50 & 50 \\
    \hline
    155 & 1.6 & 0.25 & 50 & 50 \\
    \hline
    150 & 0.0 & 0.42 & 50 & 40 \\
    \hline
    150 (sc) & 0.4 & 0.80 & 50 & 40 \\
    \hline
    145 & 0.0 & 0.36 & 50 & 50 \\
    \hline
    140 & 0.0 & 0.36 & 50 & 50 \\
    \hline
  \end{tabular}
\end{table}

The following sections summarize the medoids identified from DBSCAN clustering. Structures shown and the knot classifications by direct closure and stochastic closure are of the energy-minimized medoids. For the stochastic closure, only knot types with probabilities greater than 10\% are shown. Knot types unable to be resolved with the HOMFLY method in Topoly are listed as `Unknown'. Contact maps are shown for each medoid with a pairwise distance cutoff of $2\sigma$. In the molecular snapshots, particle index increases from red to blue. For models that include side chains, only backbone-backbone contact pairs are shown, and knot closures are applied using the 2 terminal backbone beads.

\newpage
\subsection{Medoids for $\theta_{B,o}=155^\circ, k_t=0$} \label{SI_medoids_theta155_kt0}
\begin{figure}[H]
\includegraphics[trim=0in 6.25in 3in 0in, clip, height=4in, left]{"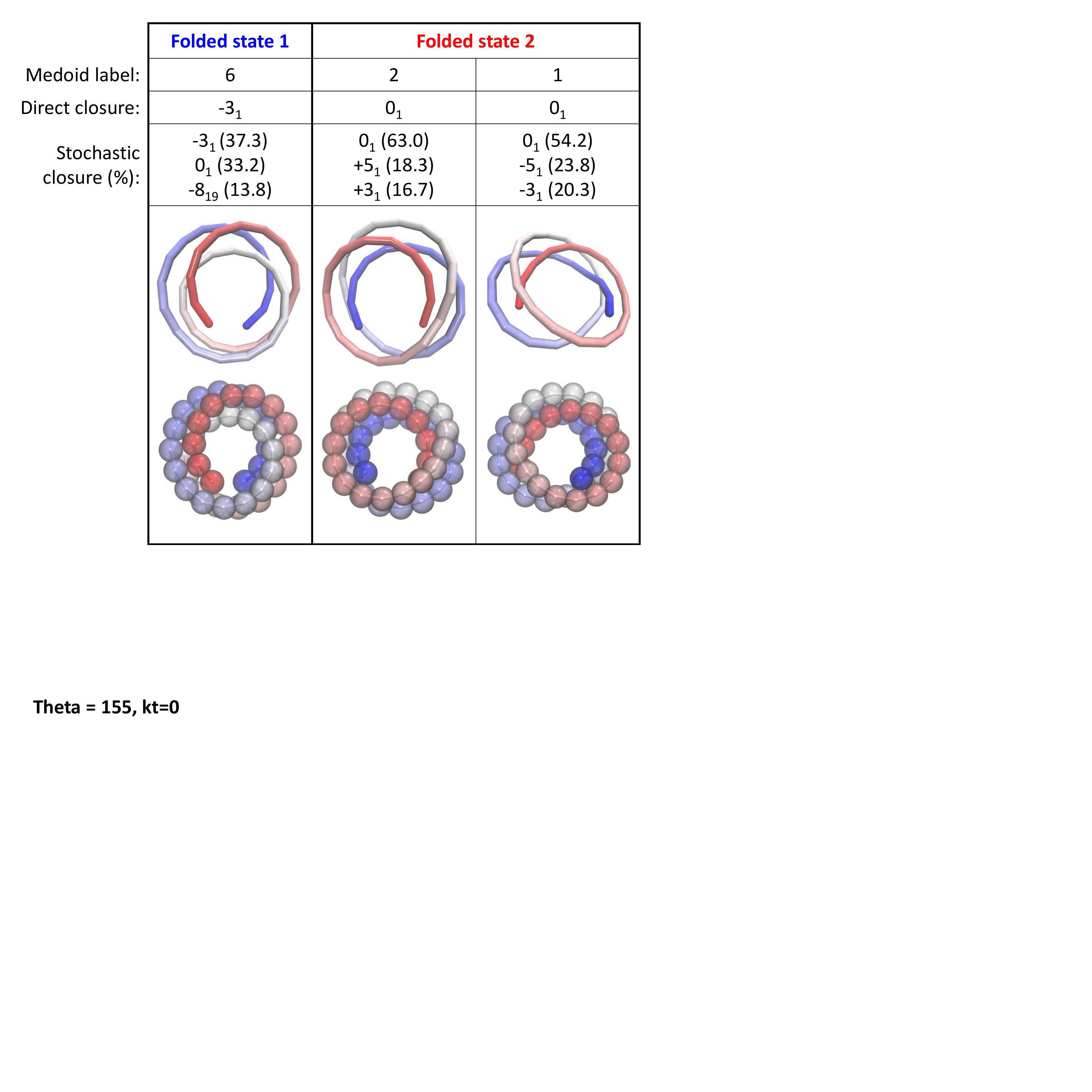"}
\includegraphics[trim=0in 6.25in 3in 0in, clip, height=4in, left]{"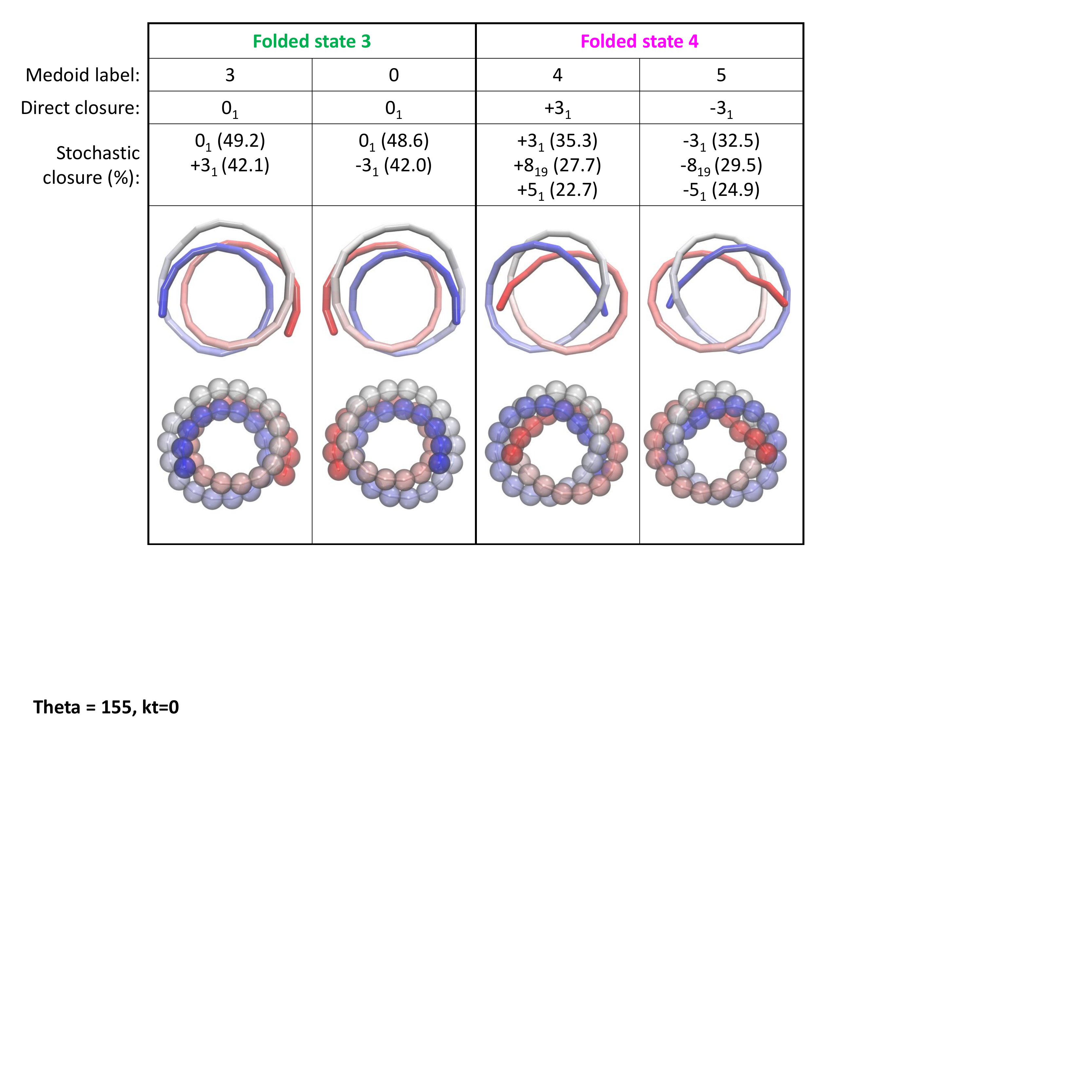"}
\label{fig:SI_medoids_theta155_kt0}
\end{figure}

\begin{figure}[H]
\includegraphics[trim=1in 0.5in 1in 0.5in, clip, height=9in]{"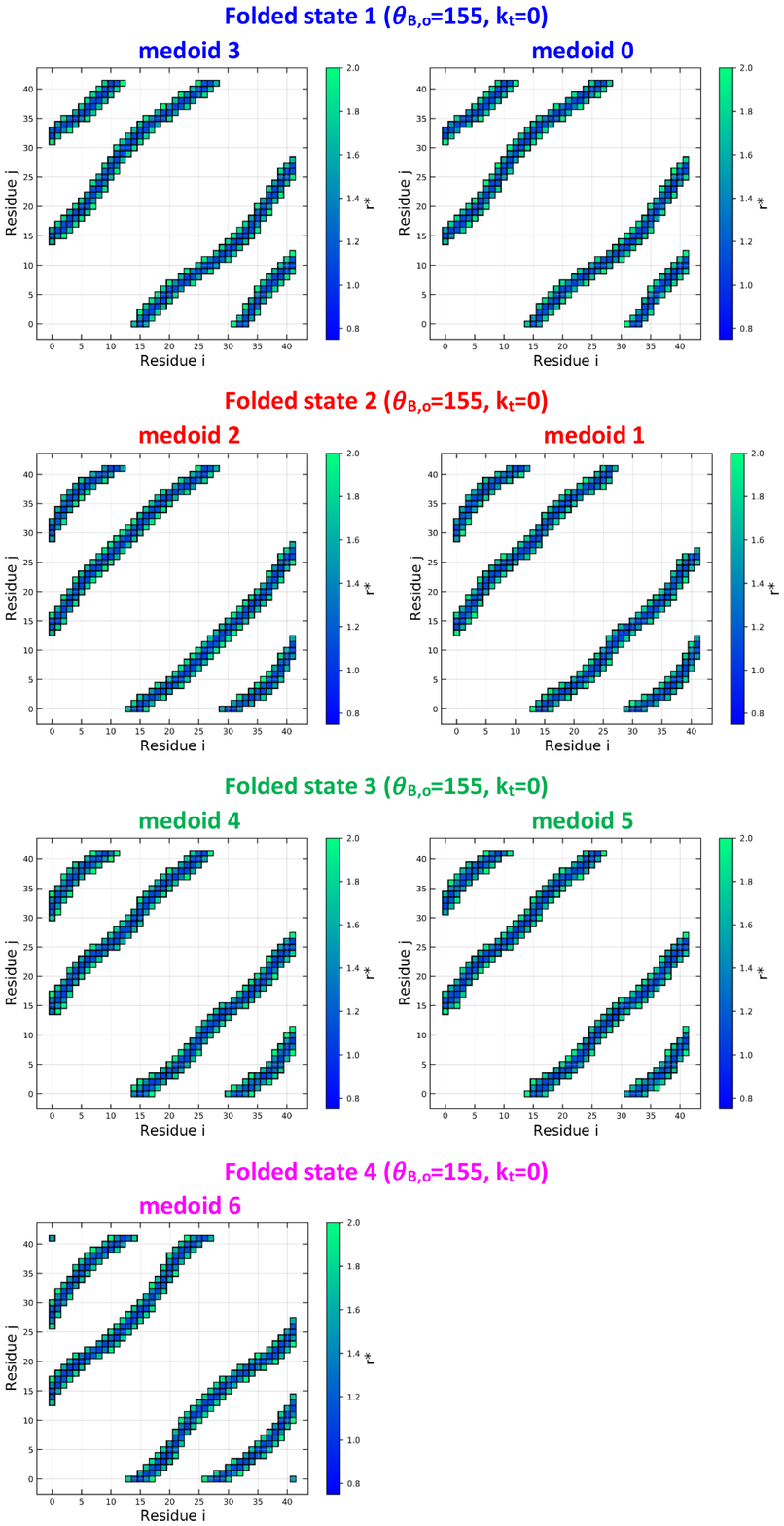"}
\label{fig:SI_contacts_theta155_kt0}
\end{figure}

\subsection{Medoids for $\theta_{B,o}=155^\circ, k_t=0.4$} \label{SI_medoids_theta155_kt04}
\begin{figure}[H]
\includegraphics[trim=0in 6.25in 3in 0in, clip, height=4in, left]{"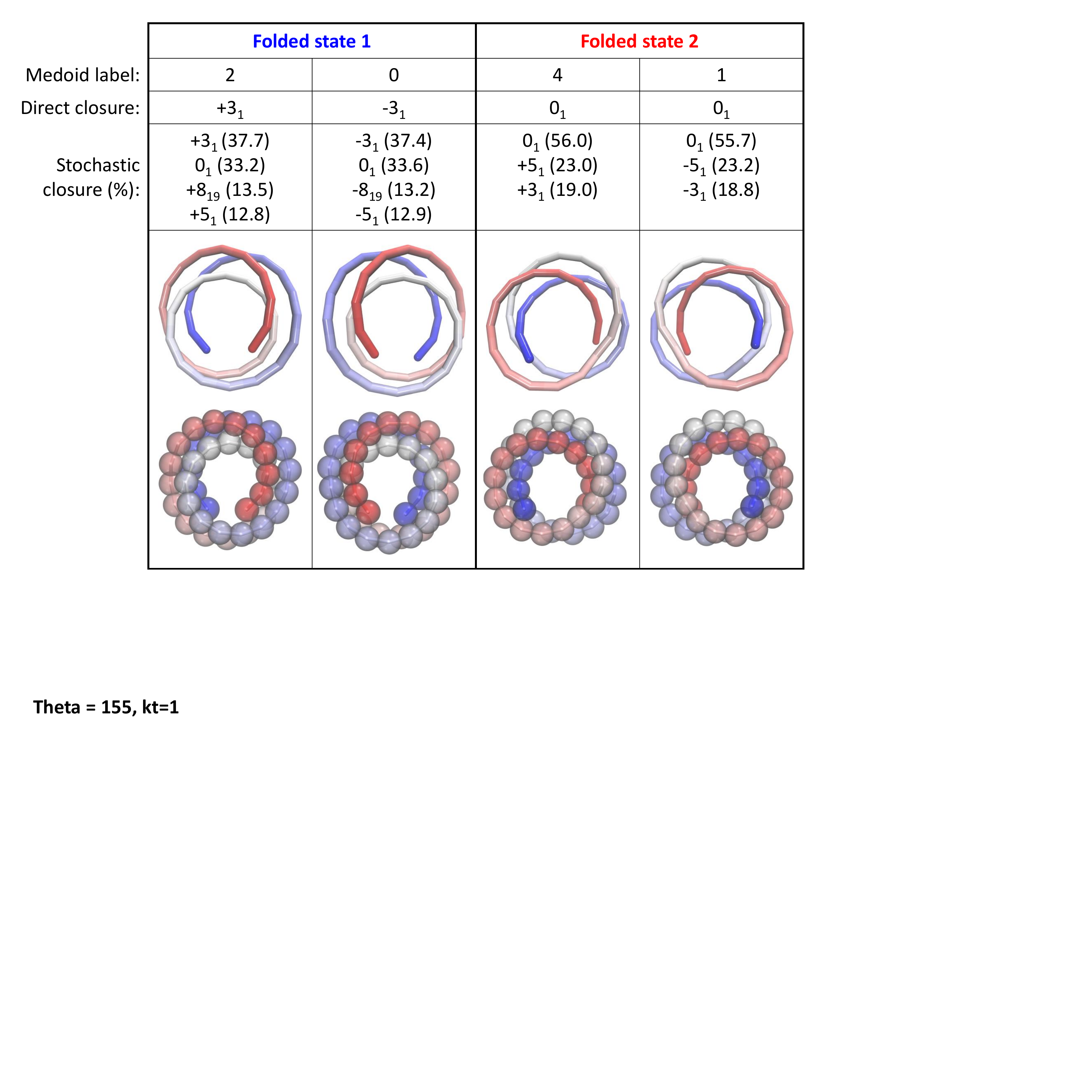"}
\includegraphics[trim=0in 6.25in 3in 0in, clip, height=4in, left]{"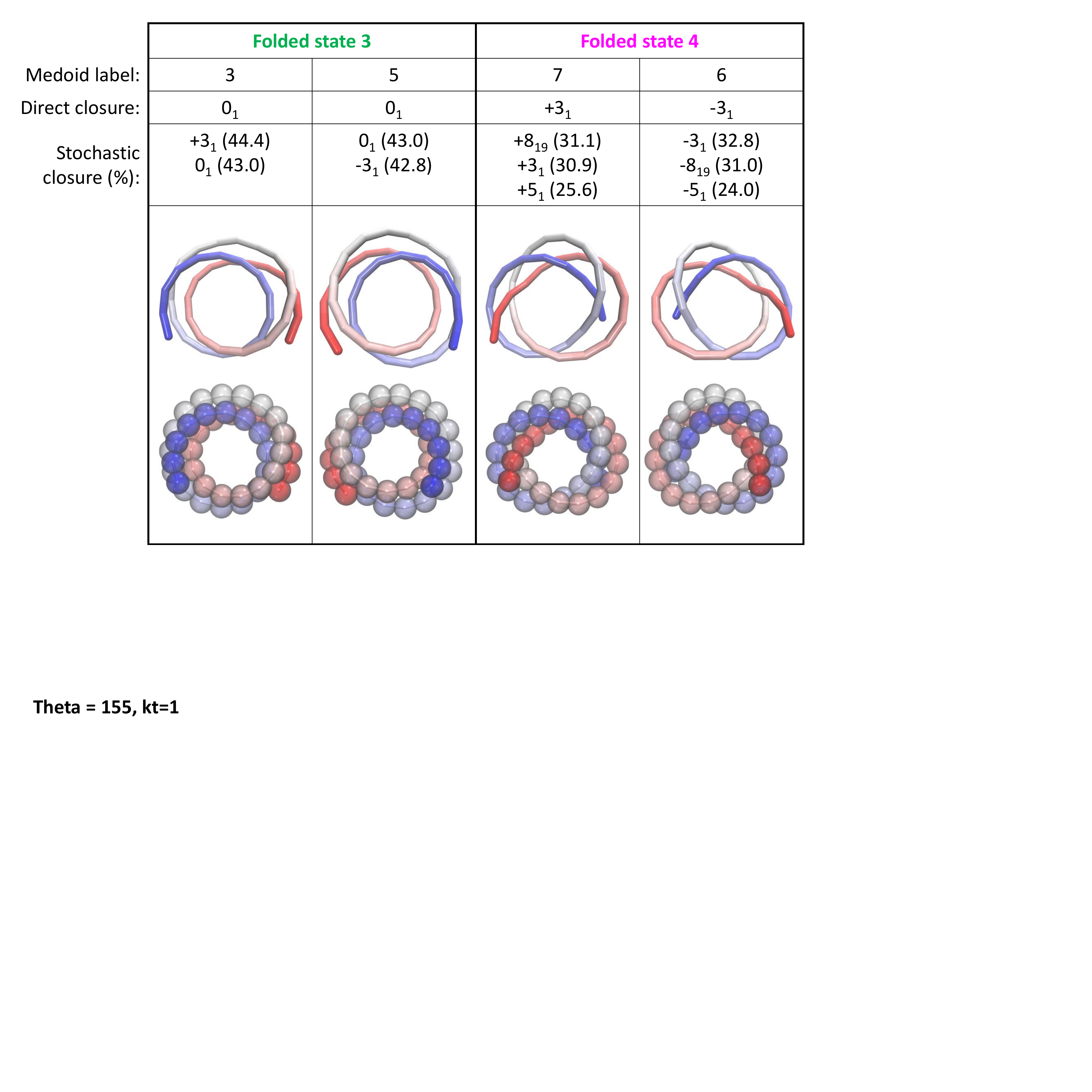"}
\label{fig:SI_medoids_theta155_kt1}
\end{figure}

\begin{figure}[H]
\includegraphics[trim=1in 0.5in 1in 0.5in, clip, height=9in]{"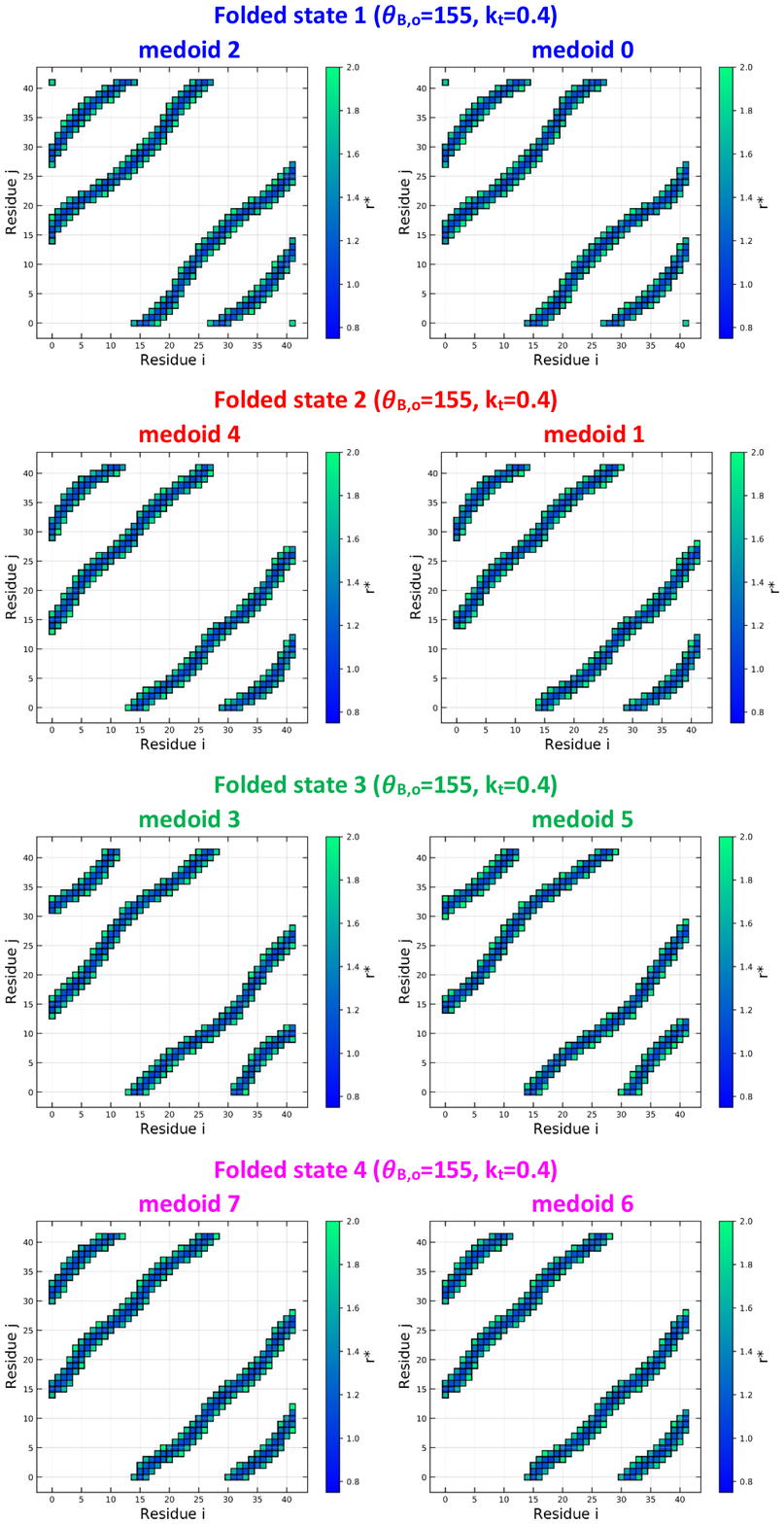"}
\label{fig:SI_contacts_theta155_kt1}
\end{figure}

\subsection{Medoids for $\theta_{B,o}=155^\circ, k_t=0.8$} \label{SI_medoids_theta155_kt08}

\begin{figure}[H]
\includegraphics[trim=0in 6.25in 3in 0in, clip, height=4in, left]{"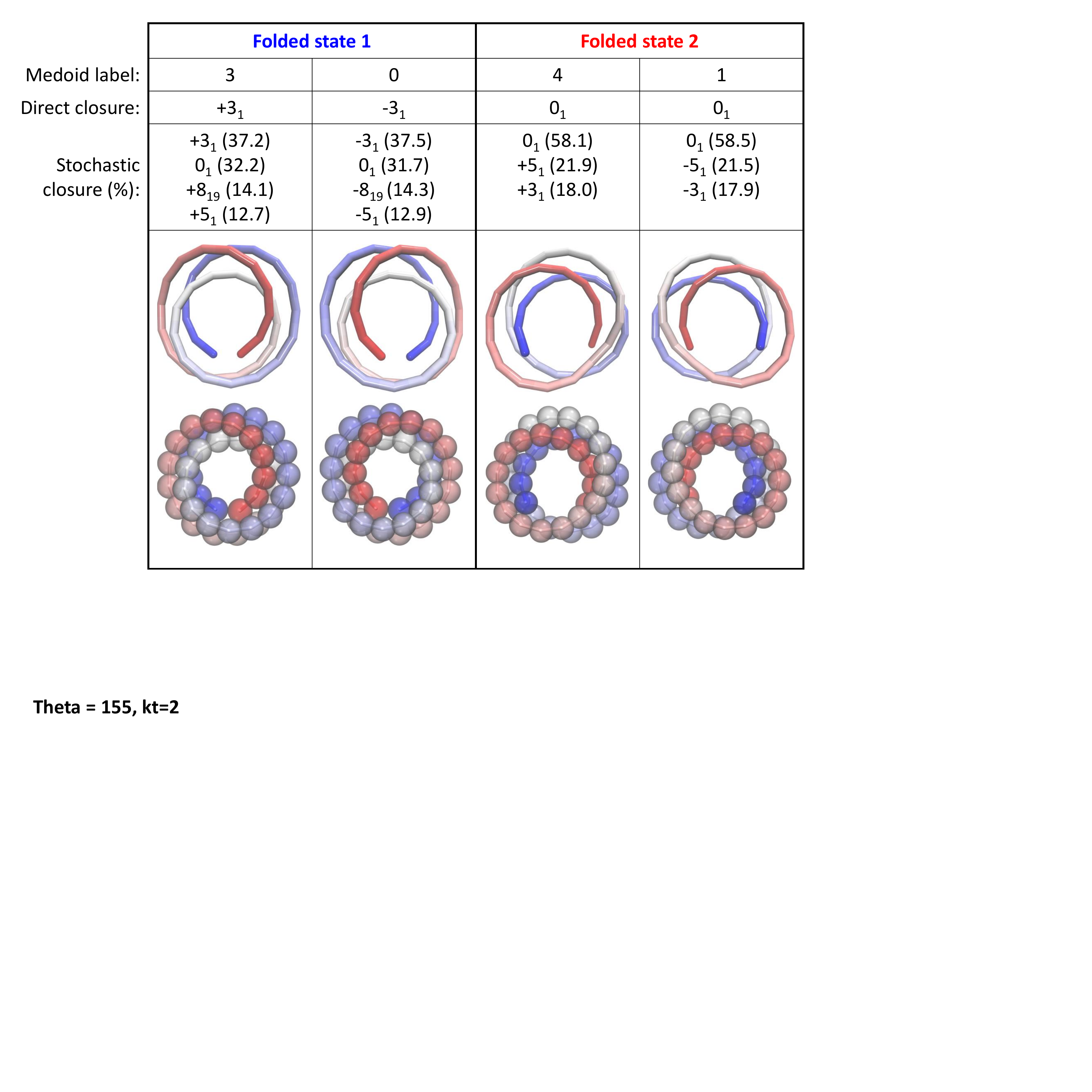"}
\includegraphics[trim=0in 6.25in 3in 0in, clip, height=4in, left]{"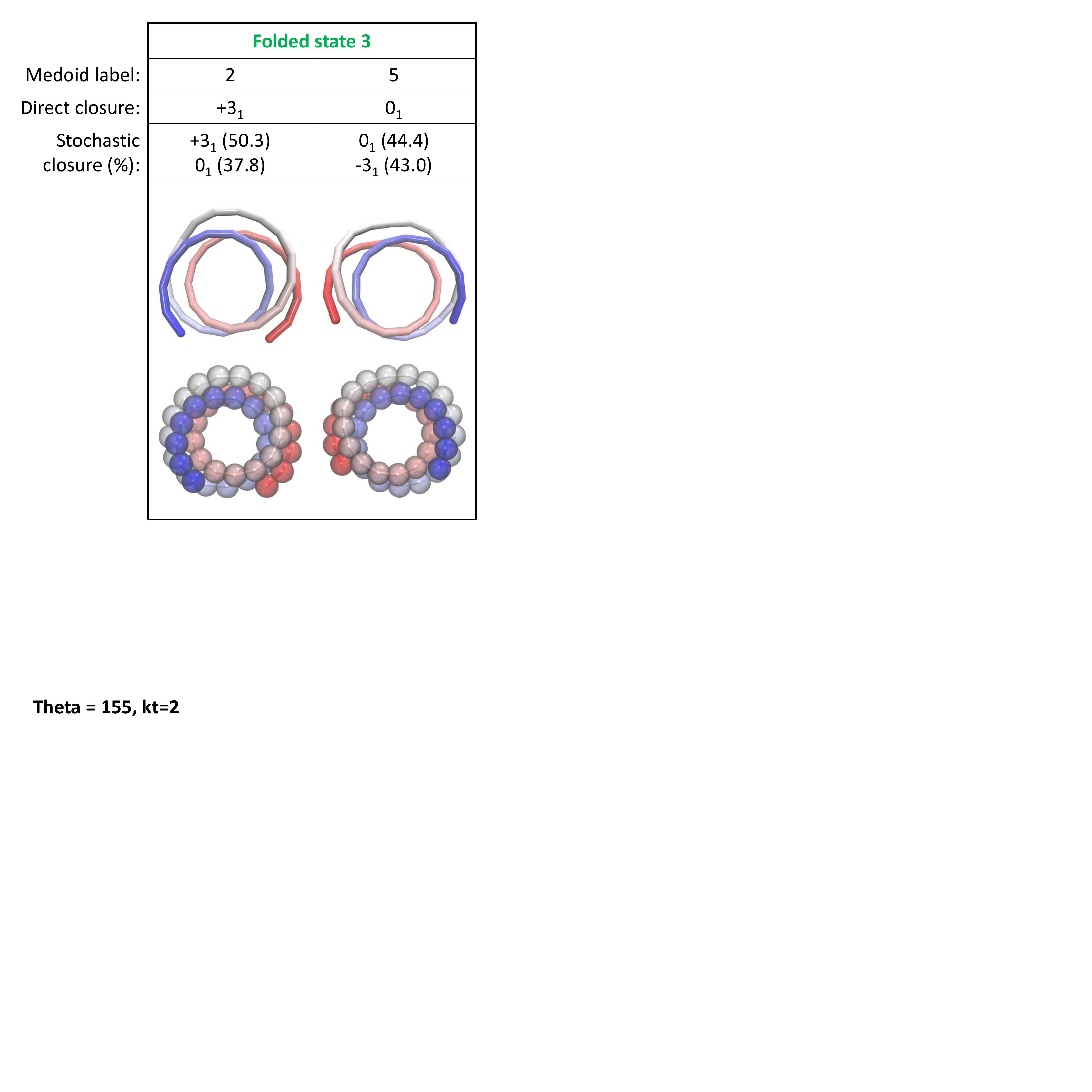"}
\label{fig:SI_medoids_theta155_kt2}
\end{figure}

\begin{figure}[H]
\includegraphics[trim=1in 0.5in 1in 0.5in, clip, height=9in]{"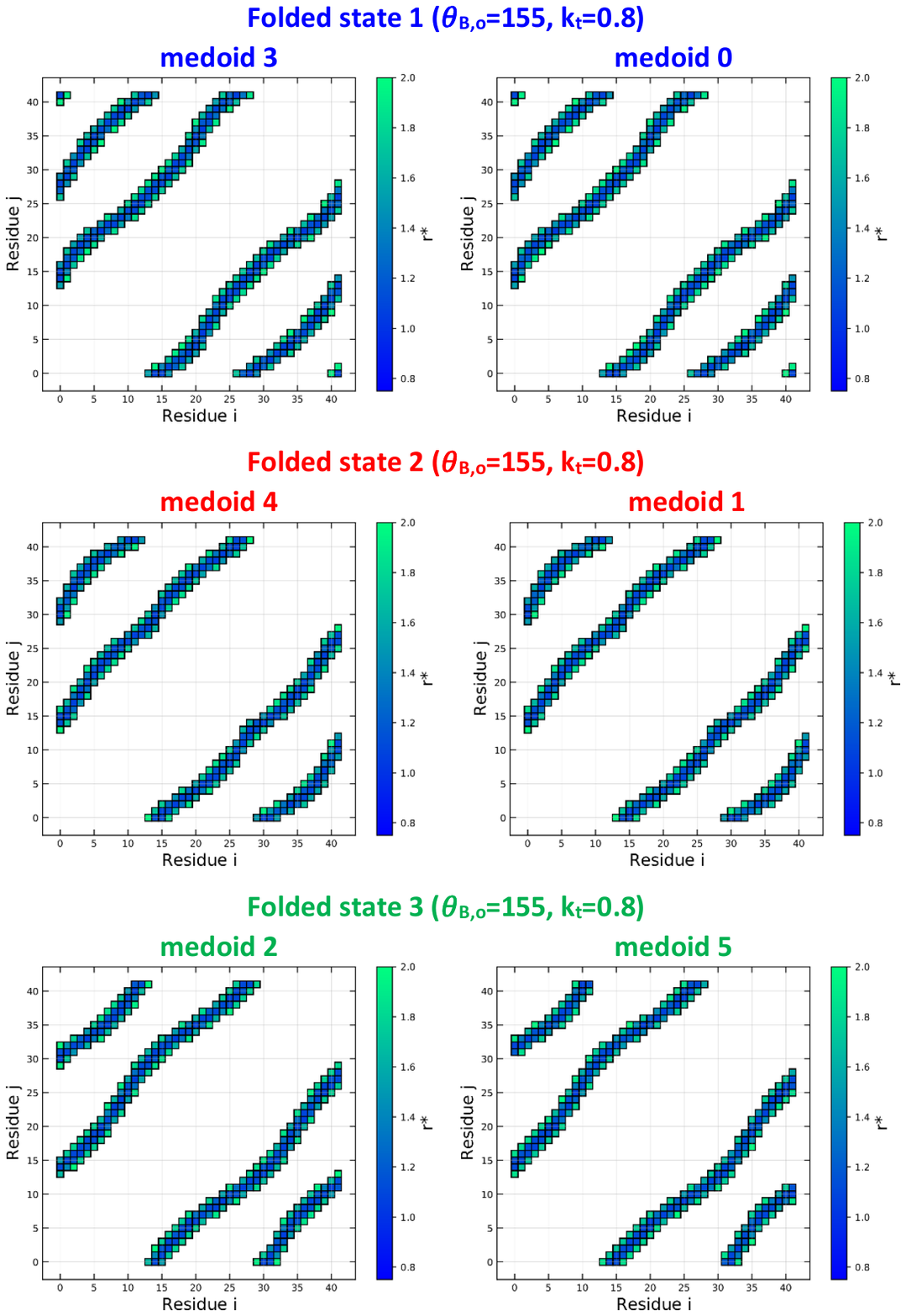"}
\label{fig:SI_contacts_theta155_kt2}
\end{figure}

\subsection{Medoids for $\theta_{B,o}=155^\circ, k_t=1.2$} \label{SI_medoids_theta155_kt12}

\begin{figure}[H]
\includegraphics[trim=0in 6.25in 3in 0in, clip, height=4in, left]{"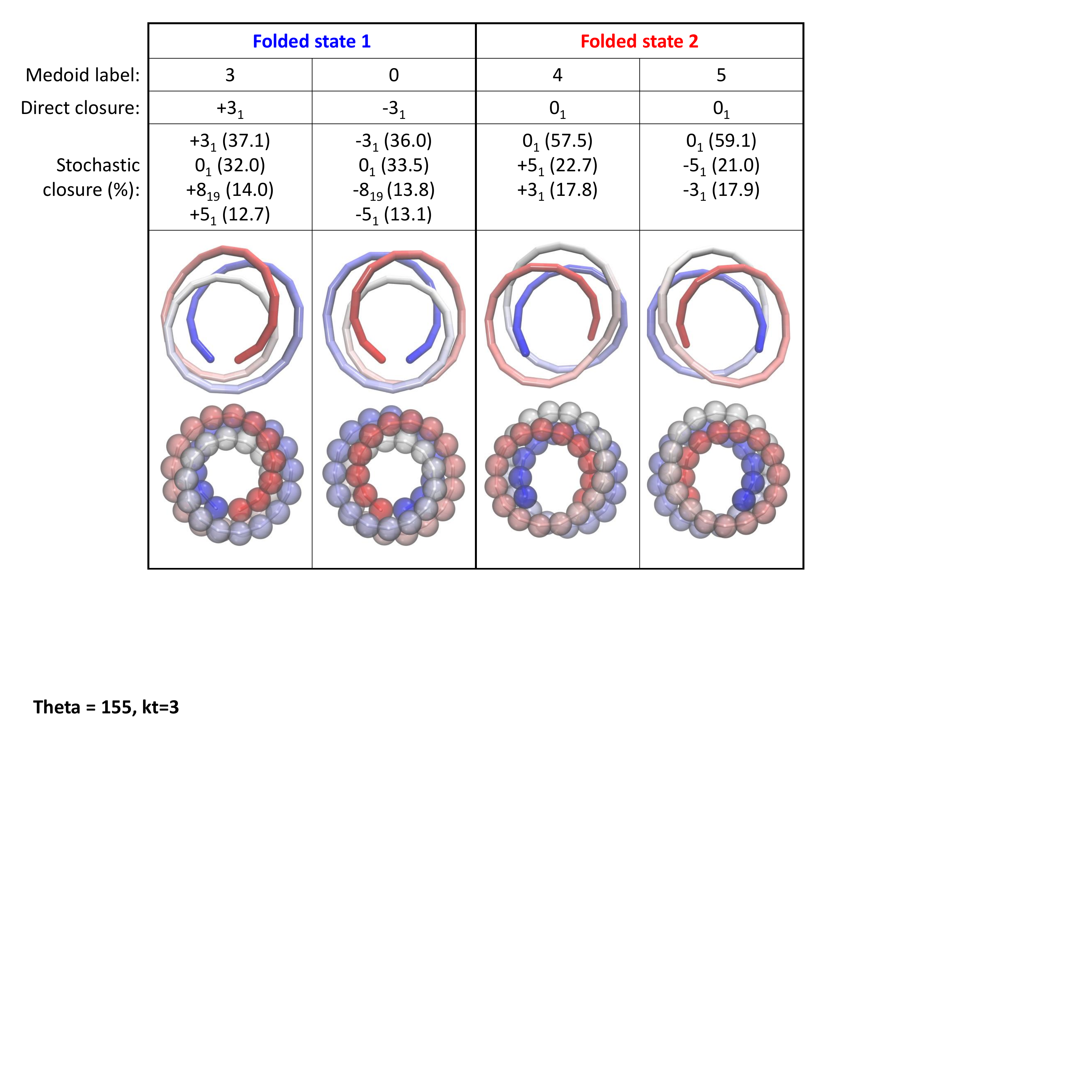"}
\includegraphics[trim=0in 6.25in 3in 0in, clip, height=4in, left]{"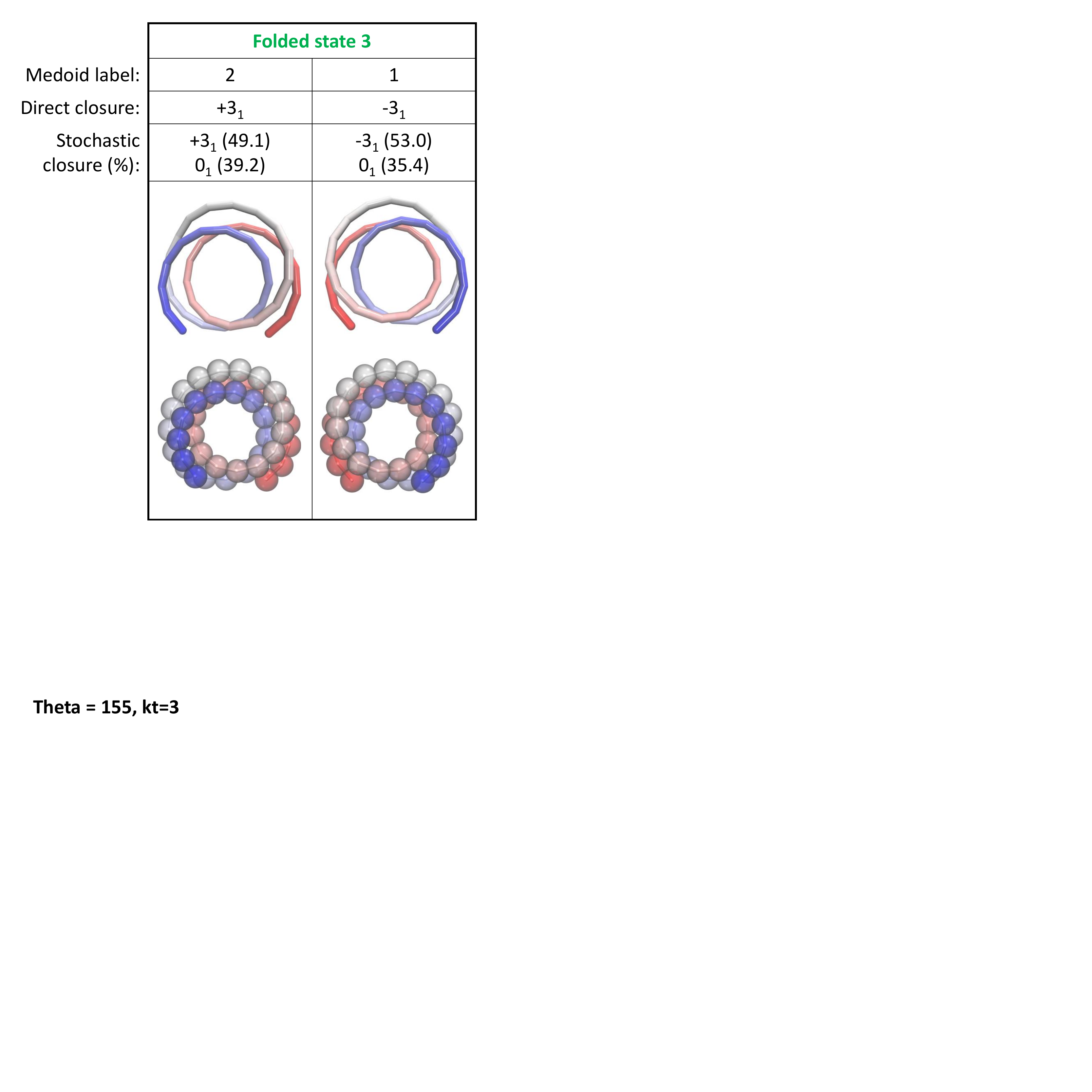"}
\label{fig:SI_medoids_theta155_kt3}
\end{figure}

\begin{figure}[H]
\includegraphics[trim=1in 0.5in 1in 0.5in, clip, height=9in]{"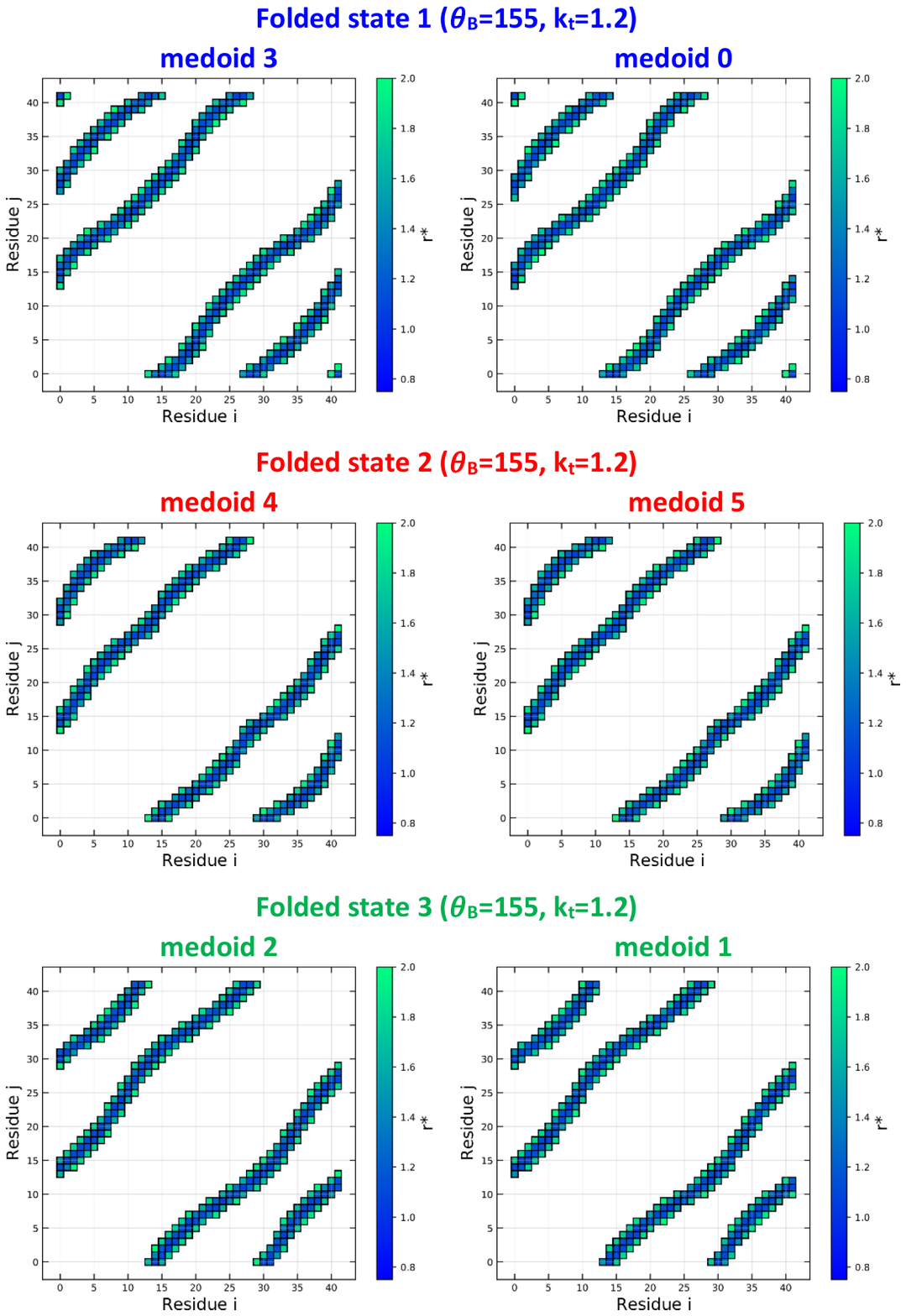"}
\label{fig:SI_contacts_theta155_kt3}
\end{figure}

\subsection{Medoids for $\theta_{B,o}=155^\circ, k_t=1.6$} \label{SI_medoids_theta155_kt16}

\begin{figure}[H]
\includegraphics[trim=0in 6.25in 3in 0in, clip, height=4in, left]{"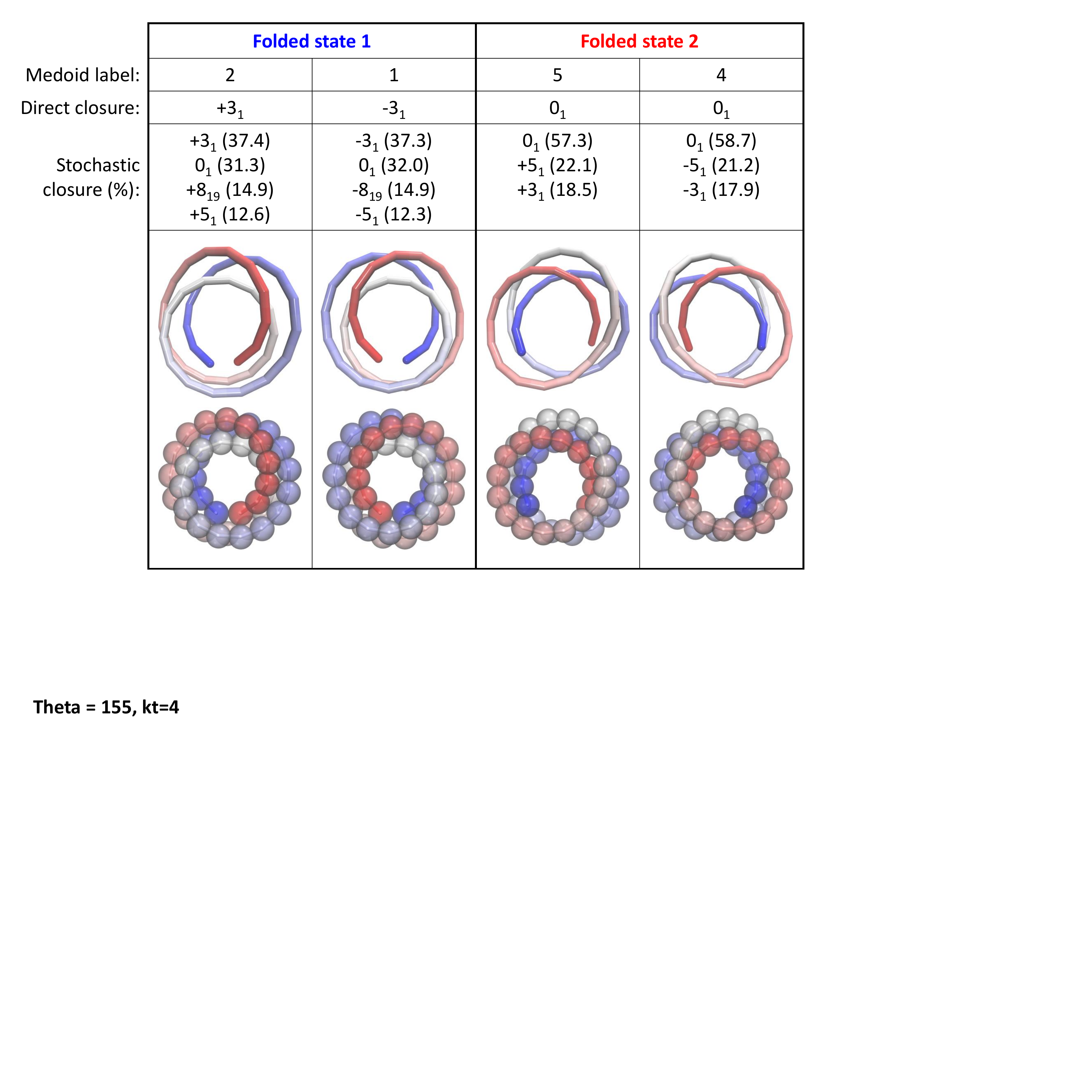"}
\includegraphics[trim=0in 6.25in 3in 0in, clip, height=4in, left]{"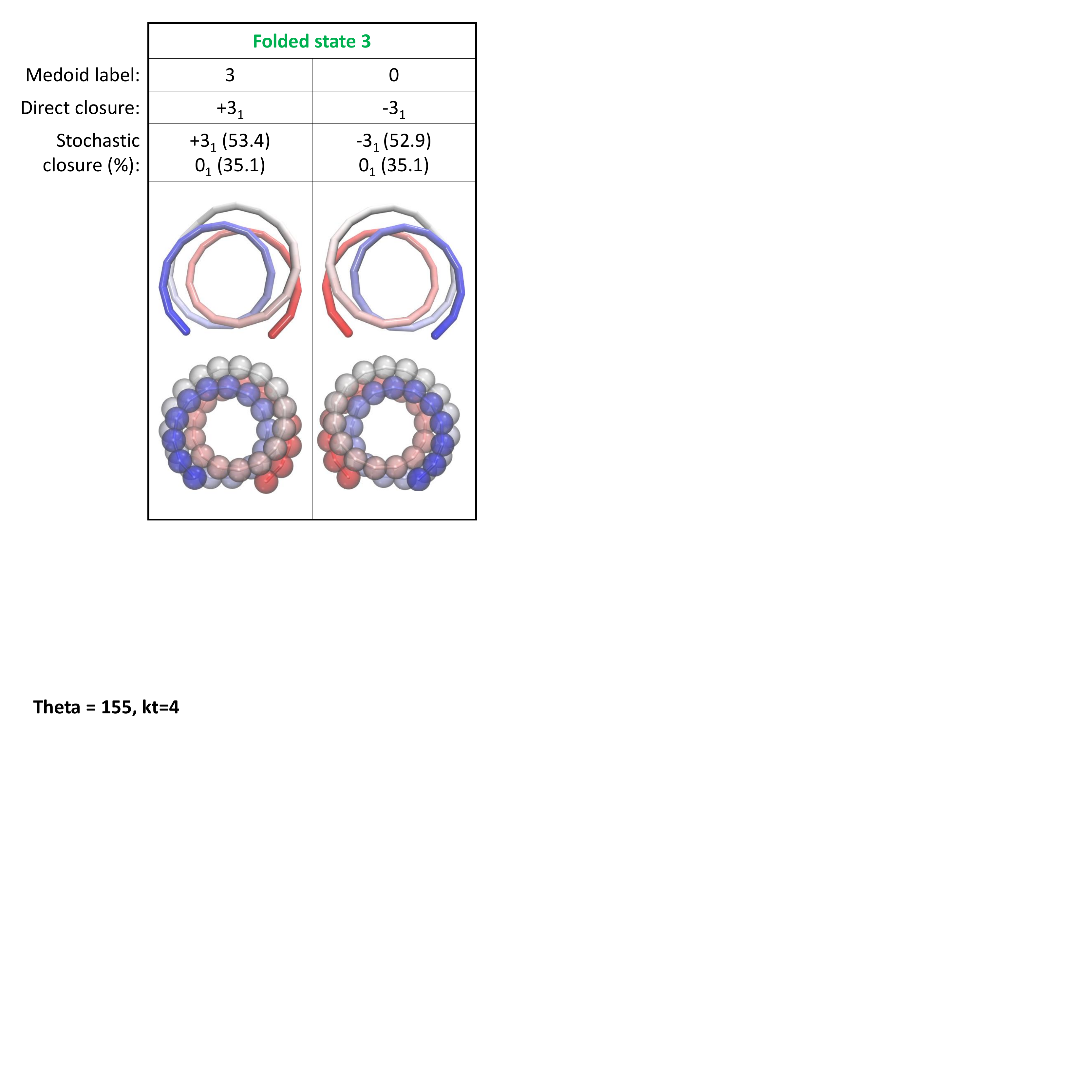"}
\label{fig:SI_medoids_theta155_kt4}
\end{figure}

\begin{figure}[H]
\includegraphics[trim=1in 0.5in 1in 0.5in, clip, height=9in]{"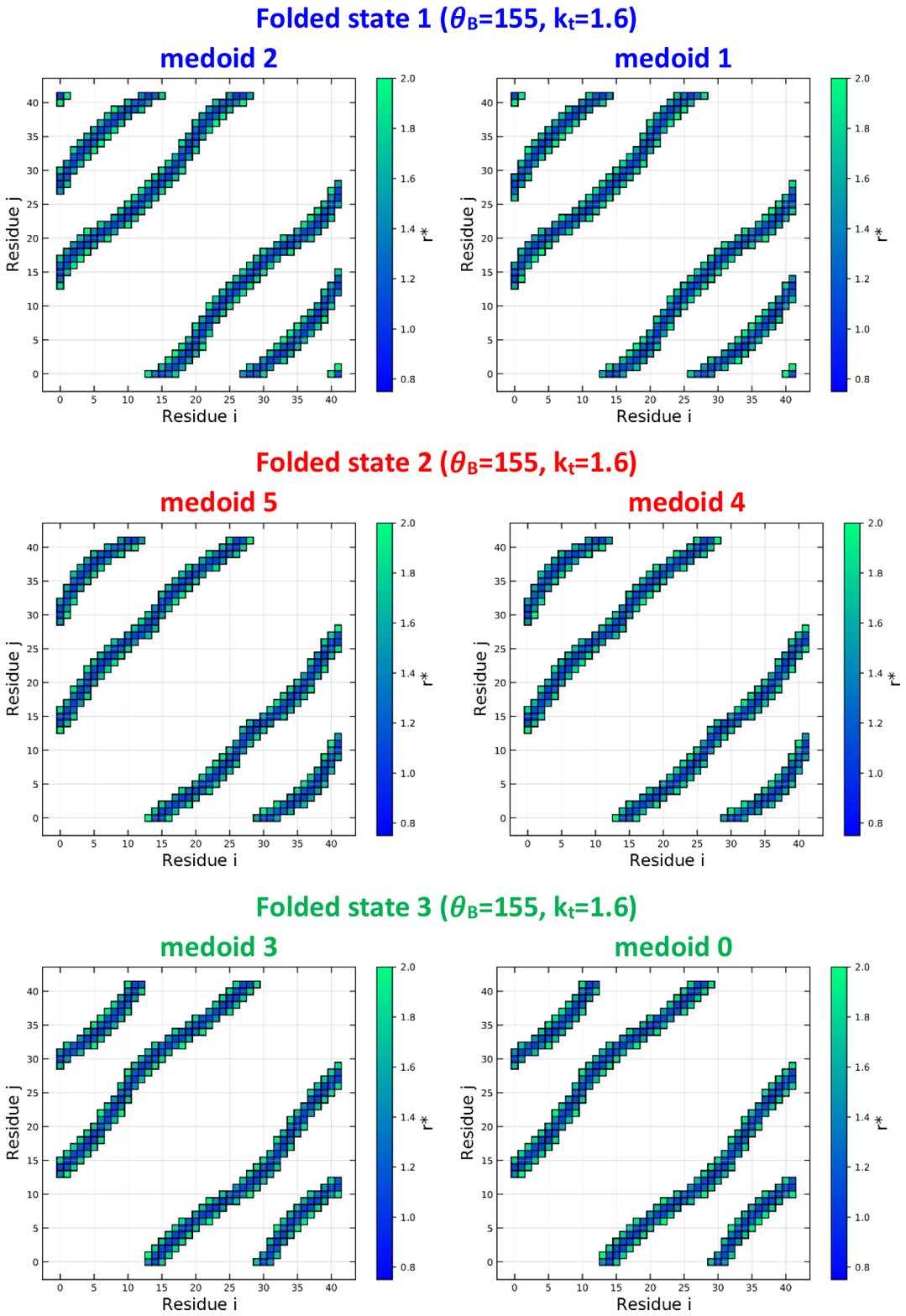"}
\label{fig:SI_contacts_theta155_kt4}
\end{figure}

\subsection{Medoids for $\theta_{B,o}=150^\circ, k_t=0$} \label{SI_medoids_theta150}

\begin{figure}[H]
\includegraphics[trim=0in 6.25in 3in 0in, clip, height=4in, left]{"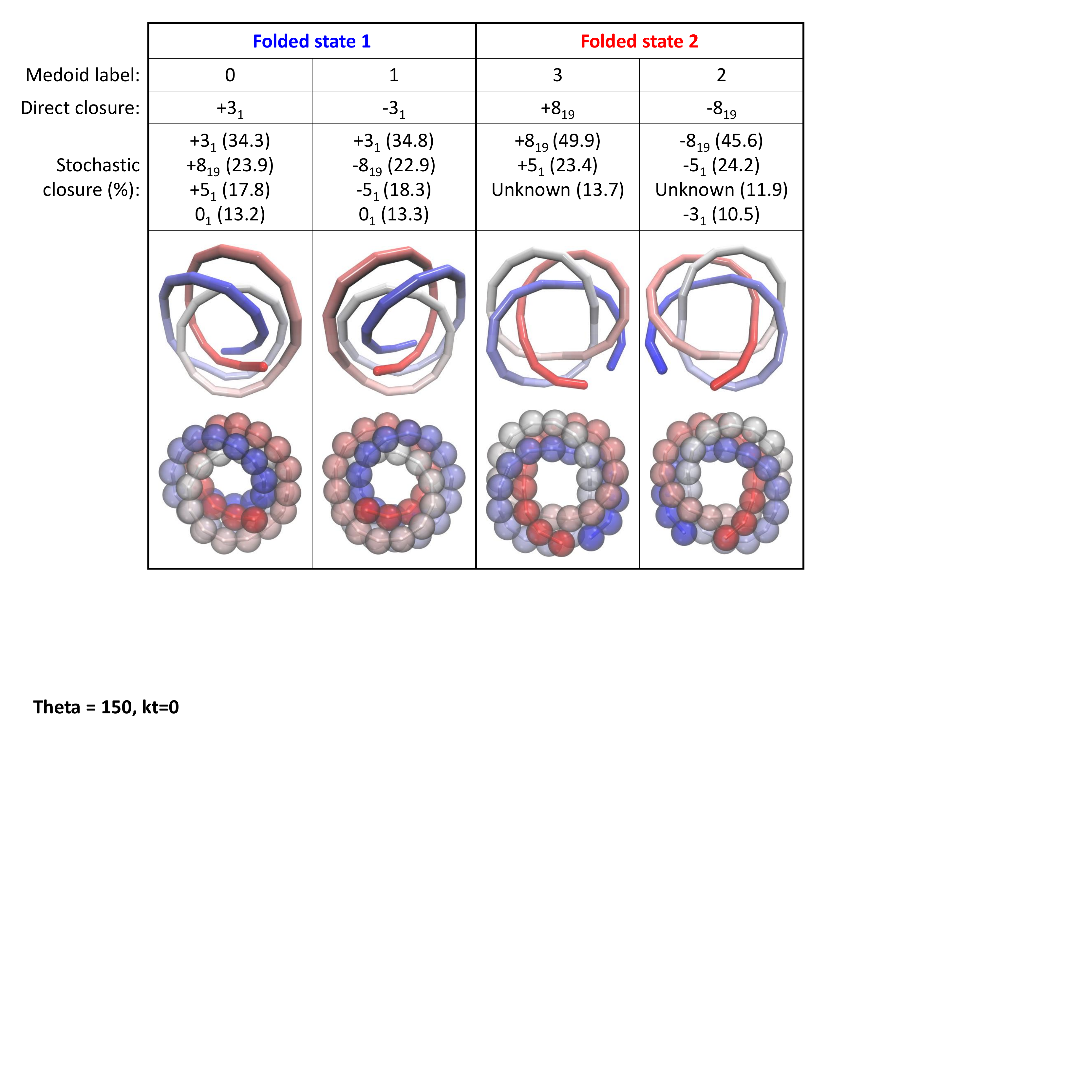"}
\label{fig:SI_medoids_theta150_kt0}
\end{figure}

\begin{figure}[H]
\includegraphics[trim=1in 0.5in 1in 0.5in, clip, width=6.5in]{"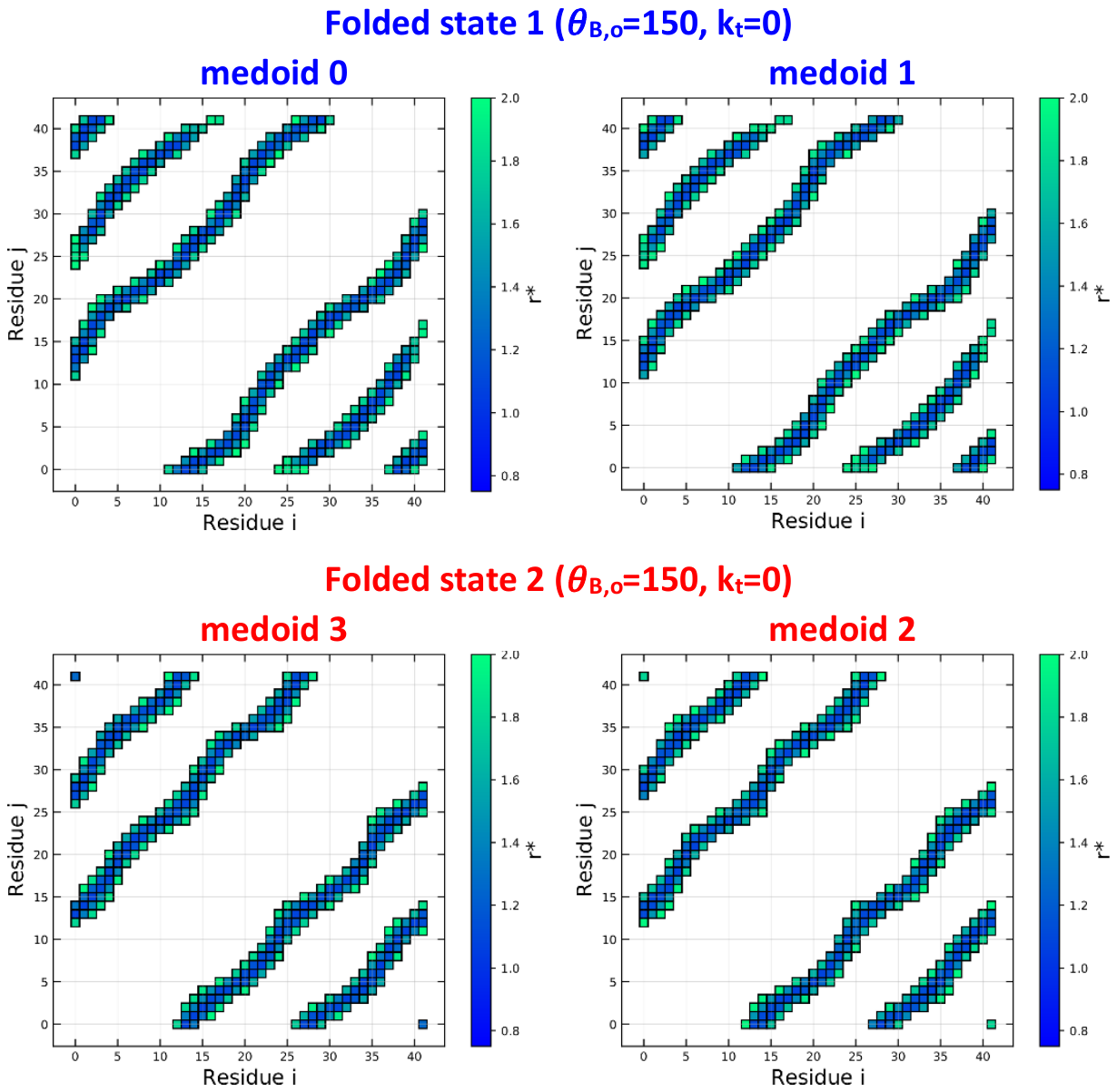"}
\label{fig:SI_contacts_theta150_kt0}
\end{figure}

\subsection{Medoids for $\theta_{B,o}=145^\circ, k_t=0$} \label{SI_medoids_theta145}

\begin{figure}[H]
\includegraphics[trim=0in 6in 1.5in 0in, clip, height=4in, left]{"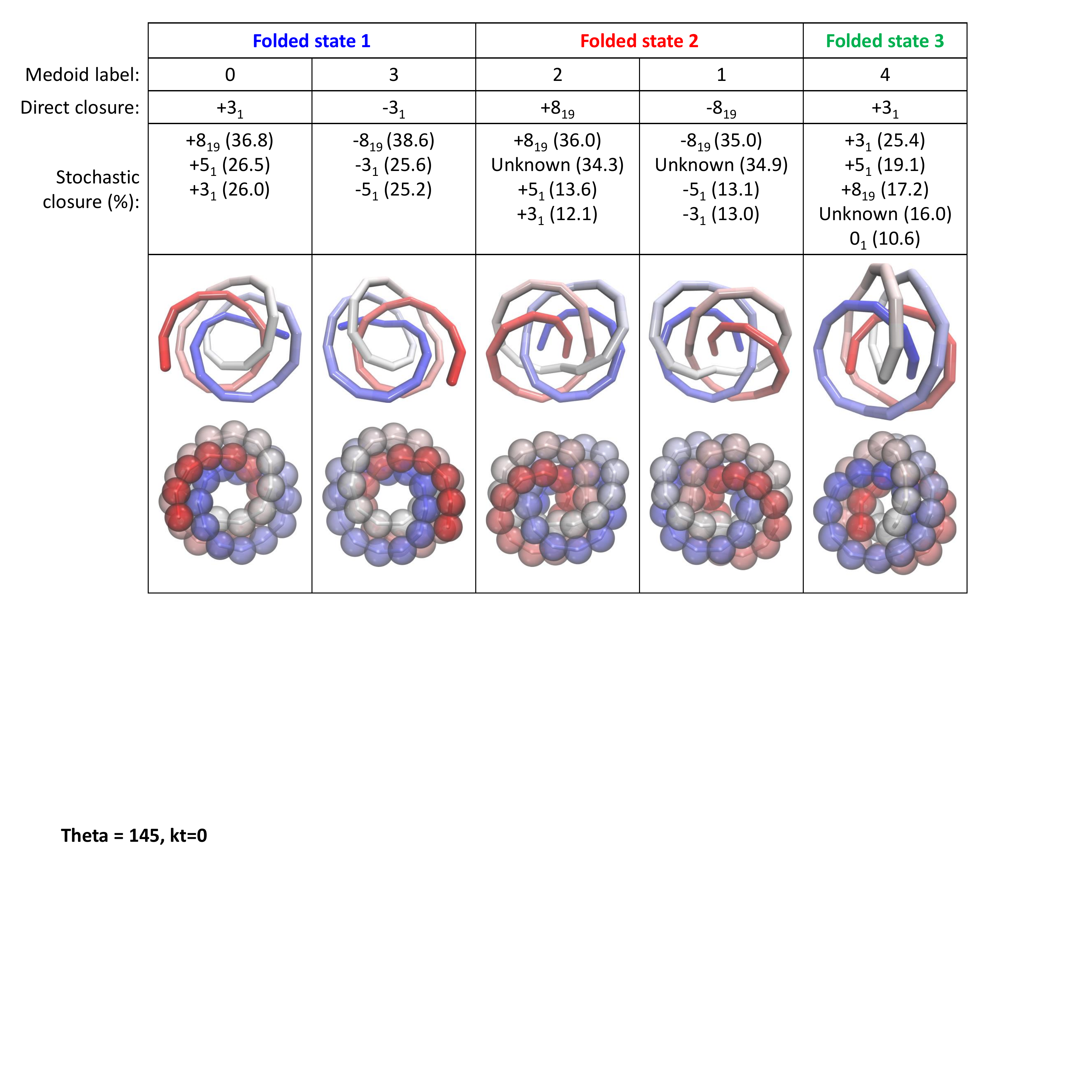"}
\label{fig:SI_medoids_theta145_kt0}
\end{figure}

\begin{figure}[H]
\includegraphics[trim=1in 0.5in 1in 0.5in, clip, width=6.5in]{"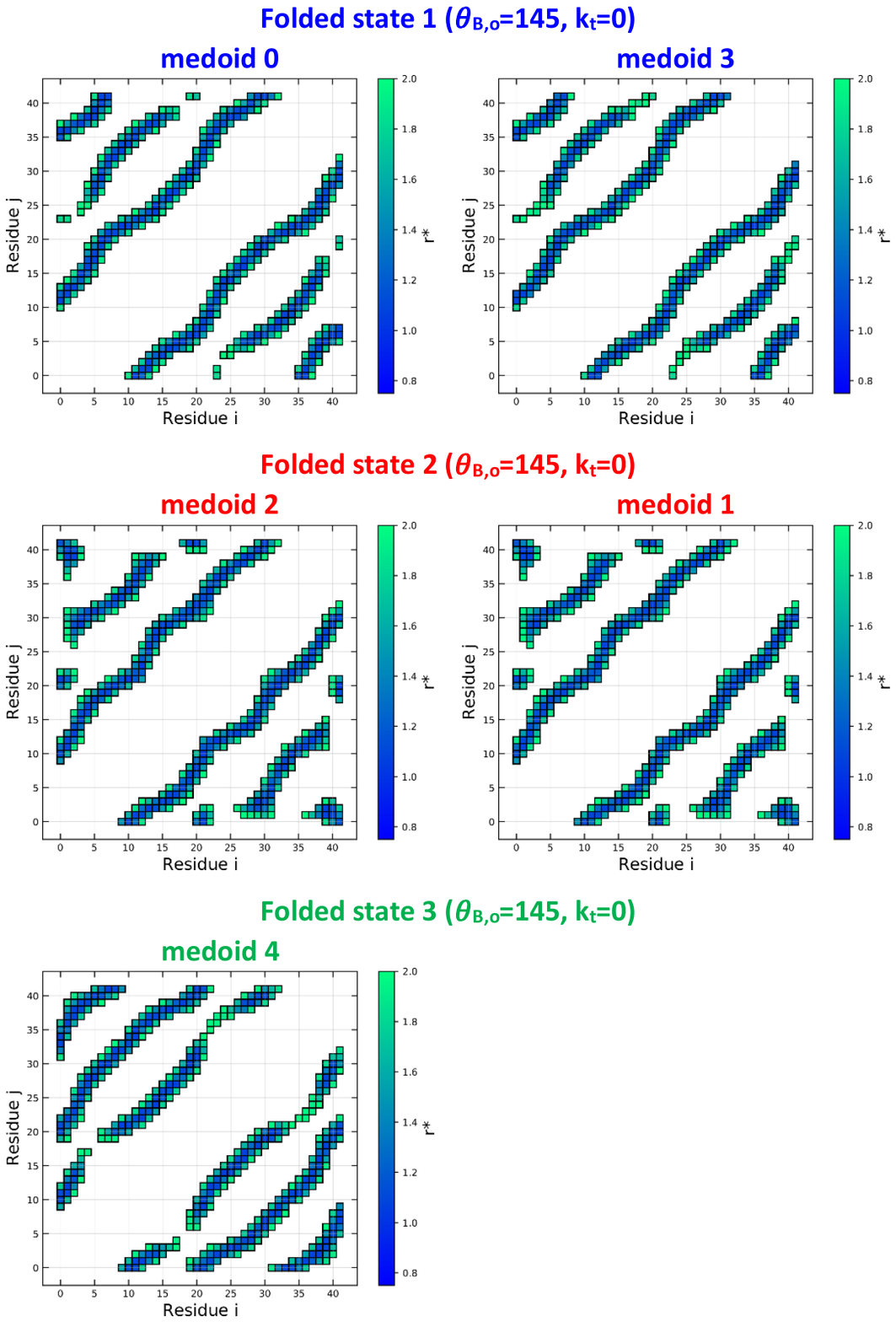"}
\label{fig:SI_contacts_theta145_kt0}
\end{figure}

\subsection{Medoids for $\theta_{B,o}=140^\circ, k_t=0$}~\label{SI_medoids_theta140}

\begin{figure}[H]
\includegraphics[trim=0in 6.25in 3in 0in, clip, height=4in, left]{"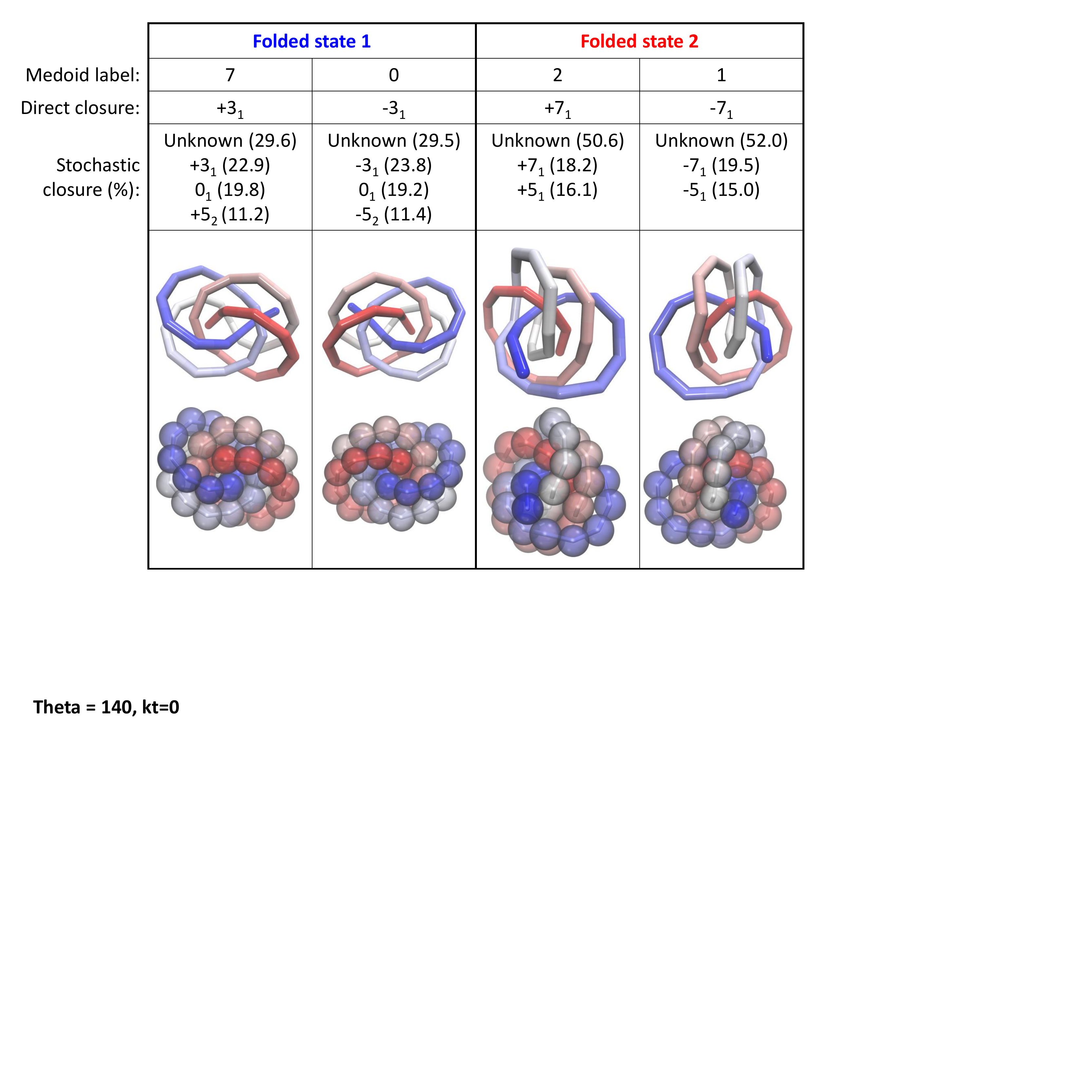"}
\includegraphics[trim=0in 6.25in 3in 0in, clip, height=4in, left]{"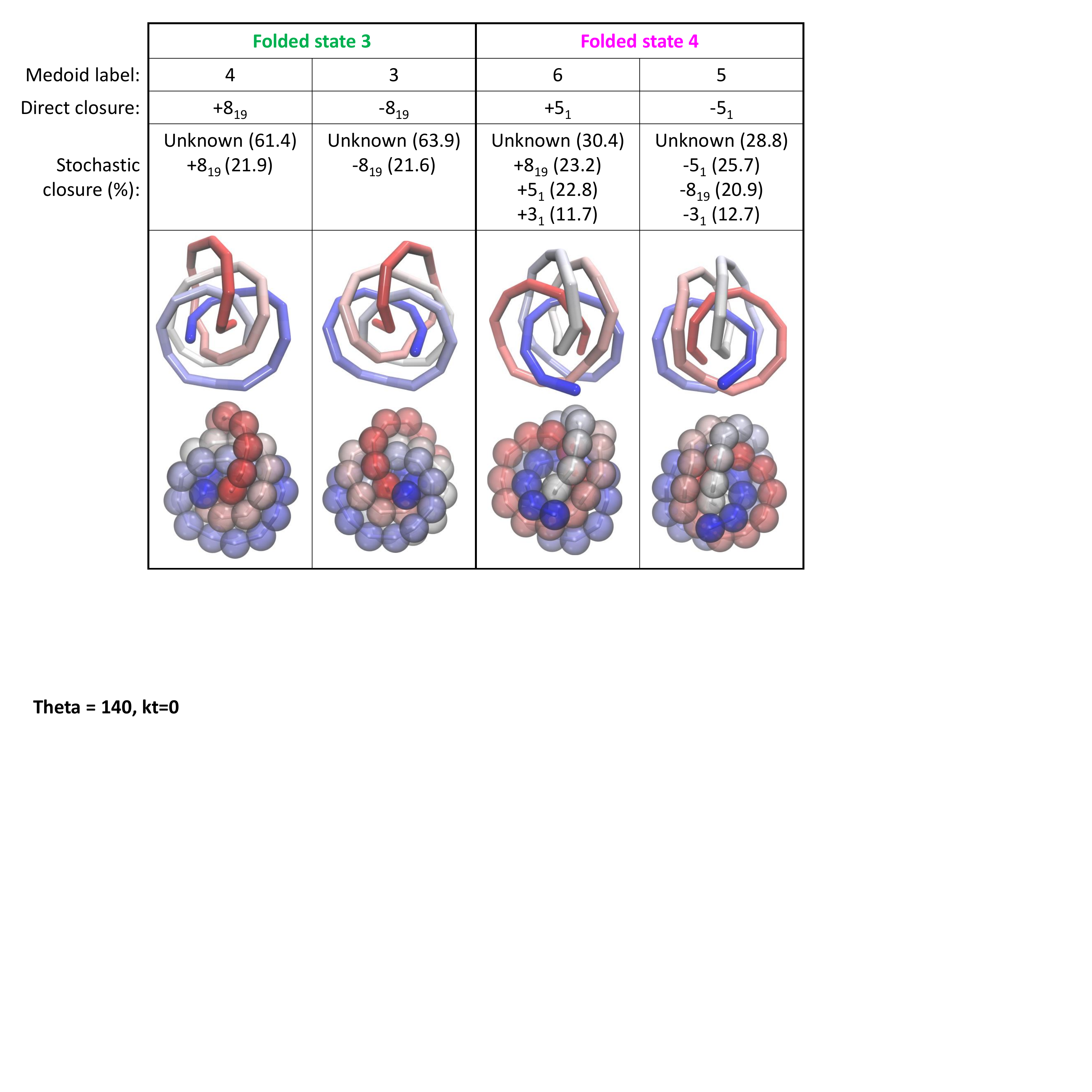"}
\label{fig:SI_medoids_theta140_kt0}
\end{figure}

\begin{figure}[H]
\includegraphics[trim=1in 0.5in 1in 0.5in, clip, height=9in]{"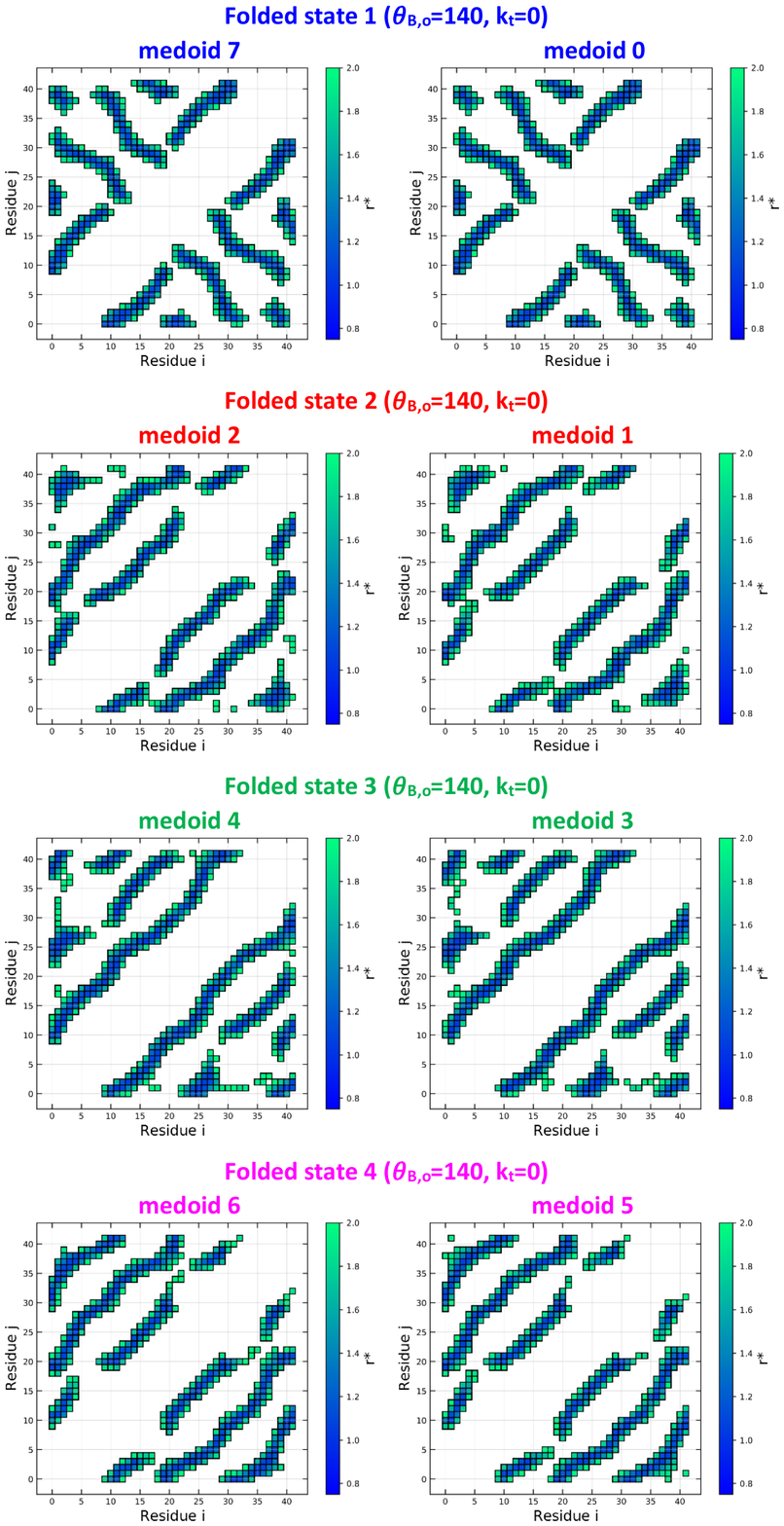"}
\label{fig:SI_contacts_theta140_kt0}
\end{figure}

\subsection{Medoids for $\theta_{B,o}=155^\circ, k_t=0.4$ (1-1 model)} \label{SI_medoids_theta155_sc}
\begin{figure}[H]
\includegraphics[trim=0in 6.25in 3in 0in, clip, height=4in, left]{"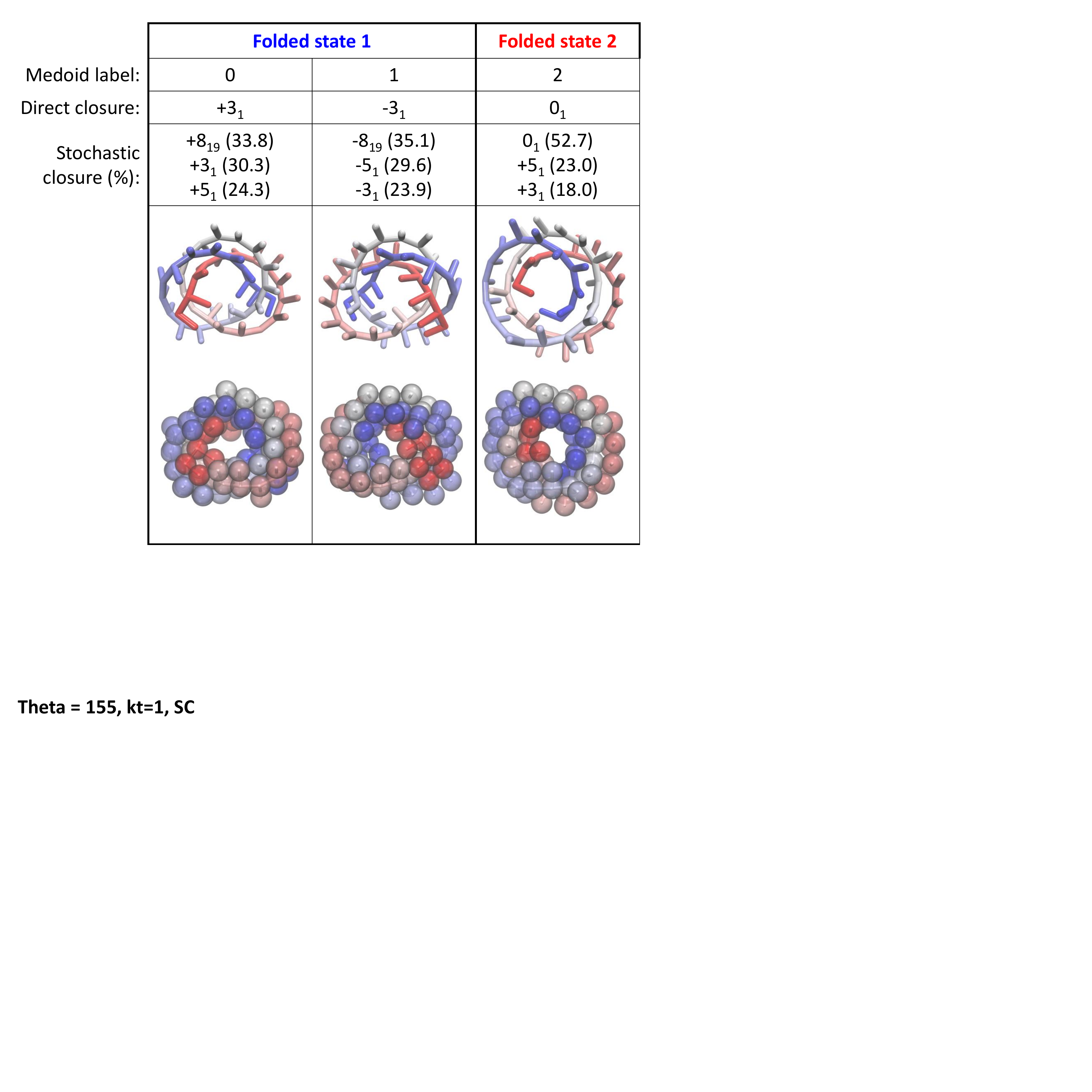"}
\includegraphics[trim=1in 0.5in 1in 0.5in, clip, width=6.5in]{"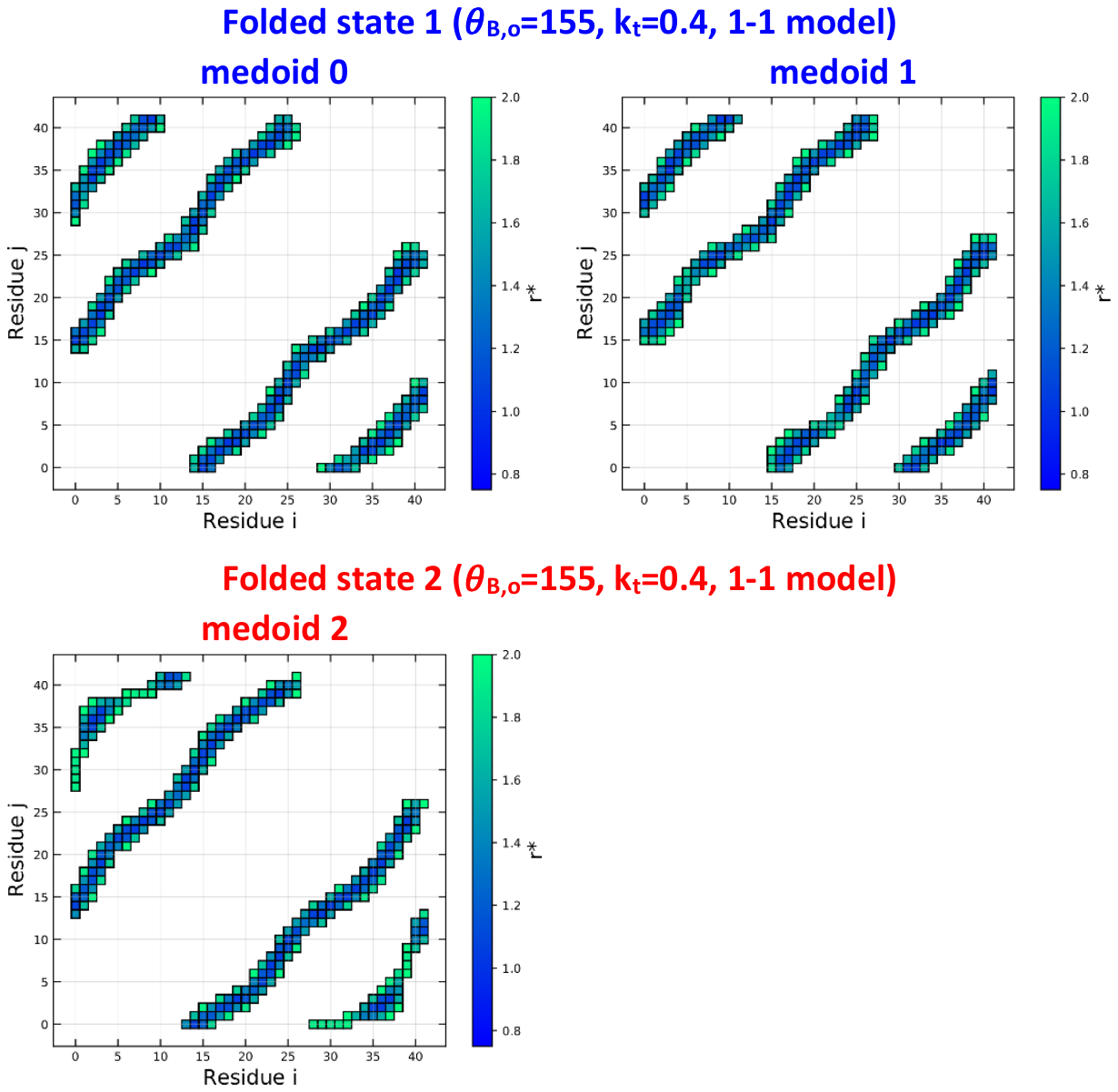"}
\label{fig:SI_contacts_theta155_kt1_SC}
\end{figure}

\subsection{Medoids for $\theta_{B,o}=150^\circ, k_t=0.4$ (1-1 model)} \label{SI_medoids_theta150_sc}
\begin{figure}[H]
\includegraphics[trim=0in 6.25in 3in 0in, clip, height=4in, left]{"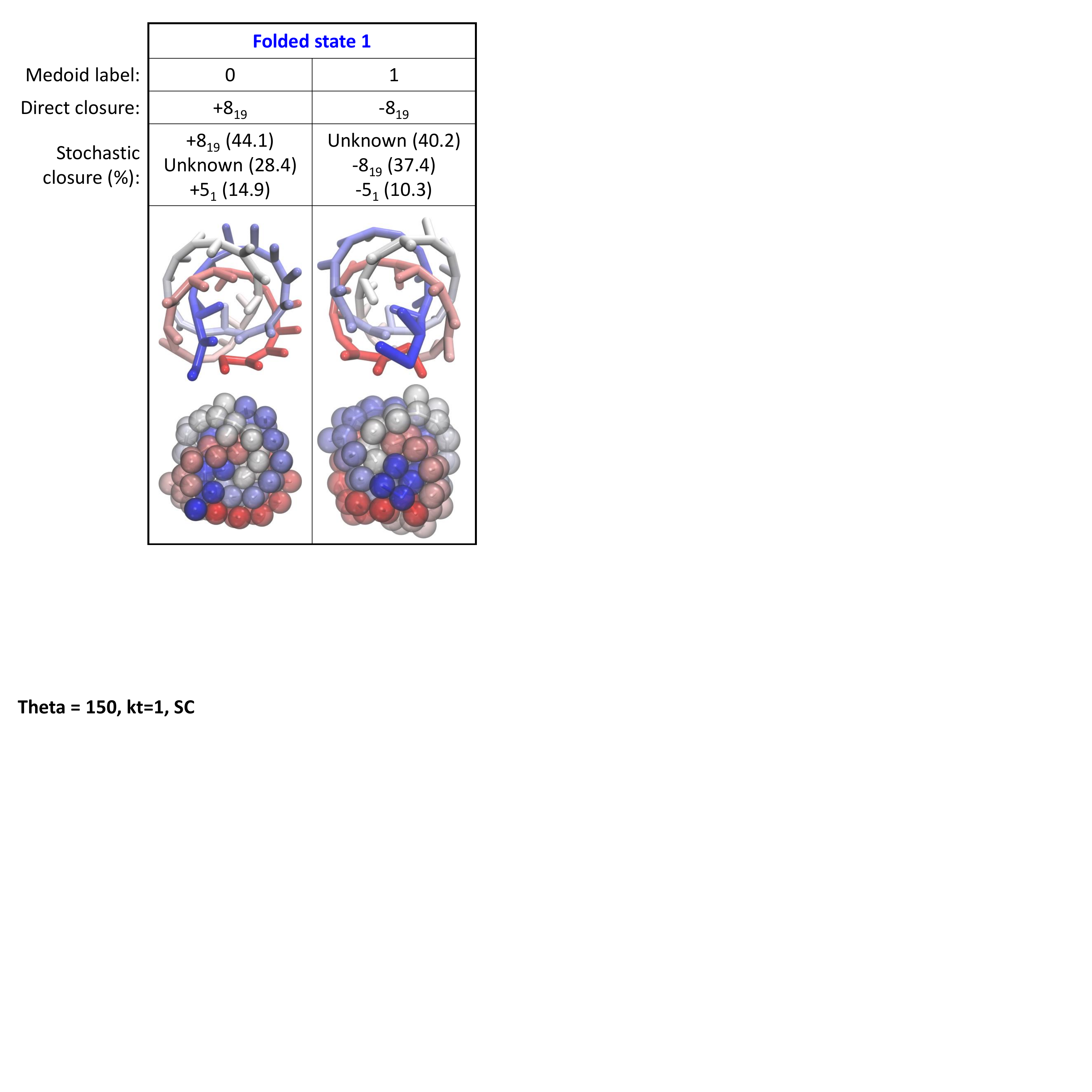"}
\includegraphics[trim=1in 0.5in 1in 0.5in, clip, width=6.5in]{"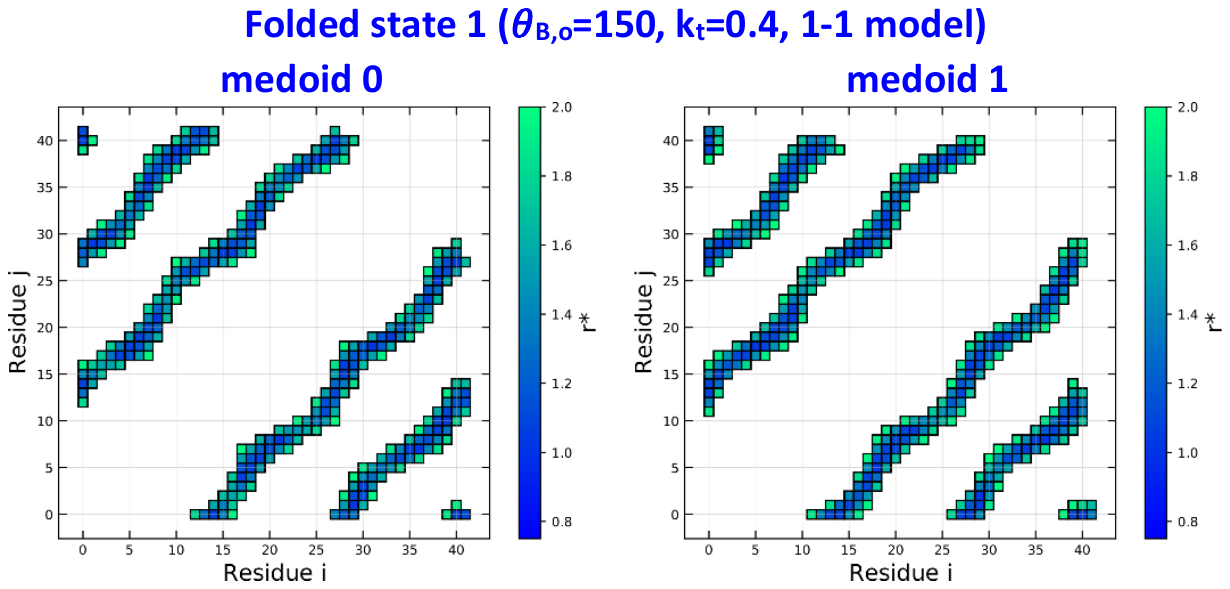"}
\label{fig:SI_contacts_theta150_kt1_SC}
\end{figure}

\newpage
\subsection{Cluster compositions by temperature state} \label{SI_cluster_populations}

\begin{figure}[H]
\includegraphics[width=6.5in]{"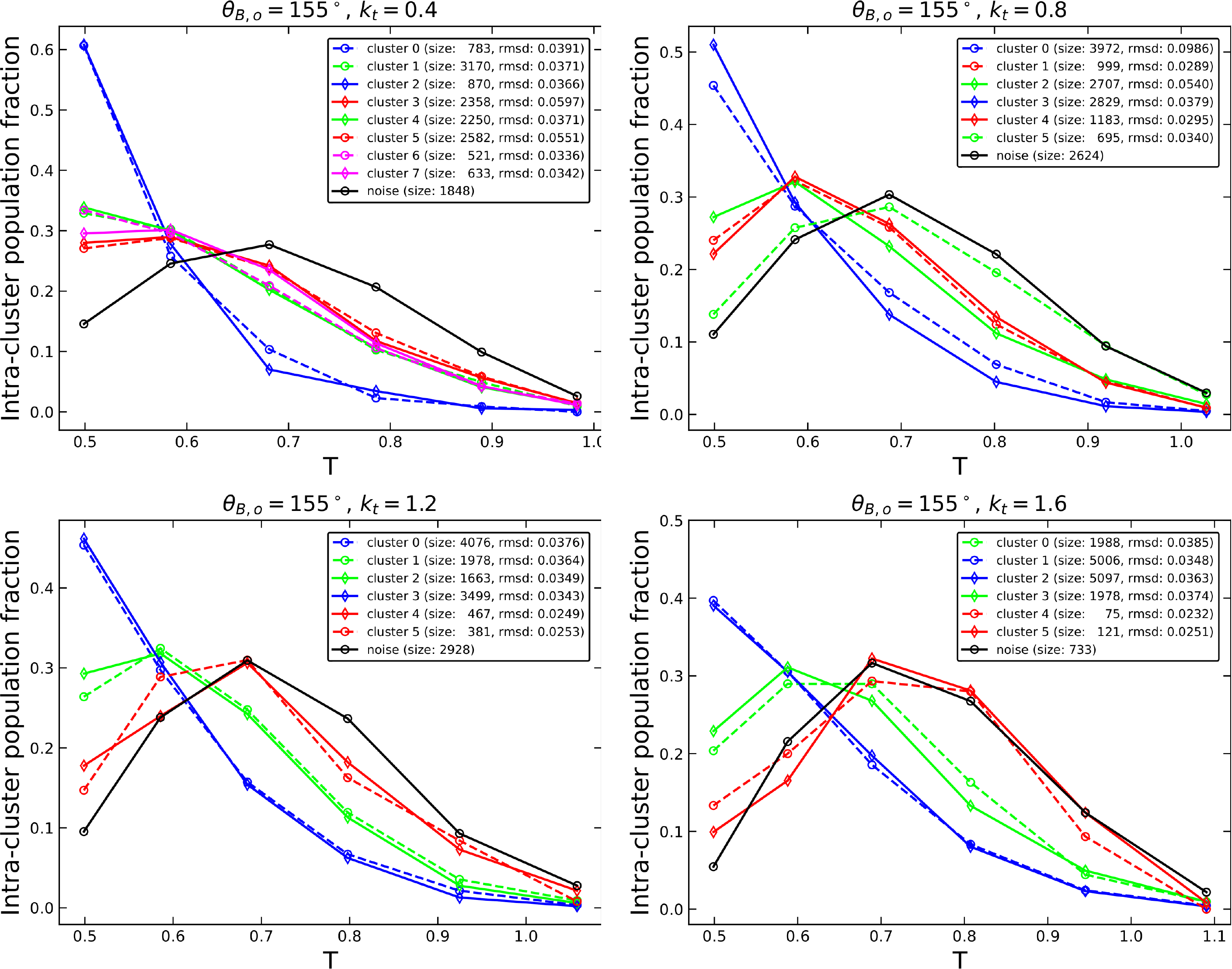"}
\caption{Fraction of structures within each cluster as a function of temperature for linear homo-oligomers with $\theta_{B,o}=155^\circ$ and varying torsion stiffness confirm that right and left-handed knots are sampled approximately equally in the REMD simulations. Note that only the lowest 6 temperature states in each case are input into the clustering algorithm. In addition, the pre-clustering filtering step removes much of the remaining higher-temperature data, resulting in the peak in DBSCAN noise (black) occurring at intermediate temperatures. Colors correspond to the folded state classifications used throughout this work. Dashed lines with circle markers correspond to left-handed knots (using direct closure), and solid lines with diamond markers correspond to right-handed knots. Unknots are assigned the handedness of the most common knot type from stochastic closure. Cluster sizes and RMSD to the medoid are shown in the legend.}
\label{fig:SI_cluster_fraction_kt}
\end{figure}

\begin{figure}[H]
\includegraphics[width=6.5in]{"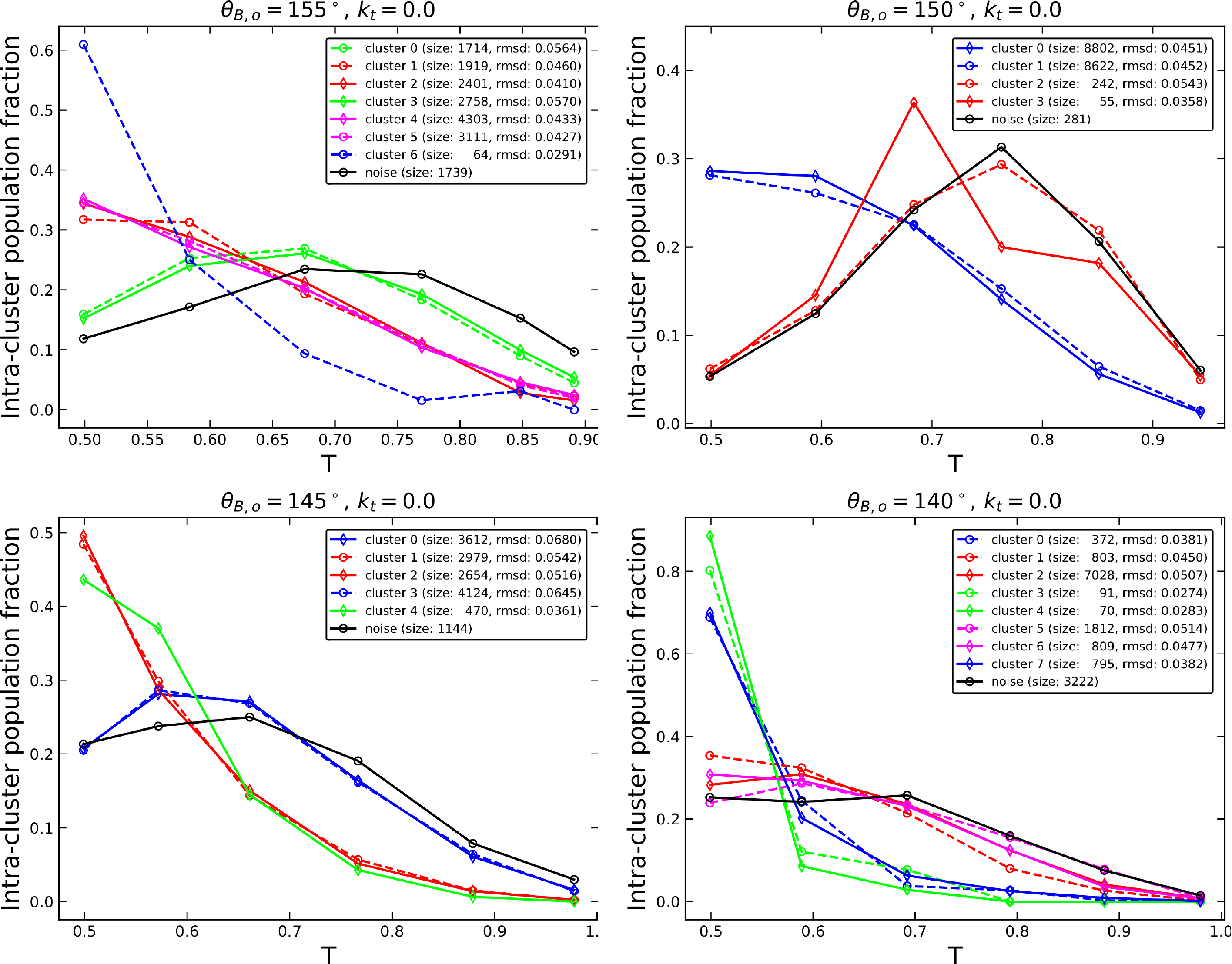"}
\caption{Fraction of structures within each cluster as a function of temperature for linear homo-oligomers with $k_t=0.0$ and varying $\theta_{B,o}$ confirm that right and left-handed knots are sampled approximately equally in the REMD simulations. Plot symbols and colors are as described in Figure \ref{fig:SI_cluster_fraction_kt}.}
\label{fig:SI_cluster_fraction_theta}
\end{figure}

\begin{figure}[H]
\includegraphics[width=6.5in]{"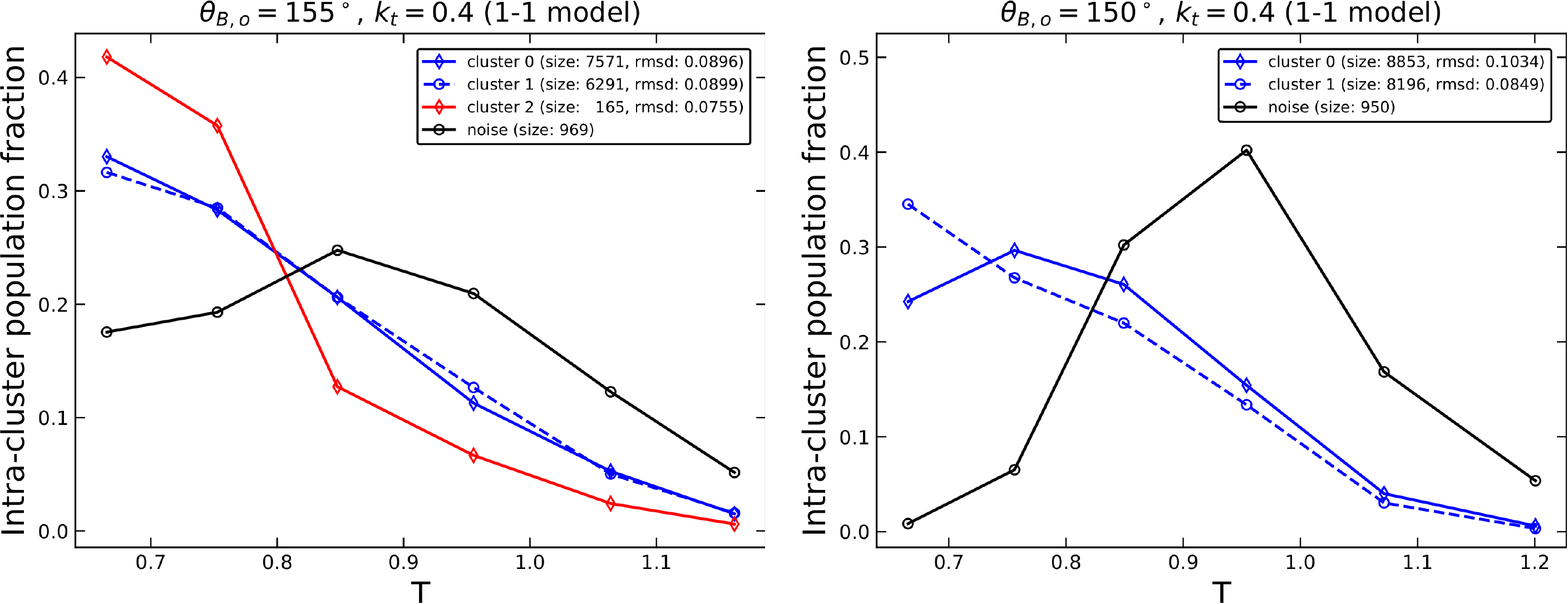"}
\caption{Fraction of structures within each cluster as a function of temperature for 1-1 homo-oligomers with $k_t=0.4$ and varying $\theta_{B,o}$ confirm that right and left-handed knots are sampled approximately equally in the REMD simulations. Plot symbols and colors are as described in Figure \ref{fig:SI_cluster_fraction_kt}.}
\label{fig:SI_cluster_fraction_sc}
\end{figure}

\section{Visualization of configurational state ensembles} \label{SI_conf_state_ensembles}

\begin{figure}[H]
\includegraphics[trim=1in 0.785in 0.5in 0in, clip, width=6.5in]{"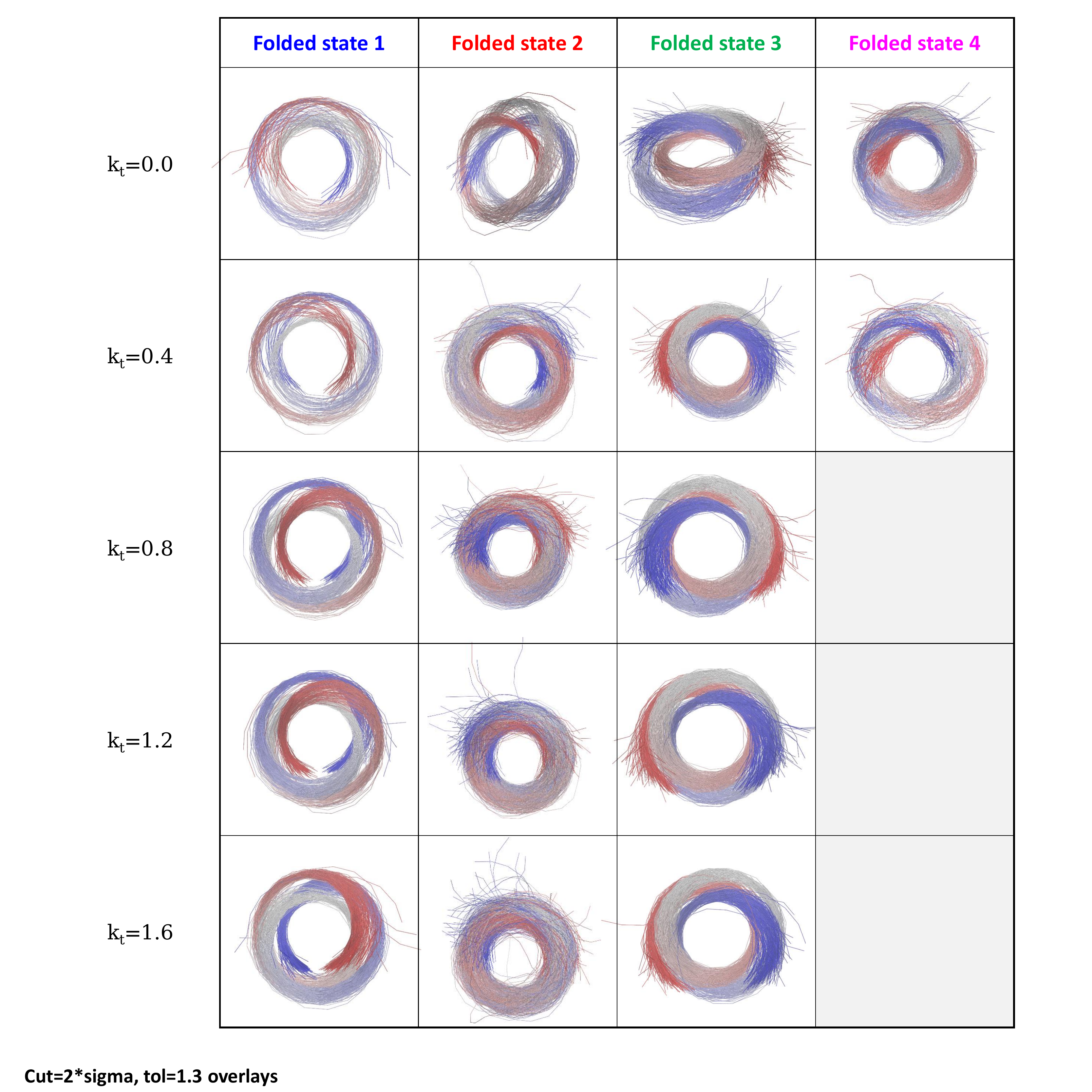"}
\label{fig:SI_tol13_cut20_vary_kt_ensembles}
\caption{Visualization of representative structures in each knotted state for linear models, with native contact cutoff of $2\sigma$, native distance tolerance factor of 1.3, $\theta_{B,o}=155^\circ$ and varying $k_t$ indicate that the states are stable, distinct, and configurationally well-defined. Structures are selected across all temperatures with a stride of 12500 frames, except for $k_t=0.0$ folded state 1, where a stride of 2500 frames is used due to the small population size. Structures are aligned with respect to end-to-end and mirror symmetry to the reference medoid in each case, which is the enantiomer whose cluster size was largest.}
\end{figure}

\begin{figure}[H]
\includegraphics[trim=1in 5.45in 0.5in 0in, clip, width=6.5in]{"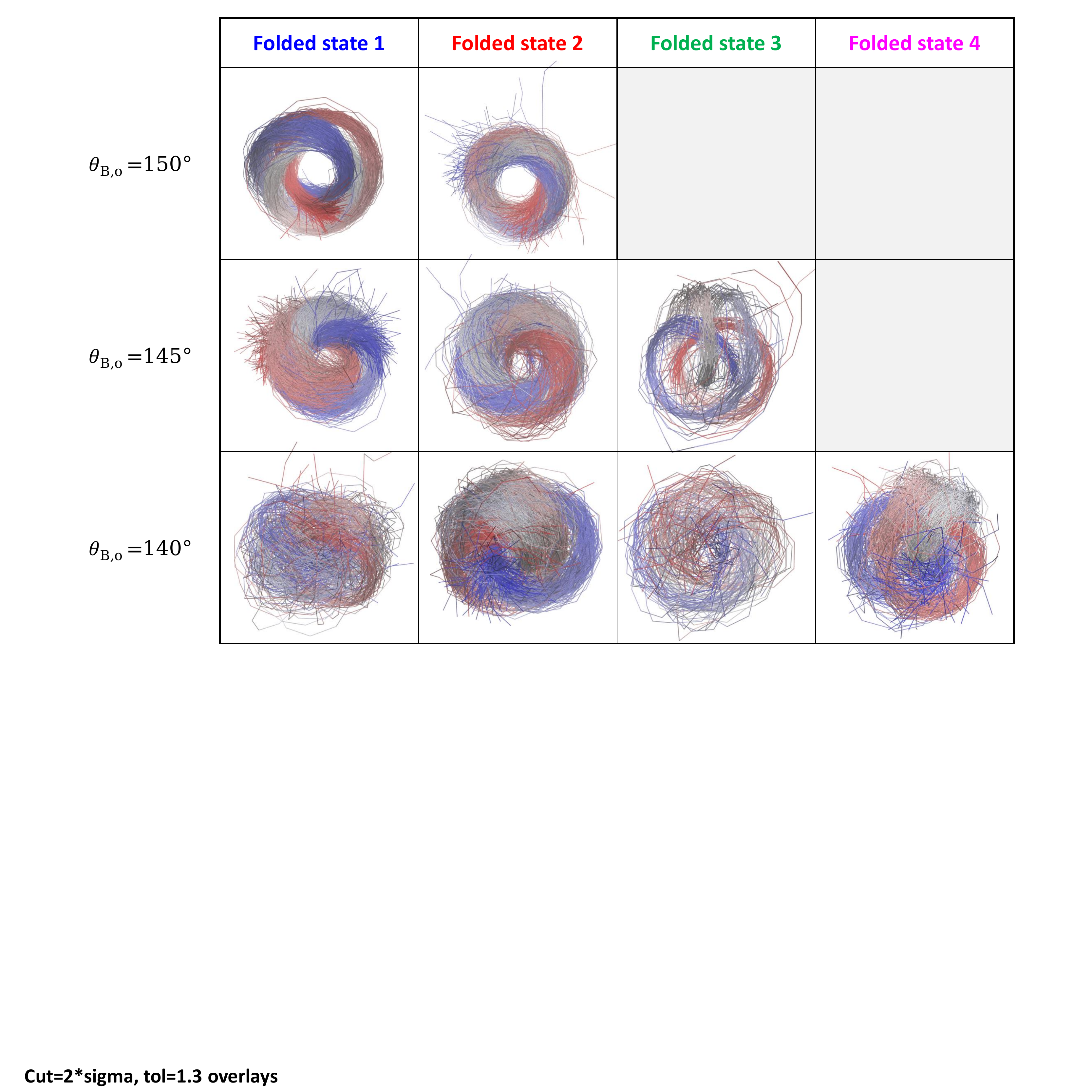"}
\label{fig:SI_tol13_cut20_vary_theta_ensembles}
\caption{Visualization of representative structures in each knotted state for linear models, with native contact cutoff of $2\sigma$, native distance tolerance factor of 1.3, $k_t=0.0$ and varying $\theta_{B,o}$ indicate that the states are stable, distinct, and configurationally well-defined, with the possible exception of $\theta_{B,o}=140^\circ$ state 3. Recall that the knotting steps in $\theta_{B,o}=140^\circ$ were found to be only marginally cooperative. Structures are selected across all temperatures with a stride of 12500 frames, except for $\theta_{B,o}=145$ folded state 3, where a stride of 2500 frames is used instead due to a small population size. Structures are aligned with respect to end-to-end and mirror symmetry to the reference medoid in each case, which is the enantiomer whose cluster size was largest.}
\end{figure}

\begin{figure}[H]
\includegraphics[trim=0.5in 7.75in 5.0in 0in, clip, width=5.5in]{"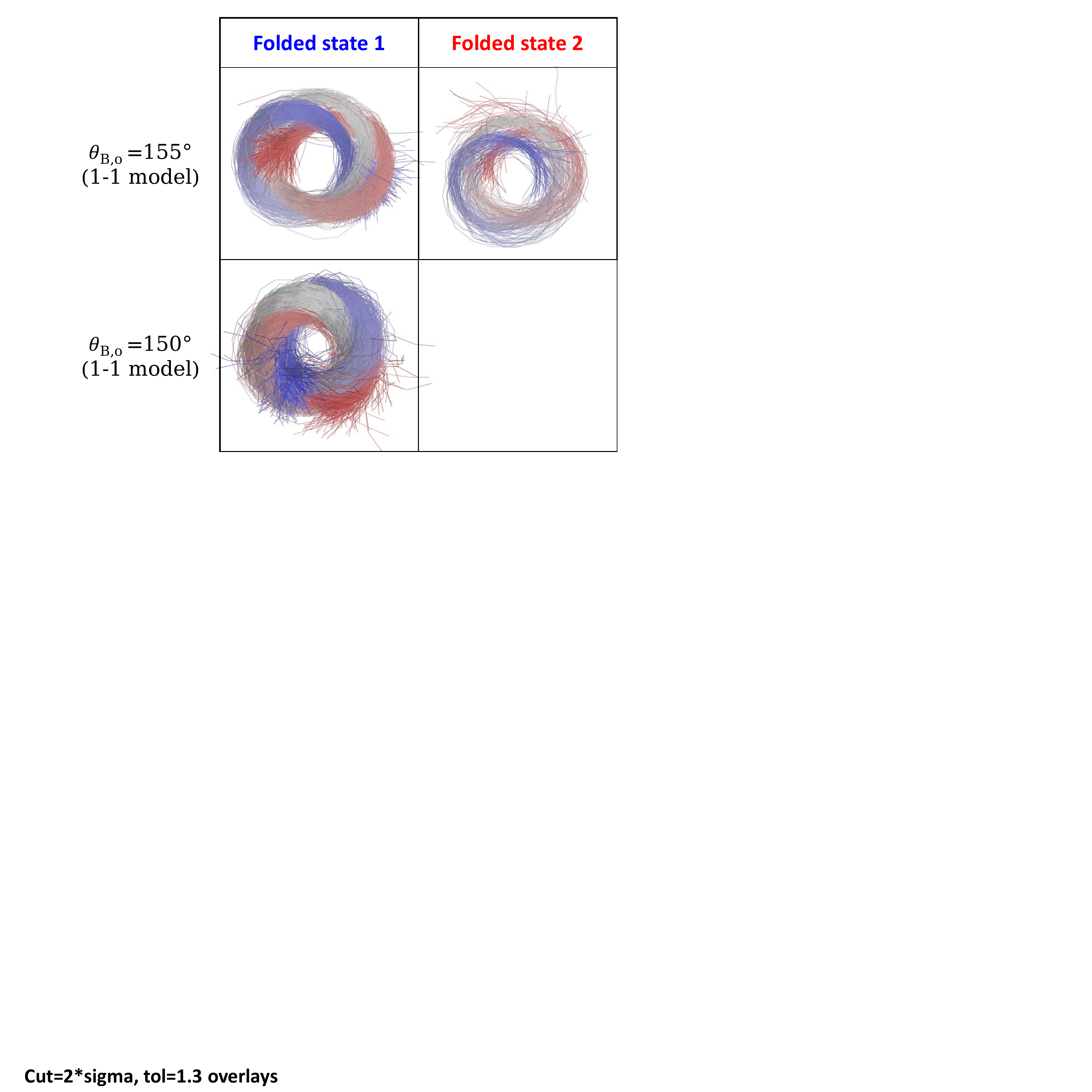"}
\label{fig:SI_tol13_cut20_vary_theta_SC_ensembles}
\caption{Visualization of representative structures in each knotted state for 1-1 models, with native contact cutoff of $2\sigma$, native distance tolerance factor of 1.3, and varying $\theta_{B,o}$ indicate that the states are stable, distinct, and configurationally well-defined. For clarity, side chains are not shown. Structures are selected across all temperatures with a stride of 12500 frames. Structures are aligned with respect to end-to-end and mirror symmetry to the reference medoid in each case, which is the enantiomer whose cluster size was largest.}
\end{figure}

\newpage
\section{Unweighted conformational state heat capacities} \label{SI_unweighted_Cv}

\begin{figure}[H]
\includegraphics[width=6.5in]{"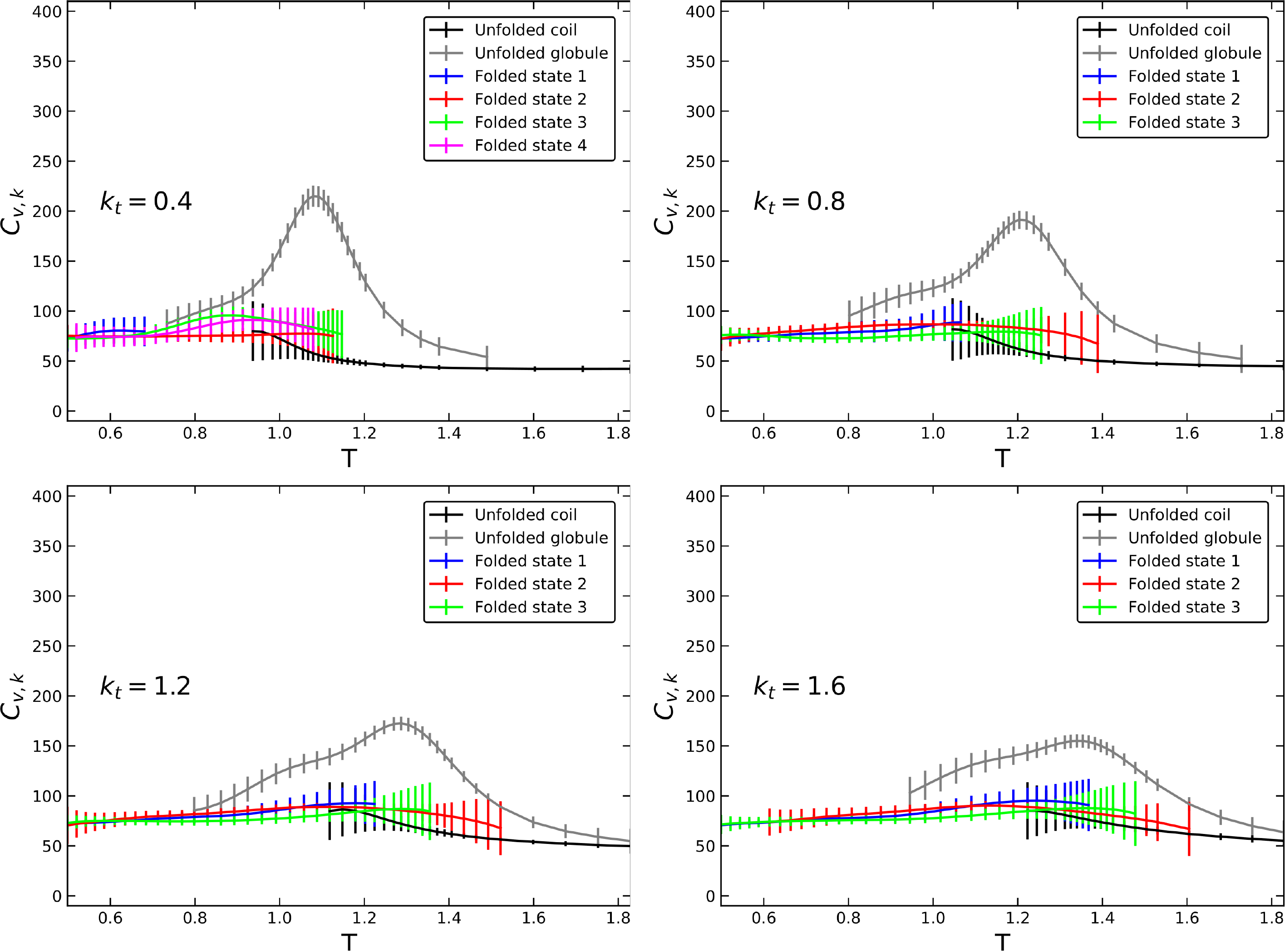"}
\caption{Unweighted heat capacities computed for each configurational state in linear homo-oligomers with $\theta_{B,o}=155^\circ$ and varying torsion stiffness show constant $C_{v,k}$ within uncertainty for all states except the unfolded globule. However, the peak within the globule state is flattened with increasing $k_t$, due to a restriction of conformational space. Colors correspond to the folded state classifications used throughout this work, with a native contact cutoff of 2$\sigma$ and tolerance factor of 1.3. Only data points with an effective number of samples greater than 50, as determined from MBAR weights~\cite{shirts_statistically_2008}, are displayed. Error bars are the standard deviation from bootstrapping the energies within each configurational state.}
\label{fig:SI_unweighted_Cv_vary_kt}
\end{figure}

\begin{figure}[H]
\includegraphics[width=6.5in]{"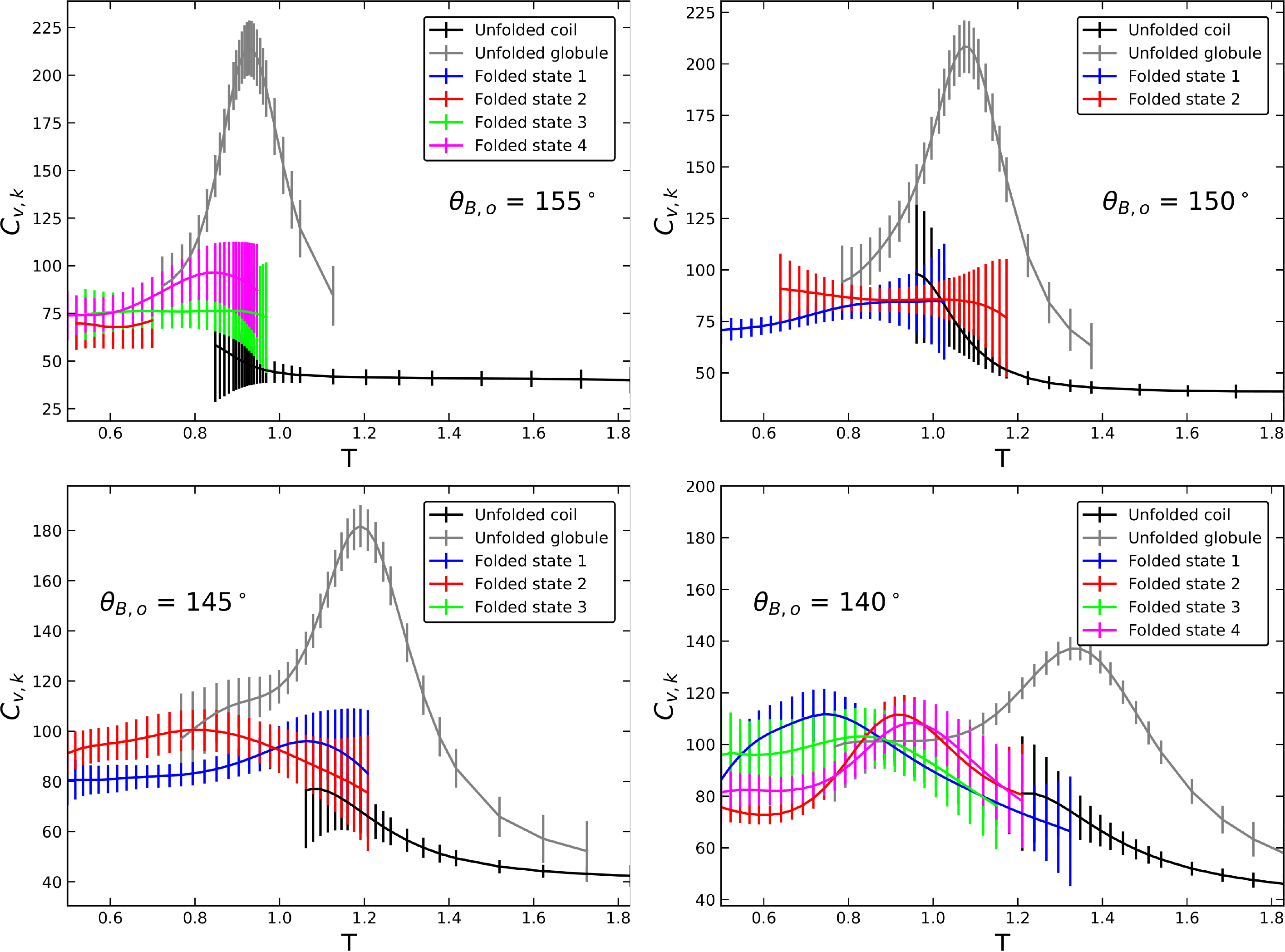"}
\caption{Unweighted heat capacities computed for each configurational state in linear homo-oligomers with $k_t=0.0$ and varying $\theta_{B,o}$ show approximately constant $C_{v,k}$ within uncertainty for all states except the unfolded globule. Only data points with an effective number of samples greater than 50, as determined from MBAR weights, are displayed.}
\label{fig:SI_unweighted_Cv_vary_theta}
\end{figure}

\begin{figure}[H]
\includegraphics[width=6.5in]{"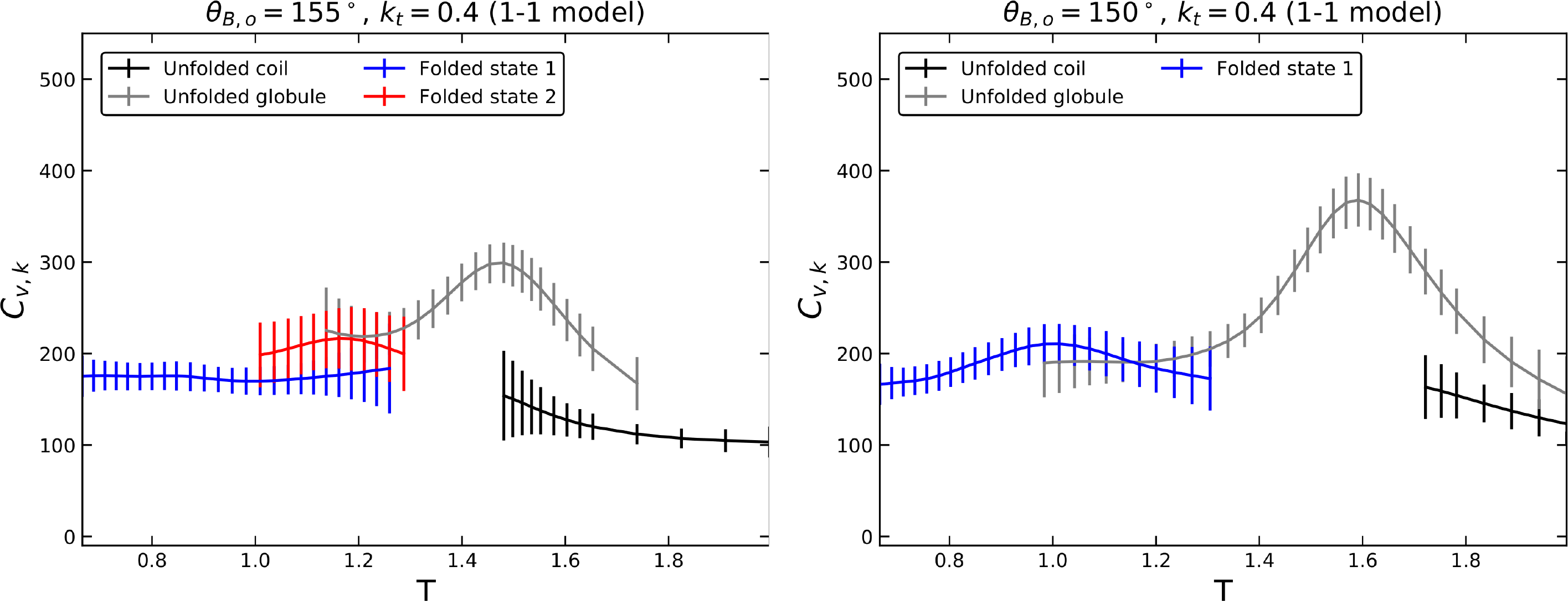"}
\caption{Unweighted heat capacities computed for each configurational state in 1-1 homo-oligomers with $k_t=0.4$ and varying $\theta_{B,o}$ show constant $C_{v,k}$ within uncertainty for all states except the unfolded globule. Only data points with an effective number of samples greater than 50, as determined from MBAR weights, are displayed.}
\label{fig:SI_unweighted_Cv_vary_theta_SC}
\end{figure}

\newpage
\section{Effect of contact definitions on state populations} \label{SI_contact_tolerance}

\begin{figure}[H]
\includegraphics[width=6.5in]{"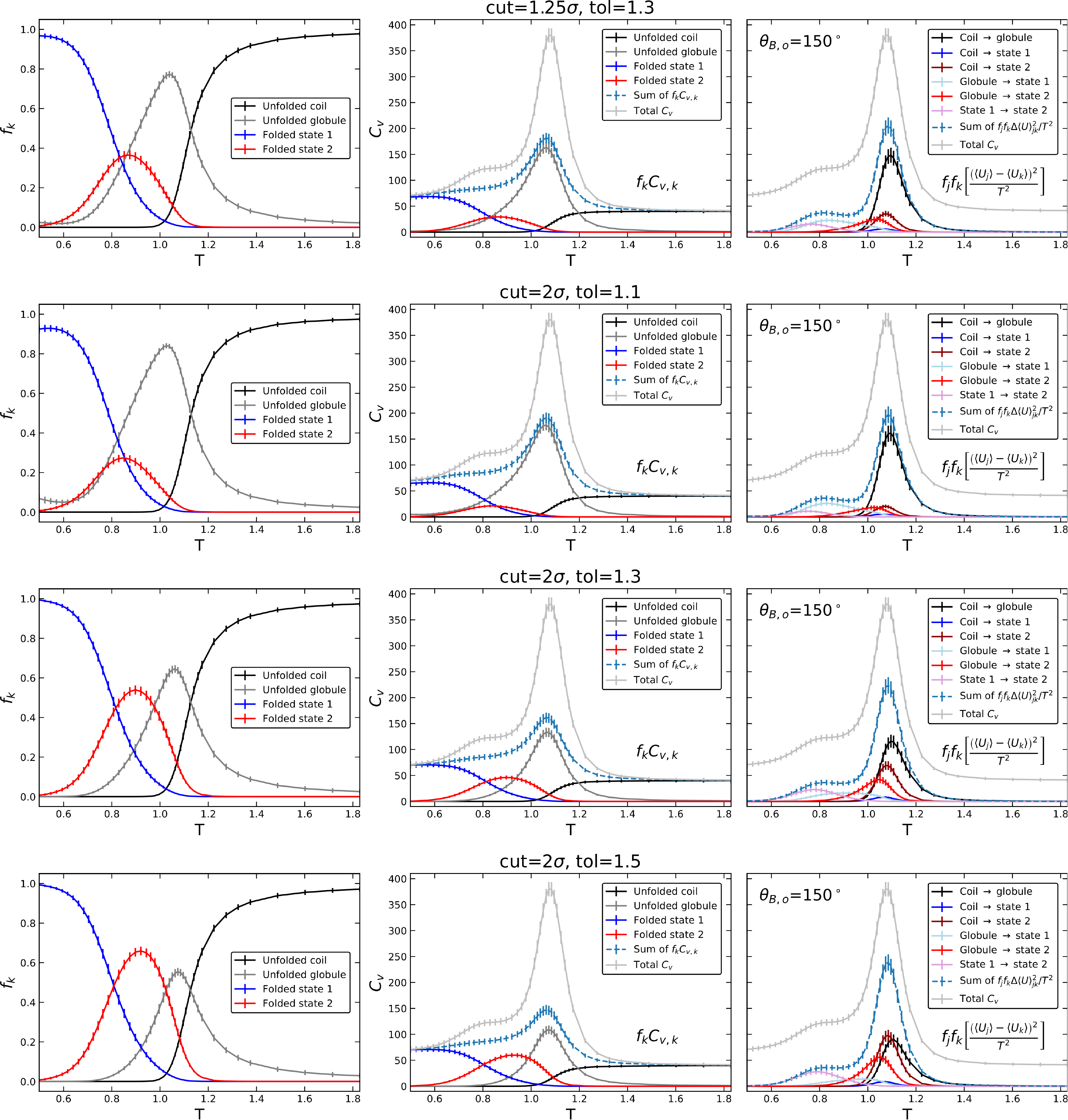"}
\caption{The population fraction versus temperature curves and heat capacity decompositions for the $\theta_{B,o}=150^\circ$ linear model with $k_t=0$ are compared for contact distance multiplicative tolerance factors of 1.1, 1.3, and 1.5 with a contact distance cutoff of $2\sigma$, and tolerance factor of 1.3 with distance cutoff of $1.25\sigma$. The population distribution of folded state 1, the native $3_1$ knot, at low temperature is only marginally affected by the contact tolerance. However, at intermediate temperatures, the contact tolerance significantly shifts the populations of globule and folded state 2. Too strict of a contact criteria incorrectly classifies the knotted states as random globules, and thereby suppresses the state-to-state transition peaks and increases the intra-state contribution from the random globule.}
\label{fig:SI_theta150_vary_tol}
\end{figure}

\begin{figure}[H]
\includegraphics[width=6.5in]{"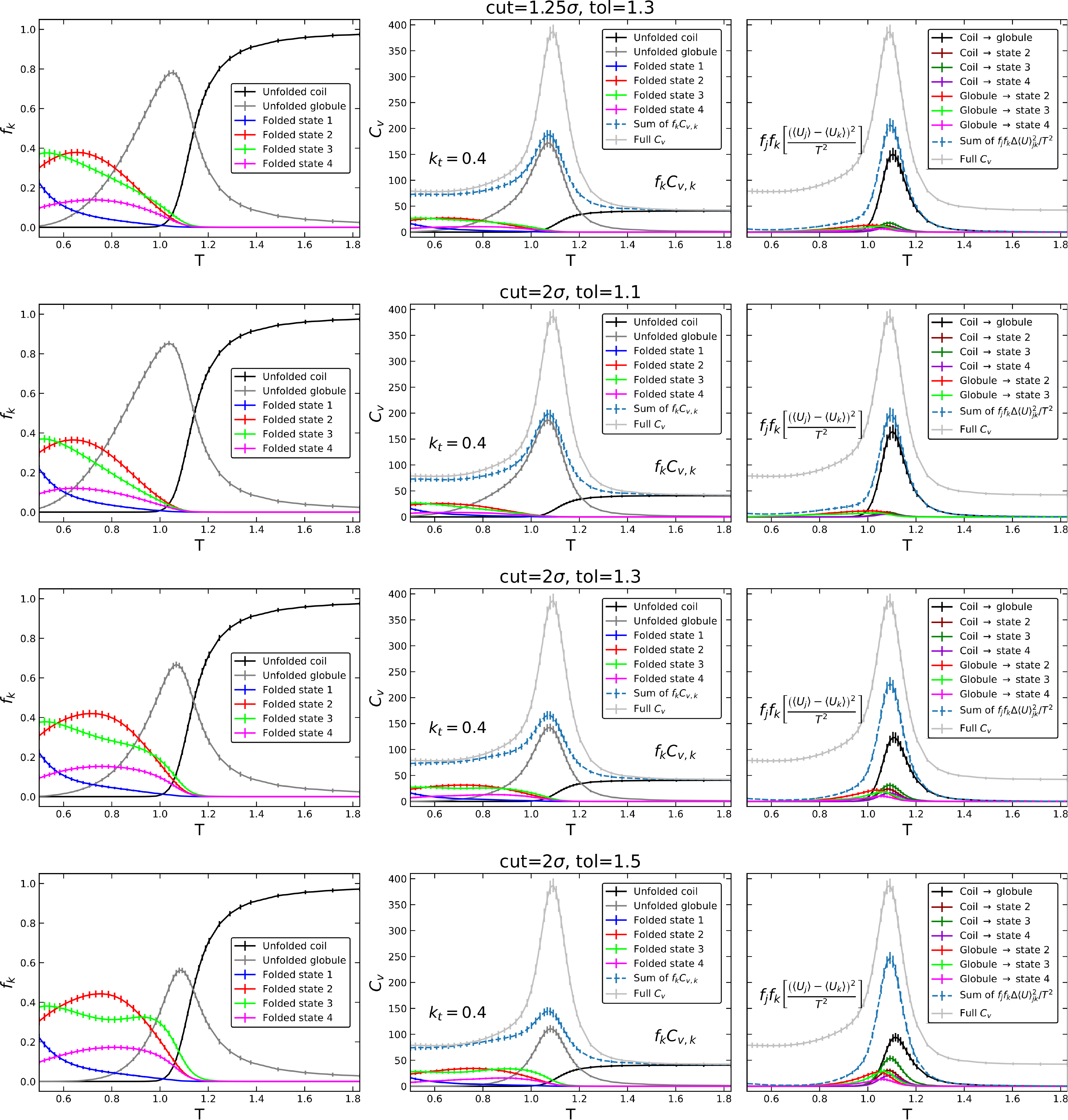"}
\caption{The population fraction versus temperature curves and heat capacity decompositions for the $\theta_{B,o}=155^\circ$ linear model with $k_t=0.4$ are compared for contact distance multiplicative tolerance factors of 1.1, 1.3, and 1.5 with a contact distance cutoff of $2\sigma$, and tolerance factor of 1.3 with distance cutoff of $1.25\sigma$. The relative populations of the 4 knotted states are not significantly affected by the contact definitions. However, too strict of a contact criteria incorrectly classifies the knotted states as random globules, and thereby suppresses the state-to-state transition peaks and increases the intra-state contribution from the random globule.}
\label{fig:SI_theta155_vary_tol}
\end{figure}

\newpage
\section{Radius of gyration by conformational state} \label{SI_Rg_vs_T}

\begin{figure}[H]
\includegraphics[width=6.5in]{"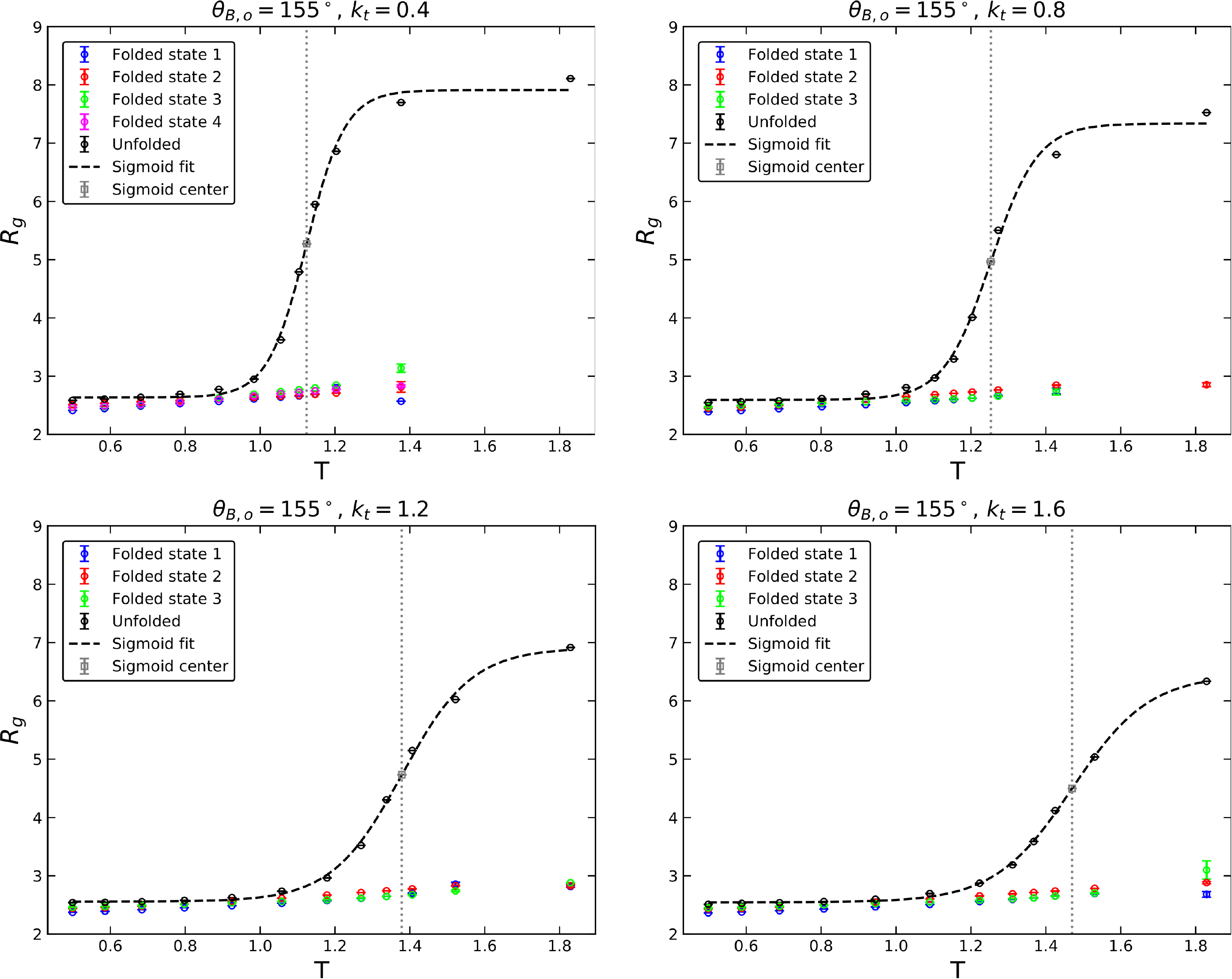"}
\caption{Mean radius of gyration as a function of temperature for each conformational state in linear homo-oligomers with $\theta_{B,o}=155^\circ$ and varying torsion stiffness. Colors correspond to the folded state classifications used throughout this work, with a native contact cutoff of 2$\sigma$ and tolerance factor of 1.3. Error bars are the standard error of the mean.}
\label{fig:SI_rg_all_states_vary_kt}
\end{figure}

\begin{figure}[H]
\includegraphics[width=6.5in]{"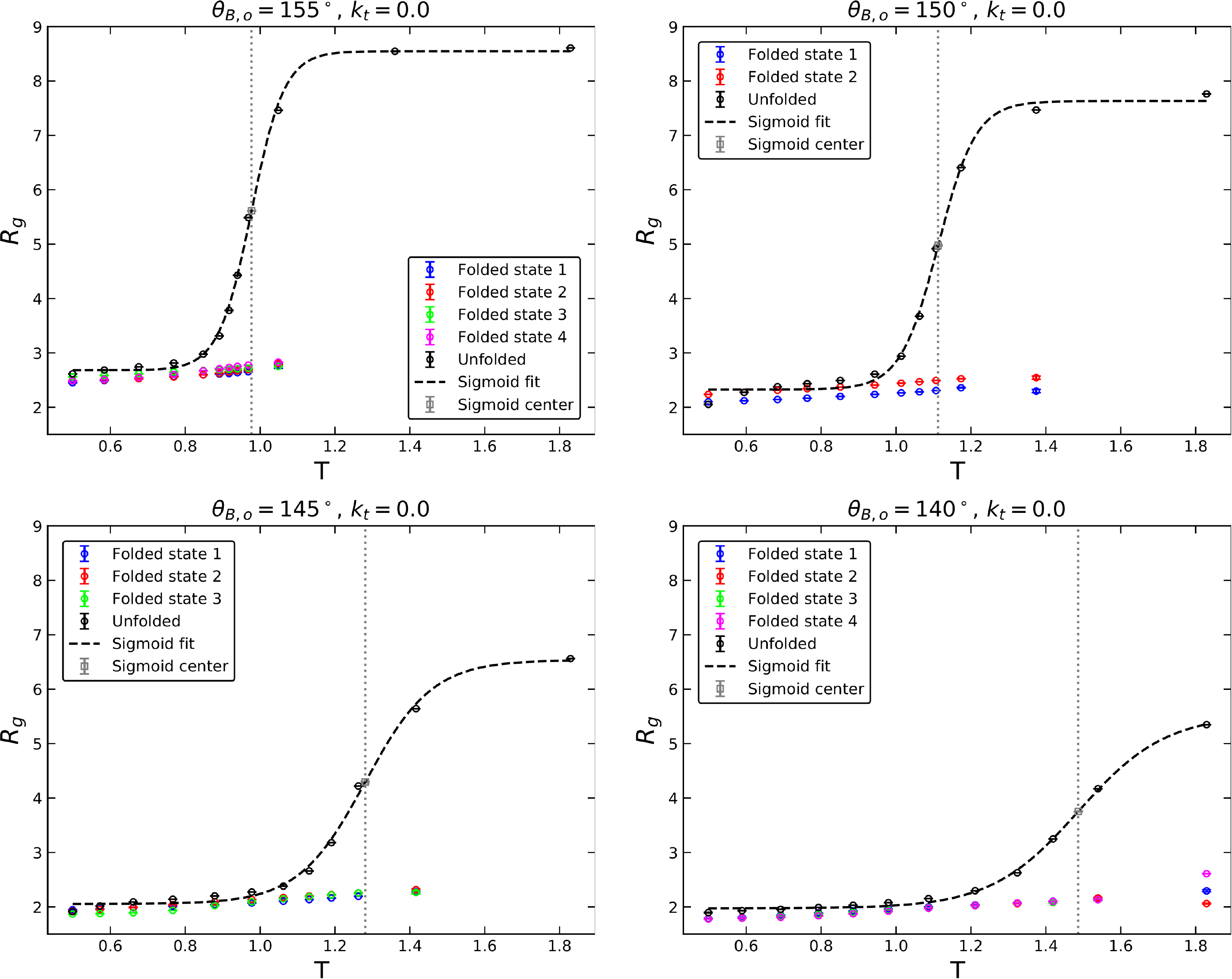"}
\caption{Mean radius of gyration as a function of temperature for each conformational state in linear homo-oligomers with $k_t=0.0$ and varying $\theta_{B,o}$.}
\label{fig:SI_rg_all_states_vary_theta}
\end{figure}

\begin{figure}[H]
\includegraphics[width=6.5in]{"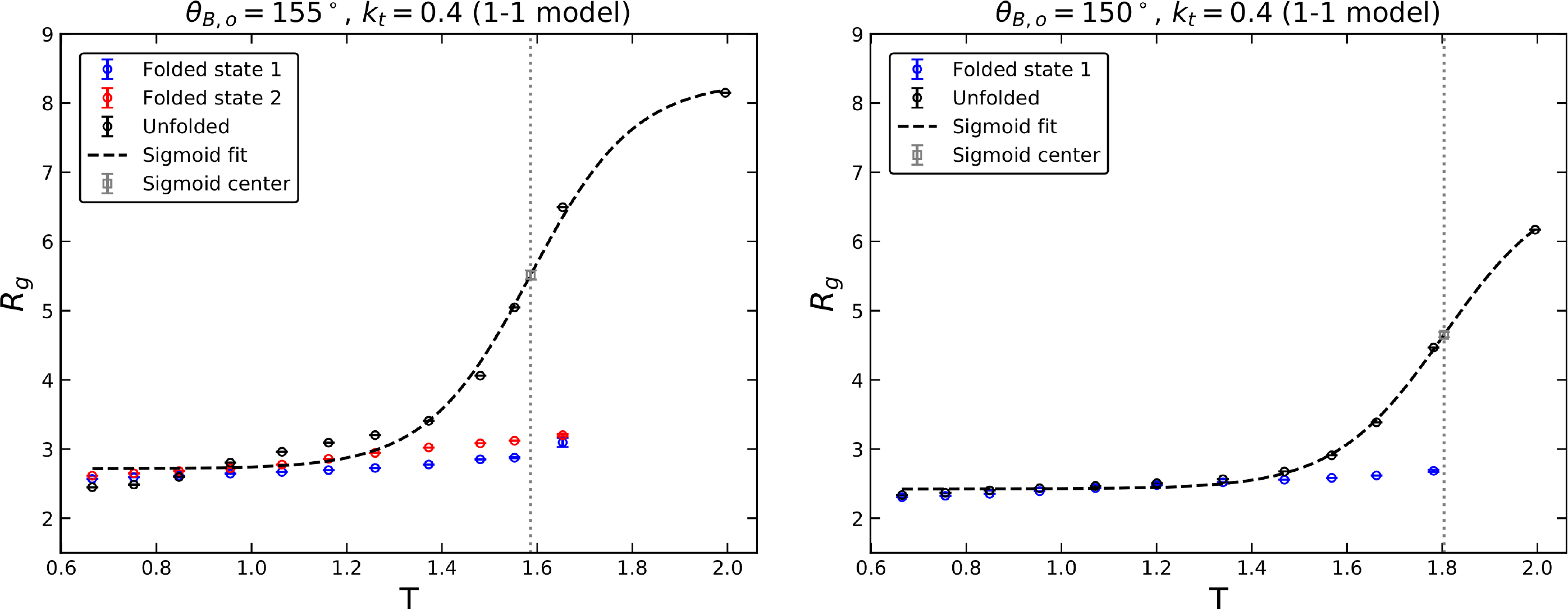"}
\caption{Mean radius of gyration as a function of temperature for each conformational state in 1-1 homo-oligomers with $k_t=0.4$ and varying $\theta_{B,o}$.}
\label{fig:SI_rg_all_states_sc}
\end{figure}

\begin{equation}
    R_g(T) = \frac{R_{g,o}+R_{g,1}}{2}-\frac{R_{g,o}-R_{g,1}}{2}\left[ \tanh{\left( \frac{T-T_0}{d} \right) } \right]
    \label{eqn:rg_sigmoid_fit}
\end{equation}

\begin{table}[H]
  \caption{$R_g$ fitting parameters}
  \label{table:rg_fitting_params}
  \begin{tabular}{||cc||ccccc||}
    \hline
    $\mathbf{\theta_B}$ \textbf{(degrees)} & $\mathbf{k_t}$ & $\mathbf{T_o}$ & $\mathbf{R_{g,o}}$ & 
    $\mathbf{R_{g,1}}$ & $\mathbf{d \times 1000}$ & $\mathbf{R_g(T_o)}$ \\
    \hline
    155 & 0.0 & 0.9771 & 2.6793 & 8.5484 & 1.5334 & 5.6139 \\
    \hline
    150 & 0.0 & 1.1125 & 2.3277 & 7.6305 & 1.7531 & 4.9791 \\
    \hline
    155 & 0.4 & 1.1243 & 2.6349 &	7.9097 & 1.8082	& 5.2723 \\
    \hline
    155 & 0.8 & 1.2530 & 2.5934 & 7.3368 & 2.1745 & 4.9651 \\
    \hline
    145 & 0.0 & 1.2808 & 2.0543 & 6.5358 & 3.1167 & 4.2950 \\
    \hline    
    155 & 1.2 & 1.3786 & 2.5570 & 6.9084 & 3.2675 & 4.7327 \\
    \hline
    155 & 1.6 & 1.4702 & 2.5468 & 6.4489 & 3.5940 & 4.4978 \\
    \hline
    155 (sc) & 0.4 & 1.5865 & 2.7152 & 8.3194 & 3.8140 & 5.5173 \\
    \hline
    150 (sc) & 0.4 & 1.8041 & 2.4227 & 6.8843 & 4.0100 & 4.6535 \\
    \hline
    140 & 0.0 & 1.4867 & 1.9736 & 5.5362 & 4.0734 &	3.7549 \\
    \hline
  \end{tabular}
\end{table}

Table~\ref{table:rg_fitting_params} lists the sigmoid fitting parameters for each parameter set. Data is sorted by sigmoid width parameter $d$, which characterizes the sharpness/cooperativity of the coil-to-globule transition. A smaller $d$ signifies a narrower characteristic temperature range and thus higher cooperativity. 

\newpage
\bibliography{references}
\end{singlespace}